\documentclass[fleqn,usenatbib]{mnras}
\usepackage{newtxtext,newtxmath}
\usepackage[T1]{fontenc}

\usepackage{graphicx}	
\usepackage{amsmath}	
\usepackage[british]{babel} 
\usepackage{placeins}
\usepackage{xcolor}
\usepackage{natbib}
\usepackage{appendix}
\usepackage{diagbox}

\usepackage{ae,aecompl}
\usepackage{lscape}

\newcommand{\pddd}{\,$P(q_{3D},r_{3D})$}
\newcommand{\pdd}{\,$P(q_{2D})$}
\newcommand{\qddd}{\,$q_{3D}$}
\newcommand{\rddd}{\,$r_{3D}$}
\newcommand{\sddd}{\,$s_{3D}$}
\newcommand{\qdd}{\,$q_{2D}$}
\newcommand{\vek}[1]{\mbox{\boldmath $#1$}}
\newcommand{\bc}[1]{\left\{ #1 \right\}}
\newcommand{\br}[1]{\left( #1 \right)}
\newcommand{\eq}[1]{\begin{equation}  #1 \end{equation}}
\newcommand{\eqa}[1]{\begin{eqnarray}   #1 \end{eqnarray}}

\newcommand{\hst}{{\it HST}}
\newcommand{\sersic}{S\'{e}rsic}
\newcommand{\galfit}{{\sc Galfit}}
\newcommand{\sextractor}{{\sc SExtractor}}
\newcommand{\galapagos}{{\sc Galapagos}}

\title[Shapes of disc-dominated galaxies]{Redshift and stellar mass dependence of intrinsic shapes
of disc-dominated galaxies from COSMOS observations below $z = 1.0$}

\author[Hoffmann et al.]
{K. Hoffmann$^{1,2}$\thanks{E-mail: kai.d.hoffmann@gmail.com},
C. Laigle$^{3}$,
N. E. Chisari$^{4}$, 
P. Tallada-Cresp\'{i}$^{5,\dagger}$,
R. Teyssier$^2$,
Y. Dubois$^3$, \newauthor
J. Devriendt$^6$
\\
$^{1}$ Institute of Space Sciences (ICE, CSIC), Campus UAB, Carrer de Can Magrans, s/n, 08193 Barcelona, Spain\\
$^{2}$ Institute for Computational Science, University of Zurich, Winterthurerstr. 190, 8057 Zürich, Switzerland\\
$^{3}$ CNRS and UPMC Univ. Paris 06, UMR 7095, Institut d'Astrophysique de Paris, 98 bis Boulevard Arago, F-75014 Paris, France \\
$^{4}$ Institute for Theoretical Physics, Utrecht University, Princetonplein 5, 3584 CC Utrecht, The Netherlands \\
$^{5}$ Centro de Investigaciones Energ\'eticas, Medioambientales y Tecnol\'ogicas (CIEMAT), Avenida Complutense 40, 28040 Madrid, Spain \\
$^{6}$ Astrophysics, University of Oxford, Denys Wilkinson Building, Keble Road, Oxford, OX1 3RH, UK \\
$^{\dagger}$ Also at Port d'Informaci\'{o} Cient\'{i}fica (PIC), Campus UAB, C. Albareda s/n, 08193 Bellaterra (Barcelona), Spain
}

\date{Accepted XXX. Received YYY; in original form ZZZ}

\pubyear{2020}

\begin{document}
\label{firstpage}
\pagerange{\pageref{firstpage}--\pageref{lastpage}}
\maketitle

\begin{abstract}
The high abundance of disc galaxies without a large central bulge challenges predictions of current
hydrodynamic simulations of galaxy formation. We aim to shed light on the formation of these objects
by studying the redshift and mass dependence of their intrinsic 3D shape distributions
in the COSMOS galaxy survey below redshift $z=1.0$.
This distribution is inferred from the observed distribution of 2D shapes, using a reconstruction
method which we test using hydrodynamic simulations. Our tests reveal a moderate bias for the
inferred average disc circularity and relative thickness, but a large bias on the dispersion of
these quantities.
Applying the reconstruction method on COSMOS data, we find variations of the average disc
circularity and relative thickness with redshift of around $\sim1\%$ and $\sim10\%$ respectively,
which is comparable to the error estimates on these quantities.
The average relative disc thickness shows a significant mass dependence which can be accounted
for by the scaling of disc radius with galaxy mass.
We conclude that our data provides no evidence for a strong dependence of the average circularity
and absolute thickness of disc-dominated galaxies on redshift and mass that is significant
with respect to the statistical uncertainties in our analysis.
These findings are expected in the absence of disruptive merging or feedback events that would
affect galaxy shapes. They hence support a scenario where
present-day discs form early ($z>1.0$) and subsequently undergo a tranquil evolution
in isolation. However, more data and a better understanding of systematics are needed to
reaffirm our results.
\end{abstract}




\section{Introduction}
The structure of late-type galaxies is comprised of different
components with distinct kinematic and morphological characteristics.
The defining component is the thin stellar disc with its spiral arm over densities
\citep[e.g.][]{Hubble26, Hubble36, Vaucouleurs59a, Vaucouleurs59b, Freeman70, vanderKruit82}.
Observations at low redshifts and in our own Milky Way revealed that this thin disc is often
enclosed by a thick stellar disc as well as a stellar halo with relatively low densities
\citep[e.g.][among others]{Carollo10,Trujillo13,Martinez19}.
Near the galactic center, the spiral arms of most discs transition either into a bar or into a bulge.
The contribution of these components to the overall mass and morphology of a given galaxy is
determined by the galaxies' formation history \citep[e.g.][]{Binney08, Vogelsberger20}.
Therefore, studying how the shapes of late-type galaxies (hereafter referred to as disc galaxies)
are distributed at different epochs of the Universe can provide significant insights into the
processes which dominate galaxy formation, as detailed below.

During the recent years it was pointed out that the observed abundance of a particular type of objects,
namely disc-dominated galaxies with only a small or no central bulge, challenges our current understanding
of galaxy formation in the $\Lambda$CDM model \citep[e.g.][]{Kormendy10}.
As a matter of fact, bulges are expected to form in galaxies for several reasons.
\cite{vdBosch98} suggested that bulges can form "inside out" from the low angular momentum components of the initial gas overdensities in which the galaxies are born \citep[see also][]{Kepner99}.
In another popular scenario, bulges result from the redistribution of angular momentum within the disc
by central bars, spiral arms or bending instabilities which distort the stellar orbits
\citep{Kormendy04, Debattista06}. This so-called ``secular evolution'' leads typically to
the formation of a small ``pseudo'' bulge, but may also promote the emergence of a larger
``classical'' bulge when occurring at high redshifts \citep{Elmegreen08, Bournaud16}.
A third and important scenario for bulge formation is the accretion of satellite galaxies,
globular clusters or clumpy streams of cold gas, which can move low angular momentum material to the center of the
disc (e.g. through violent disc instabilities) without destroying it \citep[e.g.][]{Walker96, Dekel09b, Hopkins09, Dubois12, Kretschmer20}.
In relatively rare occasions even major mergers can lead to the formation of a bulge-dominated discs, as 
discs can survive or reform if one of the progenitors is gas-rich \citep{Toomre77, Negroponte83, Hernquist92, Naab03, Hopkins08, Hopkins09,Jackson20},
especially in the case of prograde mergers \citep{Martin18}.
Since mergers and cold streams are believed to play a significant role in the formation and growth of disc galaxies
\citep[e.g.][]{Baugh96, Kauffmann93, Aguerri01, Moral06}
it is difficult to explain how a significant fraction of high mass discs (i.e. $M_\star \gtrsim 10^{10} M_\odot$)
could grow without forming a large bulge \citep{Kautsch06, Kautsch09, Weinzirl09, Buta15}.

One possible solution to this problem are feedback processes, driven by starbursts, supernovae
or active galactic nuclei, which can suppress the formation
of a bulge by removing low angular momentum material from the disc, in particular after gas-rich major mergers
\citep[e.g.][]{Governato10, Brook11, Brook12, Hopkins12, Ubler14, Dubois16, Grand17}.
However, simulations suggest that this suppression is only effective for discs with stellar masses of the
Milky Way ($\simeq 6 \times 10^{10} {\rm M}_\odot$) or below, but not at higher masses \citep{Brooks16}.
An additional challenge for formation scenarios including major mergers are the low density environments in which
bulgeless discs typically reside \citep[e.g.][]{Grossi18}. \citet{Kormendy10} argued
that this finding speaks for a gentle, rather than a violent mass accretion history \citep[see also][]{Kormendy16, Jackson20}.
Recently, \citet{Peebles20} therefore discussed an alternative solution to the problem, according to which small 
scale non-Gaussianities in the initial conditions of a warm or mixed dark matter universe could lead to a large
fraction of bulge-less discs. However, this approach remains to be explored with simulations of galaxy formation.

With this work we aim to shed light on the formation of bulge-less discs.
We therefore use a novel approach for discriminating the first
scenario in which feedback processes suppress bulge formation after mergers from alternative, less violent mass accretion scenarios by comparing the distributions
of shapes of observed disc-dominated galaxies at different redshift and
stellar mass ranges. This approach relies on the hypothesis that disruptive events,
such as mergers
and strong feedback should dramatically affect the morphology of the remaining galactic
disc, besides the absence of a bulge. 
Such changes can occur in the form of an increase in disc thickness
which results from vertical heating by feedback and merging events
\citep[e.g.][]{Quinn93, Grand16} or tidal debris of mergers \citep{Abadi03}.
The latter can furthermore lead to large sub-structures which decrease the
disc circularity, such as accreted satellites or spiral arms that form due to
tidal interactions during merging events \citep{Springel05, Robertson06, Peschken20}.
We therefore expect mergers and strong feedback to cause a significant mass and redshift
dependence of the discs thickness and circularity.

On the contrary, if the bulge-less discs underwent an early and regular accretion of mass,
for instance through infalling smooth streams of cold gas at high redshifts followed
by a calm evolution without major mergers, their morphologies should exhibit a much weaker
or no dependency on mass and redshift.
It has been discussed in the literature that a secular redshift evolution of the
disc thickness can be expected from vertical heating 
by bars, spiral arms, stellar clumps and giant molecular clouds
\citep[e.g.][]{Bournaud09, Saha10, Aumer16}.
However, \citet{Grand16} find that the effect of such internal perturbations is weak compared
to those caused by merging events. Moreover, \citet{Park21} do not find a significant redshift
dependence of the disc thickness in an ensemble of simulated disc galaxies that
underwent a quiescent growth without significant mergers since $z=1.0$.

A challenging aspect of studying observed galaxy morphologies
is to interpret the projected two-dimensional (2D) galaxy shape distributions
in terms of 3D models of galaxy formation.
%
In this work we address this challenge by reconstructing
the distribution of 3D galaxy shapes from the observed distribution of 2D shapes.
We thereby assume a simple ellipsoidal model for the 3D light distribution within a
given galaxy. The 3D shape of the ellipsoids is thereby fully characterized by two
of the three possible ratios between the major,
intermediate and minor axes ($A_{3D}$, $B_{3D}$ and $C_{3D}$ respectively),
\eq{
\label{eq:3D_axes_ratios}
q_{3D} \equiv \frac{B_{3D}}{A_{3D}}, ~~\mbox{ }~~
r_{3D} \equiv \frac{C_{3D}}{B_{3D}}, ~~\mbox{ }~~
s_{3D} \equiv \frac{C_{3D}}{A_{3D}}.
}
For disc galaxies the \qddd\ parameter can be regarded as a measure for the circularity, while \rddd\ and \sddd\
both quantify the relative disc thickness. The 2D galaxy shapes of the projected ellipsoid are described by the axis ratio
\begin{equation}
    q_{2D} \equiv \frac{B_{2D}}{A_{2D}}.
    \label{eq:2D_axes_ratio}
\end{equation}
The main advantage of such an ellipsoidal approximation is that the distribution of 2D axis ratios can be predicted from a
given model for the 3D axis ratio distribution at low computational cost.
This allows for the reconstruction of the 3D axes
ratio distribution by tuning the corresponding model parameters such that the predicted distribution of 2D axis ratios
matches observations.
%
This reconstruction technique dates back to \citet{Hubble26} who derived first constraints
on the intrinsic 3D axis ratios of galaxies by modeling them as oblate ellipsoids. Over the last
century it has been shown that, despite its simplicity, this methodology reproduces the observed
2D axis ratio distribution of late- as well as of early-type galaxies, assuming oblate or prolate
ellipsoidal models \citep[e.g.][]{Sandage70, Binney78, Noerdlinger79}.
The agreement with observed axis ratio distributions was further improved by modeling galaxies as triaxial ellipsoids,
which allowed for more detailed interpretations of the observations \citep{Benacchio80, Binney81, Lambas92}.
These early studies where continued using larger samples to study the relation between intrinsic galaxy shapes and
properties, such as size, luminosity and color in the local universe, observed by the Sloan Digital Sky Survey
\citep[SDSS, e.g.][]{Ryden04, Vincent05, Padilla08,Rodriguez13}. The evolution of intrinsic shapes with redshift has
been studied with the same approach in the galaxy surveys SDSS, 3D-HST, GOODS, COSMOS and CANDELS \citep{Yuma11, Yuma12, Holden12, Chang13, Wel14, Takeuchi15, Zhang19, Satoh19}.
Physical interpretations drawn from these reconstructed axis ratio distributions rely on the validity of the ellipsoidal model
for the 3D galaxy shapes as well as on the accuracy of the model for the 3D axis ratio distribution. In addition,
observational sources of systematics on the observed 2D axis ratio distribution, for example those induced by dust extinction within the galaxy, need to be taken into account in the analysis.

The objective of our analysis is therefore two fold.
Our main goal is to constrain potential explanations for the absence of large bulges in disc galaxies based on the reconstructed 3D shape distribution
of disc-dominated galaxies as outlined earlier in this section. Our investigation is based on data from the COSMOS galaxy
survey \citep{scoville07}, which provides excellent space-based imaging of galaxies over a wide range of redshifts and stellar masses, and is
therefore ideal for our analysis.

However, the assumptions on which the reconstruction of the 3D shape distribution is based are highly
simplistic which may induce uncertainties in our as well as in previous analyses.
Our second goal is therefore to test several of these assumptions and to assess the overall performance
of the reconstruction method. For that purpose we employ for the first time two state-of-the-art
cosmological hydrodynamic simulations of galaxy formation, Horizon AGN and Illustris TNG.
These simulations provide 3D as well as projected 2D galaxy morphologies
which allows for an examination of the reconstruction method under controlled conditions.

This paper is structured as follows.
In Section \ref{sec:data}, we present the galaxy catalogues from the COSMOS survey and the hydrodynamic
simulations and  explain our sample selection.
Section \ref{sec:method} provides details on the shape reconstruction method together with validations
of the model assumptions and accuracy tests.
The method is then applied on the COSMOS data in Section \ref{sec:results}.
A summary of our results can be found together with our conclusions in Section \ref{sec:summary}.

\section{Data}
\label{sec:data}

\subsection{COSMOS observations}
\label{sec:data:cosmos}

The $2 \deg^2$ COSMOS field \citep{scoville07} has been observed extensively by different ground and space based
telescopes, including Hubble, Spitzer, VISTA, CFHT and Subaru. The joint analysis of these observations
led to different catalogues providing estimates of galaxy properties such as stellar masses, star formation rates and
morphological characteristics over a large range in redshift and luminosity. Our analysis is based on three of
these catalogues, which are described below.
%
\begin{figure*}
\centering\includegraphics[width=17.5 cm, angle=0]{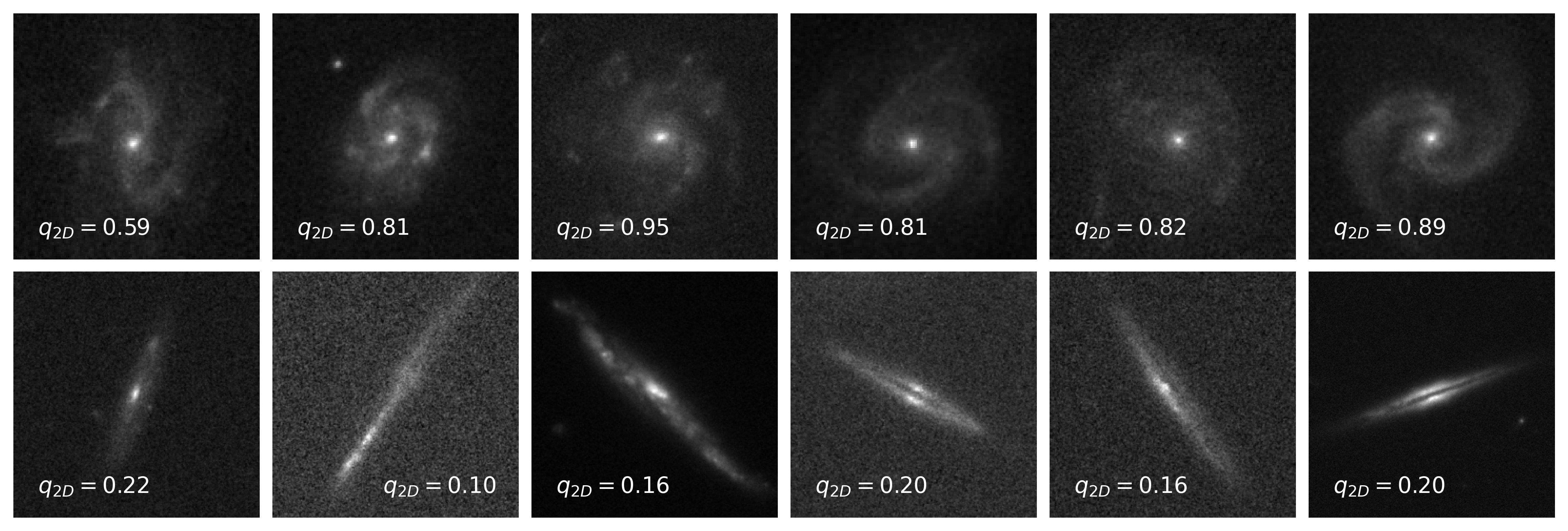}
\caption{Random examples of ACS images of late-type galaxies in our volume limited COSMOS sample in the
redshift range $0.2<z<0.4$. The galaxies are selected to be disc-dominated according to the ZEST morphological
classification scheme (Section \ref{sec:data:cosmos:morph}). Galaxies in the top and bottom panels
are further classified by ZEST as being face-on and edge-on oriented respectively.
The three columns on the left (right) show galaxies in our low (high) mass sample,
with stellar masses below (above) $M_\star^{cut}=10^{10.35}M_\odot$.  The 2D axis ratios $q_{2D}$,
provided in the ACS-GC catalogue, were obtained from single \sersic\ profile
fits (Section \ref{sec:data:cosmos:shapes}). The image sizes are adjusted for each galaxy.}
\label{fig:acs_images}
\end{figure*}

\subsubsection{Photometry}
\label{sec:data:cosmos:photo}

The public COSMOS2015 catalogue \footnote{\url{https://www.eso.org/qi/}} \citep{Laigle16}
comprises photometry in $30$ bands, covering ultra-violet to mid-infrared wavelengths.
%
In our analysis we use redshift,  stellar mass and specific star formation rate (sSFR)\footnote{We are aware that this sSFR estimates is not very robust, but we checked that it is sufficient for the purpose of our analysis.} estimates provided
in this catalogue, which were derived for each galaxy by fitting templates of spectral energy distributions
(SEDs) to the photometric data \citep{Ilbert06}.
%
We discard objects which are i) residing in regions flagged as ``bad'' (mostly because they are close to stars or to the edge of the field)
ii) saturated and iii) not classified as galaxies by requiring the corresponding catalogue flags to be
\verb|[flag_hjmcc, flag_peter, type] = [0,0,0]|.
We further impose cuts at the limiting AB magnitudes in the near-infrared $K_s$-band of $24.0$ and $24.7$
in the {\it deep} and ultra {\it ultra-deep} fields respectively. These magnitudes are
defined within a fixed $3''$ diameter aperture ($\verb|Ks_MAG_APER3|$) and the limits correspond to a $3 \sigma$ detection.
%
After applying these cuts the catalogue contains $252,527$ galaxies. For this sample the standard deviation
of the relative differences in redshift with respect to the \textit{z}COSMOS-bright spectroscopic control sample \citep{lilly07} is
$\sigma_{\Delta z / (1+z_s)} = 0.007$ (catastrophic failure fraction $\eta=0.5\%$). Note that this error is an optimistic estimate,
since it was inferred by comparison to spectroscopic redshifts of bright galaxies and is higher for dimmer objects.
However, due to the bright magnitude cuts used in this work (see Section \ref{sec:data:cosmos:main_sample})
we expect this inaccuracy to be a realistic estimate for the galaxies in our samples.
The accuracy on the stellar mass and star formation rate estimates is expected to be
$\sim 0.1$ dex and $\sim 0.2-0.6$ dex respectively at $z<1.5$ \citep{Laigle19}.
These values may be lower for our samples as we focus on bright objects only. Nevertheless, \citet{Laigle19}
find evidence that the strong scatter (and bimodality) in the $SFR$ estimates is mainly driven by the
inaccurate modelling of dust extinction within the galaxies at the SED-fitting stage (see their Figure~B3). 
We also find strong evidence of dust extinction in the galaxies from our sample (e.g. Fig. \ref{fig:qcm_cosmos}).
We therefore study the sSFR only for discs with high apparent axis ratios, which we consider to be inclined
towards a face-on orientation at which the impact of dust extinction (and dust extinction modelling) is expected
to be minimal.

\subsubsection{Shapes}
\label{sec:data:cosmos:shapes}

We use galaxy shape estimates from the public {\it Advanced Camera for Surveys General Catalog}
\citep[ACS-GC \footnote{\url{vizier.u-strasbg.fr/viz-bin/VizieR-3?-source=J/ApJS/200/9/acs-gc}},][]{Griffith12}.
This catalogue is based on {\it Hubble Space Telescope} (HST) imaging in the optical red $I_{AB}$ broad band filter F814W.
The absence of atmospheric distortions allows for an excellent image resolution, which is mainly limited by the width of the HST point spread function (PSF) of $0.085"$ in the F814W filter and the pixel scale of $0.03"$.
Sources were detected using the \galapagos\ software \citep{Haeussler11}, which runs \sextractor\ \citep{Bertin96} and
\galfit\ \citep{Peng02} in two subsequent steps. Galaxy shapes are described by the two-dimensional major over minor
axis ratios \qdd, which are derived by \galfit\ from fits of a single \sersic\ model to each objects image. The surface
brightness in this model is given by
\eq{
\Sigma(r) = \Sigma_e \
\exp \Biggl\{
-\kappa_n \
\Biggl[ \biggl(\frac{r}{r_e}\biggr)^{1/n}-1 \Biggr]
\Biggr\},
\label{eq:sersic_profile}
}
where $\Sigma_e$ is the surface brightness at the effective radius $r_e$ and the parameter $\kappa_n$ is
chosen such that $r_e$ encloses half of the total flux. The \sersic\ index $n$ quantifies the
concentration of the surface brightness profile. The 2D axis ratio \qdd\
(provided as \verb|BA_GALFIT_HI| in the catalogue) enters Equation (\ref{eq:sersic_profile}) via
$r= \sqrt{x^2 + (y/q_{2D})^2}$, where $x$ and $y$ are the coordinates on the major and minor axis respectively.
The modeled surface brightness profiles are convolved with the ACS PSF before being compared to the reference
observation during the fit. The morphological parameters from \galfit\ hence describe the intrinsic 2D galaxy shapes 
and do not require further PSF correction. We select objects from the catalogue which were classified by \sextractor\
as galaxies (\verb|CLASS_STAR_HI < 0.1|) and with good fits to the \galfit\ model (\verb|FLAG_GALFIT_HI=0|
and \verb|CHI2NU_HI < 2|). From the remaining sample we reject $65$ objects which have axis ratios equal to
zero or larger than unity or effective radii from \galfit\ above $750$ ACS pixels ($22.5''$). After these
cuts the final catalogue contains $128,365$ objects.

\subsubsection{Morphological classification}
\label{sec:data:cosmos:morph}

For the selection of disc-dominated galaxies we use the {\it Zurich Structure \& Morphology Catalog}\footnote{\url{irsa.ipac.caltech.edu/data/COSMOS/tables/morphology/cosmos_morph_zurich_1.0.tbl}}
which is derived from the same HST imaging data as the ACS-GC.
It provides a morphological classification for each galaxy, derived with the
{\it Zurich Estimator of Structural Type} and is referred to as ZEST catalogue in the following.
The ZEST classification is based on
a principal component analysis of five non-parametric diagnostics: asymmetry, concentration, Gini coefficient,
2nd-order moment of the brightest 20\% of galaxy pixels ($\textrm{M}_{20}$) and ellipticity
\citep[see][who also provide validations of their classification]{Scarlata06, Scarlata07, Sargent07}.
The catalogue contains galaxies in the COSMOS field brighter than $I_{AB}=24$.
We select objects, which
i) are classified as galaxies (\verb|[acs_mu_class, stellarity]==[1,0]|),
ii) do not reside in automatically or manually masked regions (\verb|[acs_mask, acs_masked]==[0,1]|)
and iii) are not flagged as unusable or spurious (\verb|[acs_clean, junkflag]=[1,0]|).
After applying these conditions the remaining sample contains $108,800$ galaxies.
The morphological classification is considered to be unreliable for galaxies
with half-light radii smaller than twice the size of the ACS F814W PSF (i.e. $0.17''$, that is $\gtrsim 5.6$ times larger than the ACS pixel size),
which we take into account in our sample selection (Section \ref{sec:data:cosmos:main_sample}).
We validate the morphological classification for the disc-dominated galaxies used in our analysis
in Section \ref{sec:data:cosmos:main_sample} and in Appendix \ref{app:discs_zest}.

\subsubsection{Matched catalogue}
\label{sec:data:cosmos:match}
%
Matching objects in these three COSMOS catalogues is not straightforward due to variations in common properties 
such as positions and magnitudes, which can lead to spurious mismatches. These variations can originate from
atmospheric distortions in the ground-based COSMOS2015 data, differences in the employed image analysis software
(i.e. {\sc SExtractor}  and \galfit) and its configuration, in the employed telescopes, cameras and filters, as well as
different quality cuts applied before matching.
In order to minimise the chance for mismatches, galaxies are matched based on angular positions as well as on magnitudes.
We start by matching objects in the COSMOS2015 and ACS-GC catalogue in three steps.
1) We select pairs of galaxies as candidate matches if their angular separation is $<0.6''$, which
is slightly below the typical seeing of the ground-based telescopes contributing to the COSMOS survey.
2) We discard candidate matches with a difference in brightness of more than $1.0$ magnitude.
3) We finally select the matches as those with the smallest difference in angular positions.
The COSMOS2015 magnitudes used in step 2) are measured in the Subaru $i+$ band within a fixed $3''$ aperture
(referred to as \verb|ip_mag_aper3| in the catalogue). They are compared to the \verb|SExtractor| magnitudes
\verb|MAG_BEST_HI| from ACS-GC which are defined as magnitudes measured within a flexible elliptical aperture
(referred to as \verb|MAG_AUTO|) or corrected isophotal magnitudes if contaminating sources are located in the vicinity.
The average wavelengths weighted by transmission in the Subaru $i+$ and ACS F814W filters
are $7683.88$ {\AA} and $8073.43$ {\AA} respectively, while the width and shape of their transmission curves
differ significantly. These differences are accounted for by the relatively large
tolerance in magnitude. We identify $98,604$ objects in the matched COSMOS2015 and ACS-GC catalogue.
From these matched objects $50 \ (95)\%$ have less than $0.07 \ (0.22)''$ and $0.12 \ (0.42)$
magnitude differences in their matched angular positions and luminosities respectively, which
is well below the chosen tolerances described above and indicates that the match is robust.
Subsequently we matched the joint COSMOS2015 and ACS-GC catalogue with the ZEST catalogue using the same three-step method
with the same tolerances for magnitudes and angular positions.
The ACS-GC \verb|MAG_BEST_HI| are now compared to the \verb|SExtractor| \verb|ACS_MAG_AUTO| provided in ZEST.

The final matched catalogue contains $70,708$ objects from which $50 \ (95)\%$ have less than $0.03 \ (0.2)''$ and
$0.03 \ (0.16)$ magnitudes differences in their matched angular positions and luminosities respectively. These smaller
differences compared to the first matching between COSMOS2015 and ACS-GC can be attributed to the fact that the positions and magnitudes used for the second matching
are all derived with \verb|SExtractor| from the same \hst\ ACS imaging data. A second
reason is the cut at $I_{AB}=24$ in ZEST, which excludes the dimmest objects with
highest uncertainties on position and luminosity estimates. This latter cut also explains the
strong drop in the number of objects in the final catalogue.

\subsubsection{Volume limited main sample of disc-dominated galaxies}
\label{sec:data:cosmos:main_sample}

\begin{figure}
\centering\includegraphics[width=8.0 cm, angle=0]{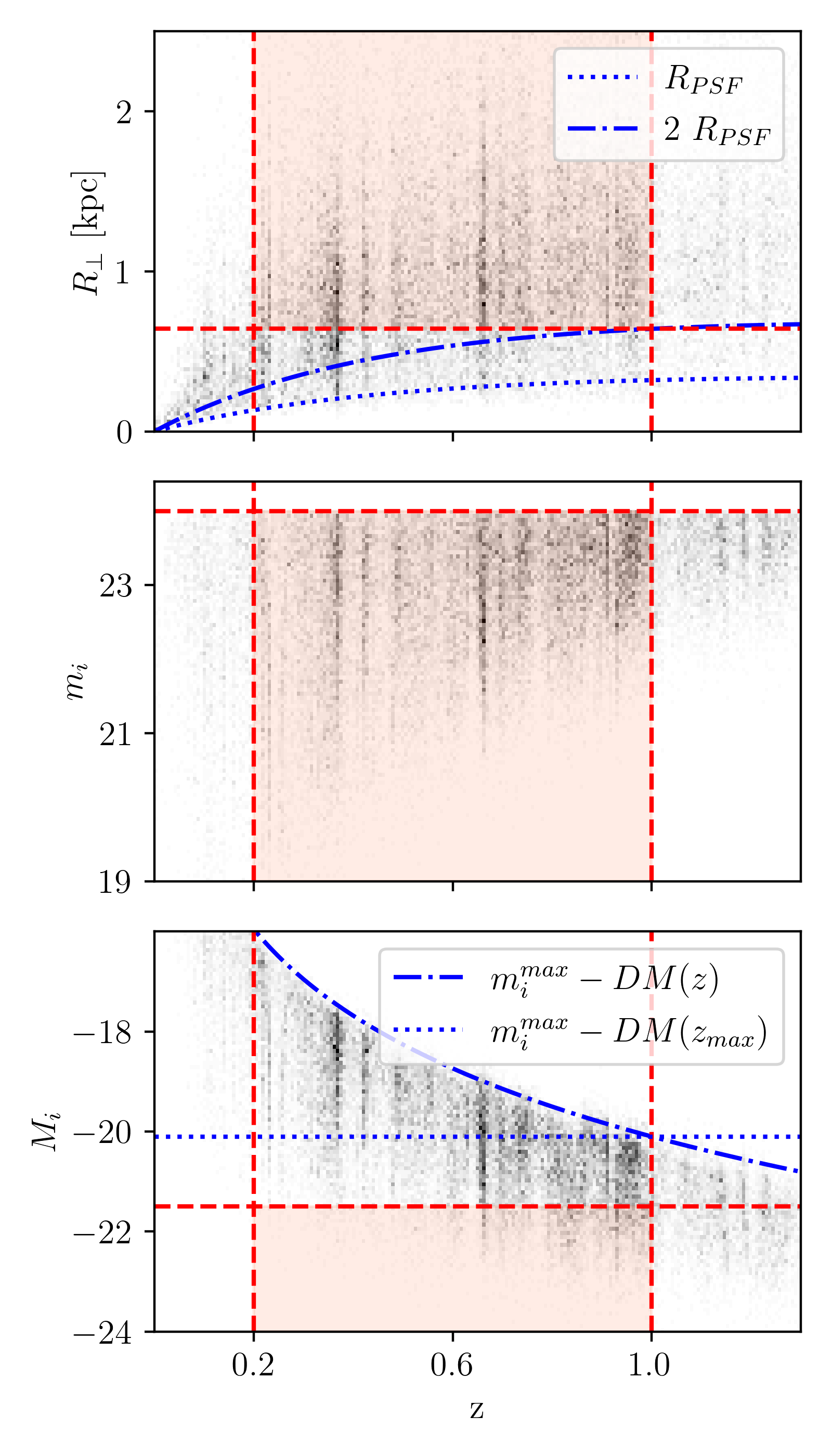}
\caption{Selection of the volume limited main sample from disc-dominated galaxies in the matched
COSMOS catalogue by photometric redshift $z$, transverse comoving radii $R_{\perp}$ and
apparent and absolute $i$ band magnitudes $m_i$ and $M_i$ respectively. The cuts on each
variable are marked by red dashed lines, enclosing the selected sample in the red area.
The blue dotted and dashed-dotted lines in the top panel show the comoving radii which
correspond to one and two times the angular size of the HST PSF respectively. The
blue dashed-dotted line in the bottom panel shows the limit on $M_i$, given by the
apparent magnitude cut $m_i^{max}$ and the distance modulus $DM$. The horizontal blue dotted
line in the same panel shows a naive cut on $M_i$ (see Section \ref{sec:data:cosmos:main_sample} for details).
The over-abundances at $z\sim 0.35$ and $z\sim 0.7$ are known large-scale structures
\citep[see e.g Fig.~4  in][]{Gozaliasl19,Guzzo07}, while some sharp vertical spikes 
in the redshift distribution might be artefacts of the photometric redshift estimation.
}
\label{fig:vol_lim_samp}
\end{figure}

Our study is focused on disc-dominated late-type galaxies in the matched catalogue, which we identify as
those with ZEST parameters \verb|type = 2| and \verb|bulg = 2 or 3|. Among them, we select a volume
limited main sample adopting cuts in photometric redshift, absolute Subaru $i+$ magnitude and comoving
effective radius. The selection is displayed in Fig. \ref{fig:vol_lim_samp} and described below.
Examples of the galaxies in our main sample are shown in Fig. \ref{fig:acs_images} and \ref{fig:acs_images2}.

\paragraph{Redshift cuts.}
The redshift range of our main sample is set to $0.2<z<1.0$ (marked by vertical red
dashed lines in Fig. \ref{fig:vol_lim_samp}). The upper limit is a compromise between
a deep selection in redshift and a sufficiently high number of galaxies in the
volume limited sample. The lower limit is defined by the end of the distribution,
below which very few objects are found due to the small volume of the light cone.
The redshifts on which the cuts are applied are the median of the likelihood
distribution from the SED template fits, which are referred to as \verb|photoz| in
the COSMOS2015 catalogue.

\paragraph{Magnitude cuts.}
Galaxies are further selected to have apparent isophotal $AB$ magnitudes
in the Subaru $i+$ band (referred to as \verb|ip_MAG_ISO| in the COSMOS2015 catalogue
and hereafter as $m_i$)
brighter than $m_i^{\rm max}=24$ above which the ZEST morphological
classification becomes unreliable \citep{Scarlata07}.
The apparent magnitude cut introduces a redshift dependent selection by the absolute magnitude
which hampers the comparison of galaxy populations at different redshifts (see bottom panel of
Fig. \ref{fig:vol_lim_samp}). We therefore require the absolute restframe Subaru $i+$ magnitudes
(hereafter referred to as $M_i$) to be brighter than $M_i^{\rm max} = -21.5$, ensuring that all
objects in our redshift range are sufficiently bright to be unaffected by the apparent magnitude cut.

\begin{figure}
\centering\includegraphics[width=8.2 cm, angle=0]{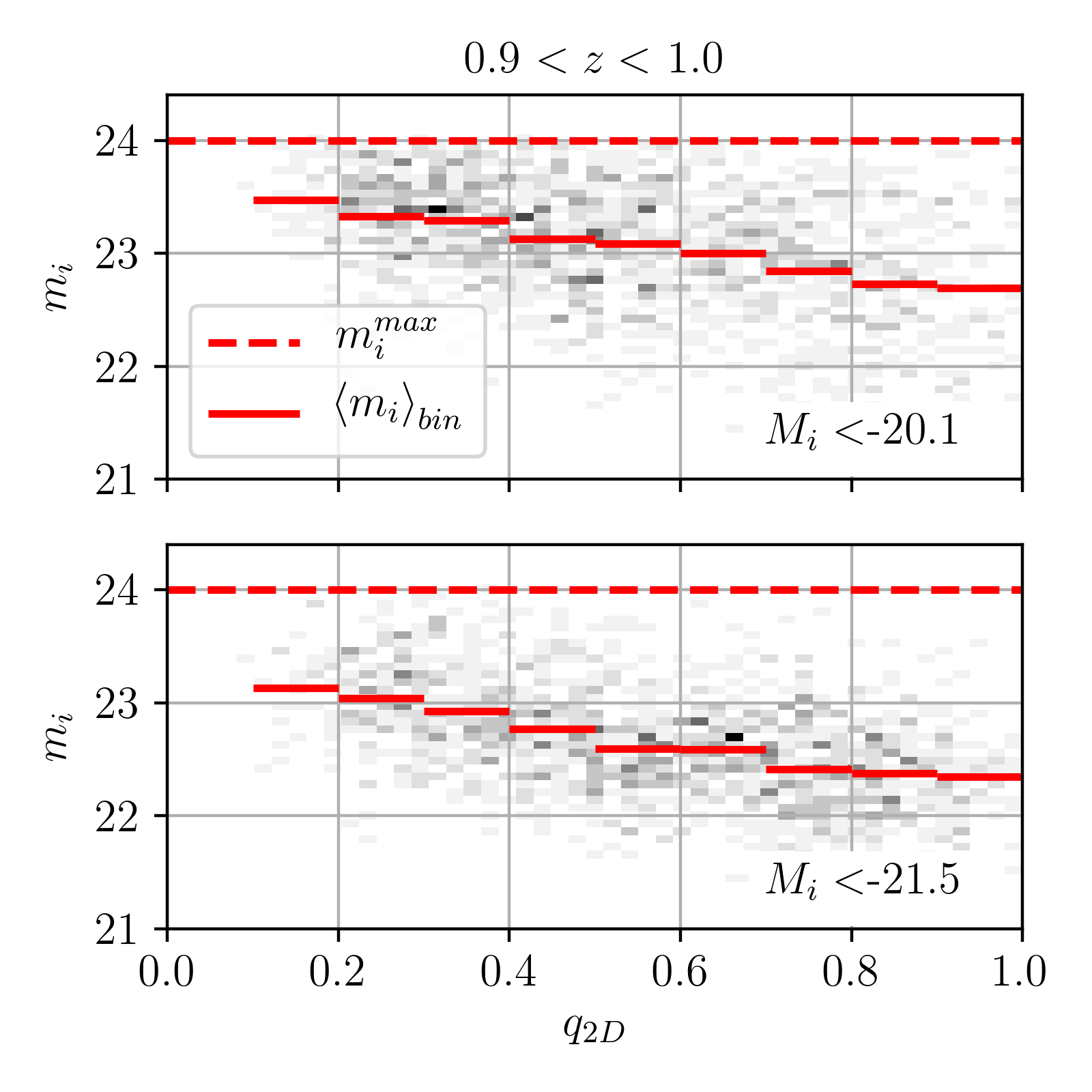}
\caption{Apparent 2D axis ratios \qdd\ of disc-dominated galaxies in COSMOS versus their apparent $i$-band magnitude
$m_i$ at the highest redshifts considered in this work. Horizontal bars show the average
$m_i$ in bins of \qdd\ to elucidate the correlation between both observables.
Face-on discs with $q_{2D} \simeq 1$ appear brighter than edge-on discs, which indicates
a dependence of dust extinction on the disc inclination relative to the observer.
Top panel shows the volume limited sample selected as shown in Fig. \ref{fig:vol_lim_samp},
but with a more conventional cut on the absolute $i$ band magnitude of $M_i<-20.1$. The
$m_i$ distribution of this sample is cut off at the apparent magnitude limit at $m_i=24$,
which biases the $q_{2D}$ distribution, as illustrated in Fig. \ref{fig:qcm_cosmos}.
We mitigated this bias with a more conservative cut of $M_i<-21.5$. The resulting
sample, shown in the bottom panel, is only weakly affected by the cut on $m_i$.}
\label{fig:q_mi}
\end{figure}

Note that the cut in absolute magnitude is chosen to be brighter than a naive cut at
$M_i^{\rm max}= m_i^{\rm max}-{\rm DM}(z_{\rm max}=1.0) = 20.1$
(indicated as blue dotted line in the bottom panel of Fig. \ref{fig:vol_lim_samp})
to mitigate effects of dust extinction on the observed axis ratio distribution.
In the top panel of Fig. \ref{fig:q_mi} we show that the apparent magnitude of discs
is fainter for objects with low axis ratios. This effect can be expected
from an increased dust extinction in objects that are inclined towards an edge-on orientation \citep[e.g.][]{Graham08}.
As a consequence, high redshift discs (at $z\simeq1.0$) that have low axis ratios ($q_{2D}\lesssim 0.5$)
and where selected by $M_i^{\rm max} = -20.1$ can fall below the apparent magnitude cut of $m_i^{\rm max}=24$,
marked as horizontal dashed line in Fig. \ref{fig:q_mi}.
Dust extinction can therefore introduce a bias in the observed axis ratio distribution
towards apparently rounder (i.e. face-on) galaxies, as we discuss in more detail
in Appendix \ref{app:mcut_qapp}.
In the bottom panel of Fig. \ref{fig:q_mi} we demonstrate that selecting galaxies with absolute
magnitudes brighter than $M_i = -21.5$ ensures that that the apparent magnitudes are brighter
than $m_i^{\rm max}=24$, even if the galaxies are seen edge-on ($q_{2D}\ll 1$) at the
highest redshifts considered in this work.

\begin{figure}
\centering\includegraphics[width=8.2 cm, angle=0]{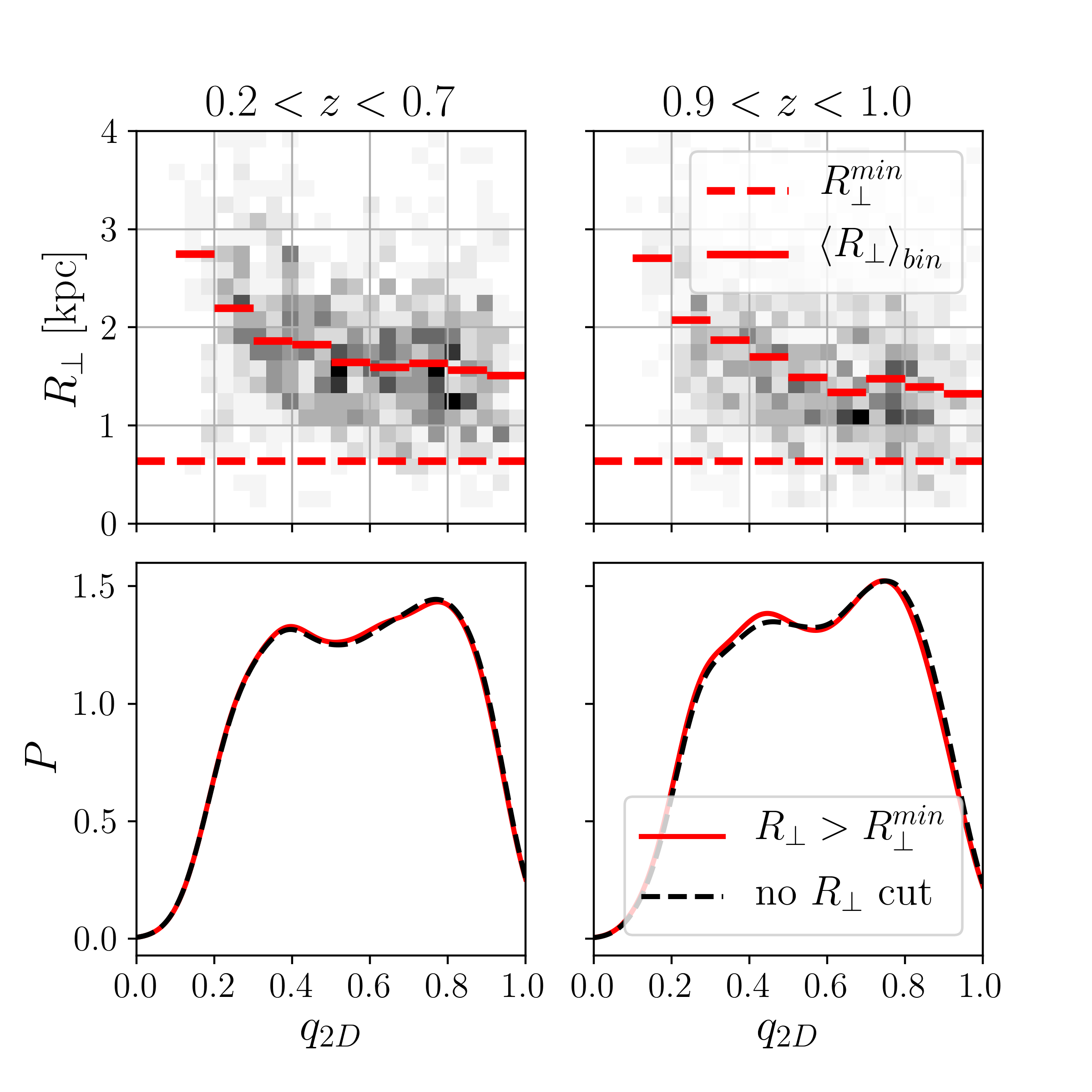}
\caption{
{
{\it Top}: Comoving transverse effective radii $R_\perp$ of disc-dominated galaxies in our volume
limited COSMOS sample verses the apparent 2D axis ratio \qdd. Red bars mark the average $R_\perp$
in bins of \qdd\ and show that the apparent size depends on the apparent axis ratio, presumably as
an effect of the disc inclination. The horizontal dashed lines show minimum value for $R_\perp$ used
as cut in our sample selection. {\it Bottom}: Distribution of \qdd\ for disc galaxies before and
after the size cut (black dashed and red solid lines respectively). This figure demonstrates that
our size cut does not bias the \qdd\ distribution of our sample.}
}
\label{fig:q_rperp}
\end{figure}

\paragraph{Size cuts.}
Our final selection addresses the sizes of galaxies. We require objects in our sample to have
angular radii $R_{\theta}$ which are at least twice as large as the width of the PSF
in the HST F814W imaging (i.e. $R_{\theta}^{\rm min} = 0.17''$) to minimize the
effect of inaccuracies in the PSF correction on shape measurements as well as PSF effects
on the morphological classifications \citep{Scarlata07, Griffith12}. 
Simply applying a cut on $R_{\theta}$ would lead to objects with smaller physical sizes entering
the main sample at lower redshifts. This effect could introduce an apparent evolution of the galaxy
shapes with redshift since the apparent physical sizes and shapes are correlated
(see top panels of Fig. \ref{fig:q_rperp} and discussion below).
We therefore apply a cut on the transverse comoving radii $R_{\perp}(z) = R_{\theta} D_A(z)$,
which ensures that the observed angular radii of the galaxies in our sample are always larger
than $R_{\theta}^{\rm min}$, even for the most distant objects at $z_{\rm max}=1.0$.
Here $D_A(z)$ is the angular diameter distance at the source redshift $z$.
Assuming a flat $\Lambda$CDM universe we obtain the condition
$R_{\perp} > R_{\theta}^{\rm min} D_A(z_{\rm max}) = 0.64 \ \text{kpc}$.
$D_A$ and ${\rm DM}$ are computed using cosmological parameters from the \citet{Planck18}, i.e.
$(\Omega_M, \Omega_b, H_0,\sigma_8,n_s) = (0.31,0.049,67.66,0.81,0.9665)$.

We show $R_\perp$ corresponding to one and two times the angular width of the HST
PSF ($0.085''$) at a given redshift as dotted and dashed-dotted blue lines respectively 
in the top panel of Fig. \ref{fig:vol_lim_samp} together with the cut on $R_\perp$
which defines the main sample (red dashed horizontal line).
The angular radii $R_{\theta}$ are the PSF corrected effective
radii from the \verb|GALFIT| single S\`{e}rsic model fits, which quantify the galaxy size along the projected
major axis (referred to as \verb|RE_GALFIT_HI| in the ACS-GC).

\begin{figure}
\centering\includegraphics[width=8.0 cm, angle=0]{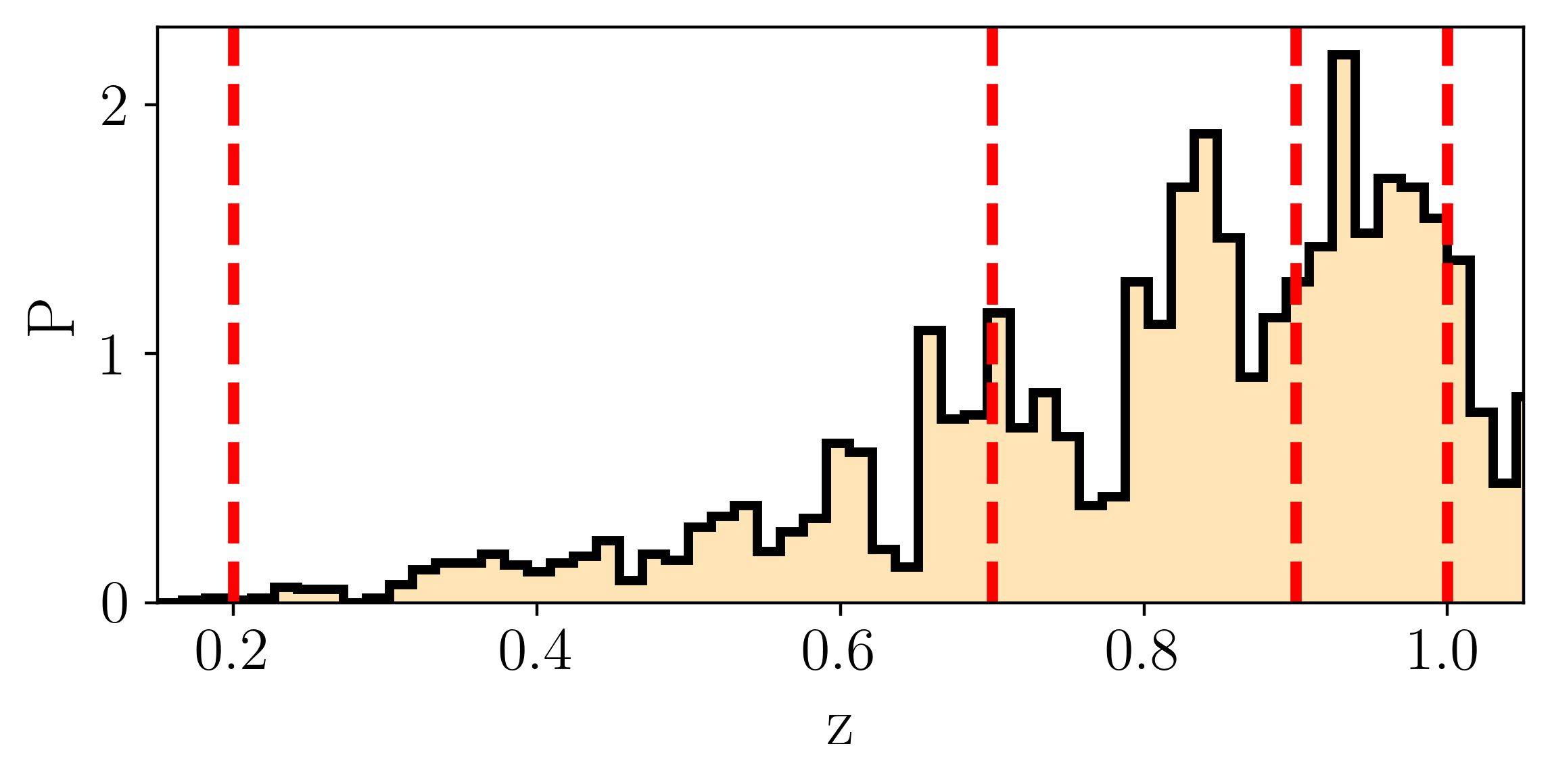}
\caption{Selection of redshift sub-samples from our volume limited main sample
of disc-dominated galaxies in COSMOS. The photometric redshift distribution
is split into three redshift bins, which are enclosed by vertical red dashed lines 
at $z=0.2,0.7,0.9$ and $1.0$.}
\label{fig:zpdf}
\end{figure}

For opaque discs with zero thickness
the size of the projected major axis would be equal to the three-dimensional major axis,
independent of its inclination with respect to the plane of projection. In that case a cut
by projected size would not introduce a cut by shape or inclination with respect to the observer.
However, real disc galaxies have a non-zero thickness and are not opaque but to a certain degree transparent.
Their observed surface brightness profile is hence affected by projection effects, which
depend on the inclination angle with respect to the observer. As a consequence the diameter
of the observed 2D isophotes is larger for edge-on than for face on discs
\citep[e.g.][]{Holmberg46, deVaucouleurs59, Heidmann72}.
In the top panels of Fig.~\ref{fig:q_rperp} we show for galaxies brighter than our absolute magnitude
cut that $R_{\perp}$ increases significantly with decreasing apparent axis ratios, indicating the expected dependence
of $R_{\perp}$ on the disc inclination.
However, for our bright sample and within the redshift range we consider, the lower limit on $R_{\perp}$ is sufficiently
small to have only a negligible impact on the observed distribution of axis ratios as shown in the bottom panels of Fig. \ref{fig:q_rperp}. We therefore do not expect a relevant impact of the size cut on the disc inclinations of our sample,
which could otherwise bias the axis ratio distribution towards edge-on objects \citep[e.g.][]{Huizinga92}.
All cuts defining the volume limited sample are summarized in Table \ref{tab:vol_lim_samp}.
After applying these cuts the final volume limited main sample contains $3,739$ disc-dominated galaxies in total.
The redshift distributions of these galaxies is shown in Fig. \ref{fig:zpdf}.

\begin{figure}
\centering\includegraphics[width=8.0 cm, angle=0]{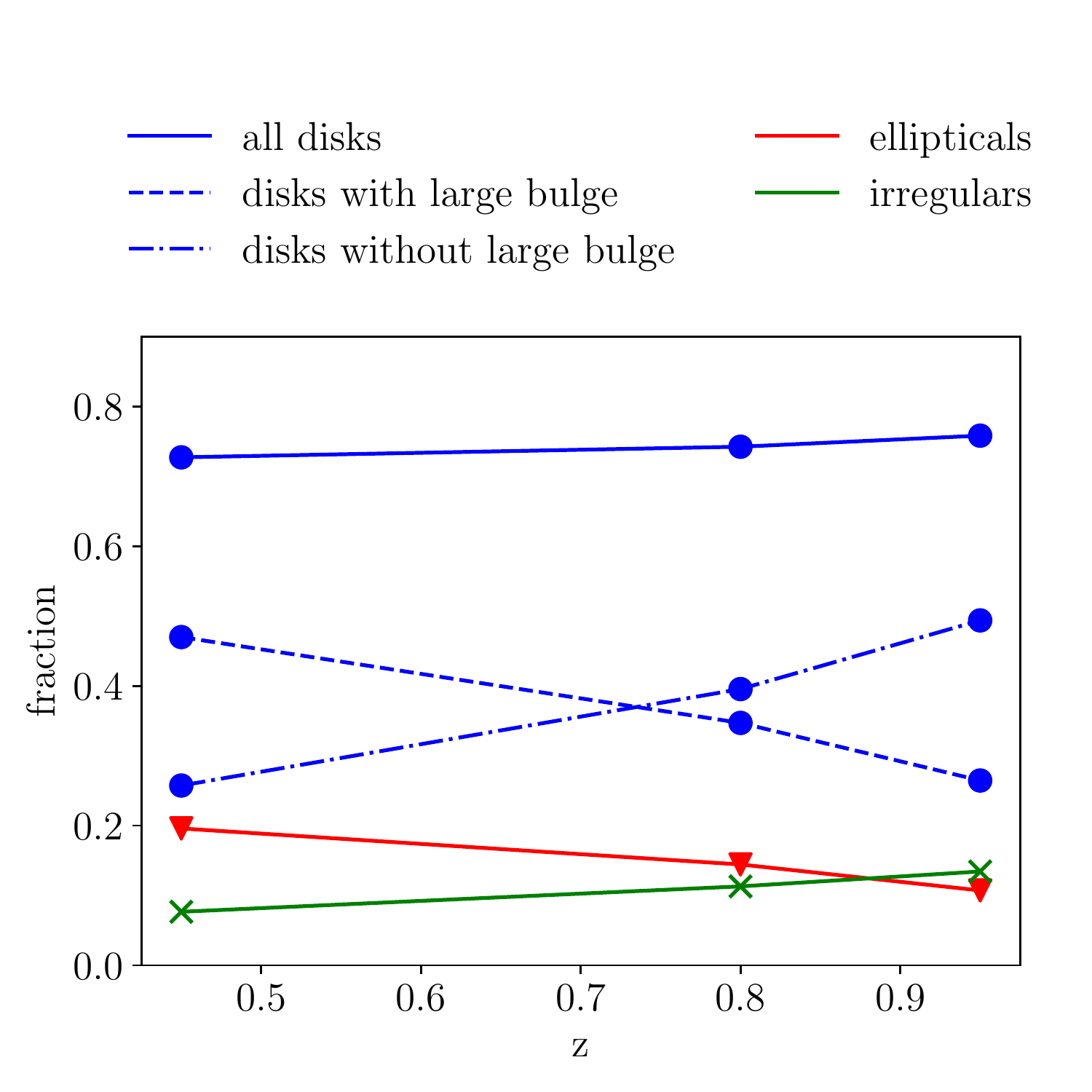}
\caption{Fractions of galaxy types in our volume limited COSMOS sample in the three redshifts bins
shown in Fig. \ref{fig:zpdf}.
Blue circles, red triangles and green crosses show results for galaxies classified as discs, ellipticals
and irregulars respectively. Results for discs with and without a large bulge are connected
by dashed-dotted and dashed lines respectively.
}
\label{fig:ztypefrac}
\end{figure}

\paragraph{Type and color distributions.}
In Fig. \ref{fig:ztypefrac} we display the relative abundance of the disc-dominated galaxies
in our main sample with respect to the total galaxy population in that sample in three
redshift bins.
These abundances are compared to those of galaxies classified as discs with large bulges
(ZEST parameter \verb|type=2| and \verb|bulg=0 or 1|), as elliptical (\verb|type=1|) and
irregular (\verb|type=3|). The figure confirms reports from the literature that a large
fraction of discs galaxies has no significant bulge. The fraction of discs remains roughly
constant at $\lesssim 80$\%, with a slight increase with redshift. It is further interesting
to note that at $z\simeq1.0$ the majority of discs has no large bulge, while the opposite is
the case at $z\simeq0.4$.
The decline of the fraction of bulgeless discs with decreasing redshift lines up with the findings
from \citet{Sachdeva13} based on Chandra Deep Field observations. 
Fig. \ref{fig:ztypefrac} further shows that the rate at which this fraction
declines is similar to the rate at which the fraction of discs with large bulges increases.
This finding is consistent with the scenario in which discs with large bulges form out of bulgeless discs,
potentially during mergers.
Note, the absolute values of the different fractions may be specific to our sample selection.
In particular the lower limit on the size separates out many elliptical galaxies, which tend
to be more compact than discs, increasing the fraction of discs in our sample.

\begin{figure}
\centering\includegraphics[width=8.0 cm, angle=0]{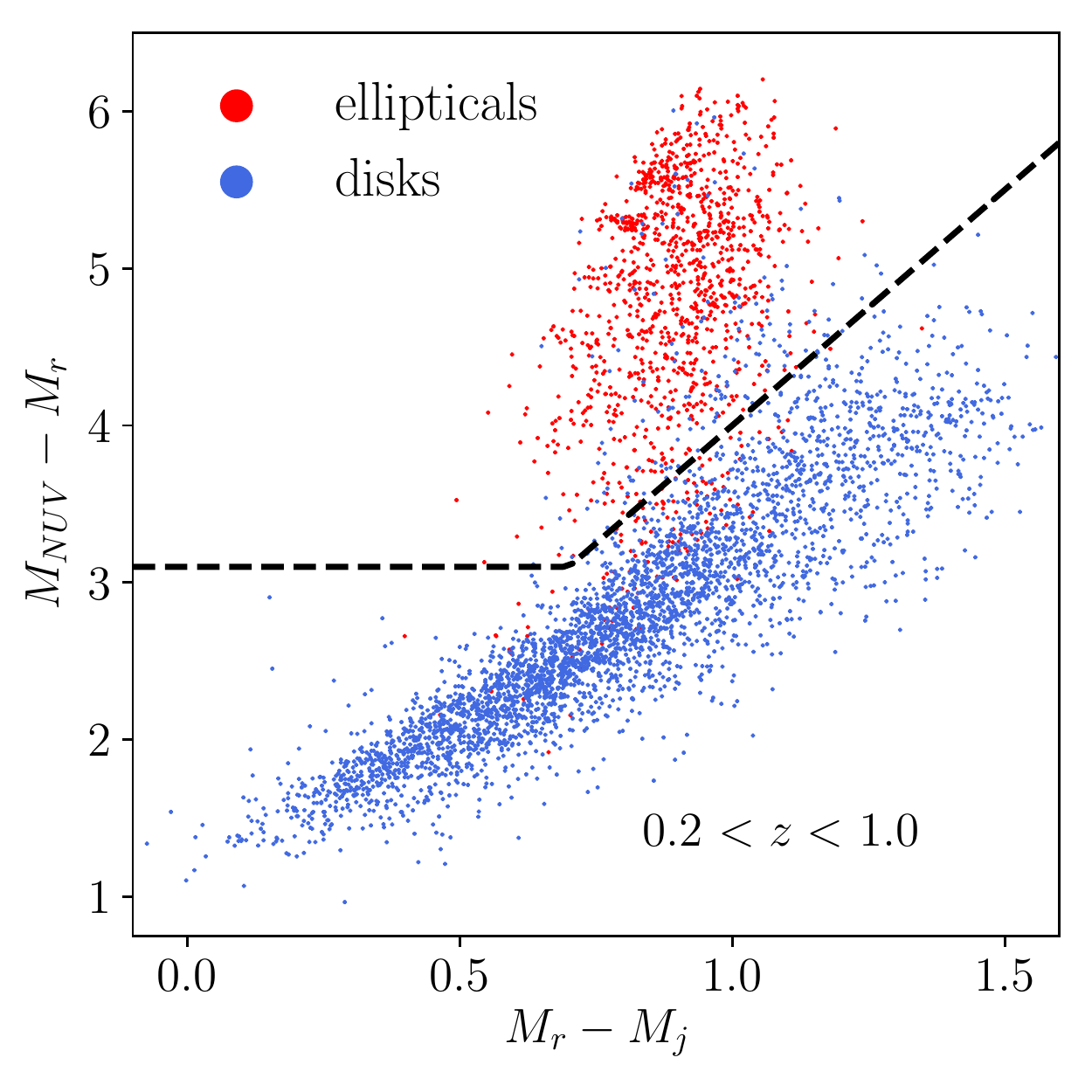}
\caption{Color-color diagram, based on estimates of the absolute restframe magnitudes
in the $r$, $j$ and $NUV$ filters from the COSMOS2015 catalogue for galaxies
which pass our main sample selection on size, magnitude and redshift. Red and blue dots
show objects which are classified in the ZEST catalogue as early-type and disc-dominated
late-type respectively. The black dashed line is taken from \citet{Laigle16} and separates
the quenched and the star forming populations.}
\label{fig:ccd}
\end{figure}

%
For further validation of our data set we compare the color-color diagrams of the disc-dominated
and elliptical galaxies in our main sample in Fig. \ref{fig:ccd}. We find that the discs and the ellipticals reside preferentially in the blue star forming
and red quenched sequence respectively, which are well separated by the color-color cuts from \citet{Laigle16}.
This result can be expected in general from the well known correlation between galaxy morphology and color
\citep[e.g.][]{Larson80, Strateva01, Baldry04, Martig09}. We find that the bulge dominated discs in our
sample populate both, the quenched as well as the star forming sequences (not shown in the figure for clarity).
This result lines up with the large bulges found in discs on the red quenched sequence, for instance in data
from COSMOS and the Sloan Digital Sky Survey \citep[see][]{Bundy10, Guo20}.
Overall, the expected correlation between color an morphology shown in Fig. \ref{fig:ccd} confirms {\it a posteriori}
that the morphological classification from ZEST and the photometric properties from the COSMOS2015 are consistent
with each other. This test shows that the matching between both catalogues is sufficiently reliable
for deriving physical interpretations from their combined morphological and photometric information.
%
\begin{table}
    \centering
    \begin{tabular}{ l l }
     galaxy property & constraint \\  \hline
     photometric redshift & $0.2 < z < 1.0$ \\  
     apparent $3''$ aperture AB Subaru $i+$ magnitude & $m_i<24$ \\
     absolute rest-frame Subaru $i+$ magnitude & $M_i<-21.5$ \\
     transverse comoving effective radius & $R_{\perp} > 0.64$ kpc \\
    \end{tabular}
    \caption{Selection cuts for the volume limited COSMOS main sample,
    shown in Fig. \ref{fig:vol_lim_samp}}
    \label{tab:vol_lim_samp}
\end{table}

\subsubsection{Stellar mass - redshift samples}
\label{sec:data:cosmos:sub_samples}

\begin{figure}
\centering\includegraphics[width=8.2 cm, angle=0]{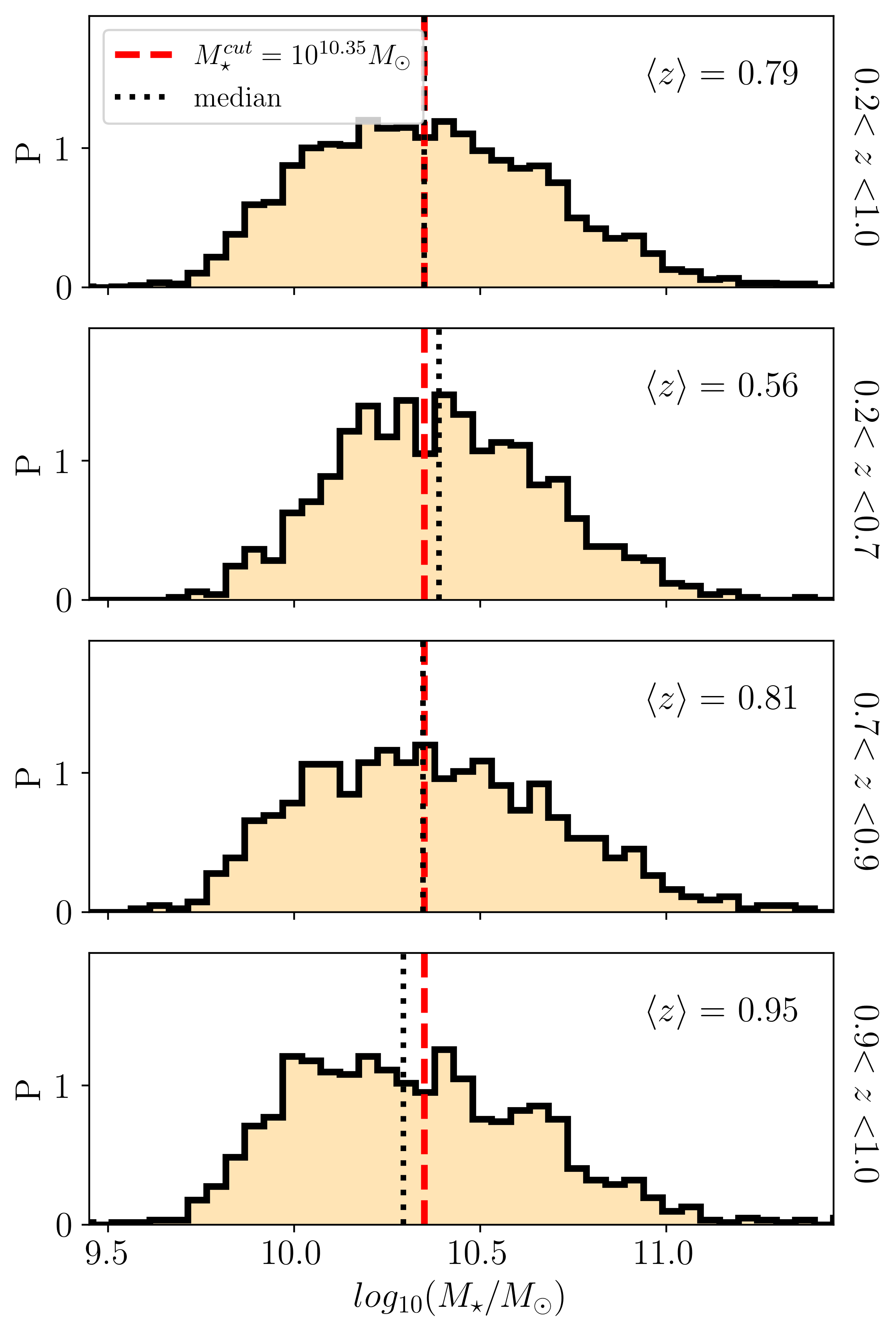}
\caption{Selection of low and high stellar mass sub-samples from the COSMOS main sample,
defined over the entire redshift range, and from the three redshift sub-samples shown in
Fig. \ref{fig:zpdf} (top and lower panels respectively). The stellar mass distributions
are split at $M_\star^{cut}=10^{10.35}$ (red dashed lines), which corresponds to the
median stellar mass of the main sample and lies close to the median stellar masses of
the redshift sub-samples (black dotted lines).}
\label{fig:mpdf}
\end{figure}

In order to investigate the redshift and stellar mass dependence of the galaxy shapes, we
select three redshift sub-samples and two stellar mass sub-samples from the COSMOS main sample
as shown in Fig. \ref{fig:zpdf} and in the top panel of Fig. \ref{fig:mpdf} respectively.
For more detailed investigations we further select six sub-samples, which are selected
by both, redshift and stellar mass, as shown in the lower panels of Fig. \ref{fig:mpdf}.
The width of the redshift bins are chosen such that each sample contains a sufficiently large
number of objects required for our statistical analysis.  The stellar mass cut at
$M_{\star}^{\rm cut}=10^{10.35} {\rm M}_{\odot}$ corresponds to the median stellar mass of the
main sample and lies close to the median stellar mass of the redshift sub-samples.
The cuts in mass and redshift are summarized in Table \ref{tab:params_rec_cosmos} together with the number
of galaxies in each sub-sample.

\begin{figure*}
\centering\includegraphics[width=17.5 cm, angle=0]{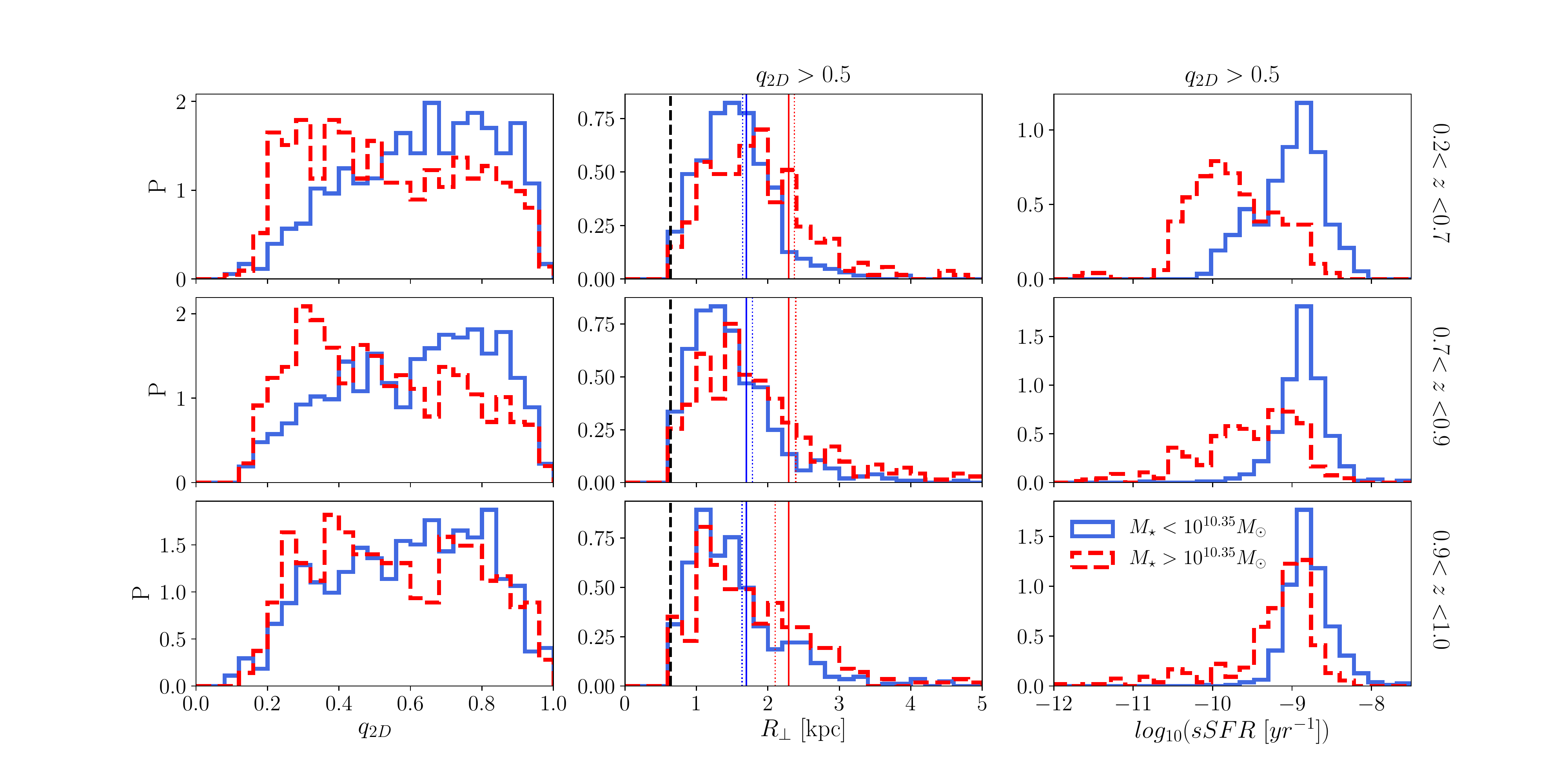}
\caption{
Probability distributions of apparent 2D axis ratios, transverse comoving radii 
and specific star formation rates (left, central and right panels respectively) for
disc-dominated galaxies in our volume limited COSMOS main sample with stellar masses below
and above $10^{10.35} {\rm M}_\odot$ (blue and red histograms respectively).
Vertical panels show results in three redshift bins with limits indicated on the right.
The comoving radii and specific star formation rates are shown for discs with apparent axis ratios
above $q_{2D}>0.5$, to minimize bias induced by projection and dust extinction
(see Fig. \ref{fig:q_rperp} and \ref{fig:qcm_cosmos}). Vertical black dashed
lines mark the minimum radius of the volume limited sample. Vertical solid and dotted
lines indicate the mean radii for each mass sample over the full redshift range
and in each redshift bin respectively.}
\label{fig:qrssfr_pdf}
\end{figure*}

We show in the left column of Fig.~\ref{fig:qrssfr_pdf} the distribution of the observed
shapes from the ACS-GC catalogue, quantified by the axis ratio \qdd, in the three redshift bins.
We find that at all redshifts the distributions of the low and high mass samples are skewed towards
high and low 2D axis ratios respectively. The difference between the mass samples increases with
decreasing redshifts, which is mainly driven by an increasing fraction of
high mass galaxies with low 2D axis ratios ($q_{2D} \simeq 0.3$). The axis ratio distribution of
low mass galaxies on the other hand shows no strong change with redshift. In fact, we argue
in Section \ref{sec:results} that the redshift dependence of both samples is not significant
when considering shot-noise errors on the binned distributions.

Nevertheless, a weak trend in redshift can also be seen in the transverse comoving radii $R_\perp$
in the central column of Fig. \ref{fig:qrssfr_pdf}. These
radii are shown for discs inclined towards a face-on orientation with apparent axis ratios $q_{2D}>0.5$
to reduce the impact of inclination on the observed size (see Fig. \ref{fig:q_rperp}). As for
the axis ratios we find that at high redshift, the size distribution for low and high mass samples
is similar. At lower redshift, the deviation between both distributions is slightly larger, mainly
due to an increase of the sizes in the high mass disc sample. This trend can be seen in the average
radii, displayed as dotted and dash-dotted vertical lines for the high and low mass sample, respectively.

A non-geometric galaxy property which follows the same behaviour is the specific star formation
rate ($sSFR$) from the COSMOS2015 catalogue, whose distribution is displayed in the right column of
Fig. \ref{fig:qrssfr_pdf}. The selection here is again restricted to galaxies with $q_{2D}>0.5$
to mitigate a potential bias in the $sSFR$ estimates induced by inaccuracies of dust attenuation model
employed in the SED fitting, as reported by \citet{Laigle19}.
We see that the $sSFR$ distribution is similar for the low and high mass samples at high redshifts.
At low redshifts, both distributions differ significantly, which is mainly driven by a strong
decrease of the $sSFR$ for high mass discs.
This result lines up with those of \citet{Grossi18}, who find that the $sSFR$ of disc
dominated galaxies in COSMOS field decreases with increasing stellar mass and decreasing with redshift.
It is worth noting that the mass and redshift dependence is weaker for the $SFR$ (not shown here)
than for the $sSFR$.
This weaker dependence is consistent with the overall weak redshift dependence of star formation in nearby
galaxies, suggested by findings of \citet{Kroupa20} and may result from a partial compensation by the mass
differences between the sub-samples.
Overall, our findings indicate a correlation of the galaxy shapes, sizes and specific star formation rates
with the stellar masses, which appear to increase as galaxy formation proceeds. We will discuss this result
in more detail together with the reconstructed 3D axis ratio distribution in Section \ref{sec:summary}.

We stress here that our discussion of differences between the $R_\perp$ and $sSFR$ distributions
from different mass and redshift samples discussed above is based on our visual impressions.
Clarifying whether or not these differences are significant would require a statistical analysis, which is beyond the scope of this work.

\subsection{disc galaxies in hydro-dynamic simulations}
\label{sec:data:sims}
We use the state-of-the-art hydrodynamic simulations of galaxy formation Horizon-AGN\footnote{www.horizon-simulation.org}
and Illustris TNG100\footnote{www.tng-project.org} (hereafter referred to as HAGN and TNG100 respectively)
to test the methods and assumptions on which our analysis of the COSMOS data is based.
These tests are performed at redshift $z=1.0$, were the available snapshots of these simulations are
closest to the maximum of the redshift distribution of our volume limited COSMOS sample.
Both simulations assume a $\Lambda$CDM cosmology with recently constrained parameters, cover similar
cosmological volumes and include sub-grid mechanisms to model the formation of galaxies at a similar degree
of complexity, as detailed below.
However, their simulation techniques differ significantly, which allows for testing the robustness of our
conclusions. The main characteristics of these simulations are compared to each other in Table \ref{tab:sim_params}.
Note that these simulations where run at relatively low resolutions in order to cover large volumes,
which has a noticeable impact on the galaxy morphologies as discussed in Section \ref{sec:method:P3Dmodel}.

\subsubsection{Horizon-AGN}
\label{sec:data:sims:HAGN}

The hydrodynamic simulation HAGN was produced with the grid-based adaptive-mesh-refinement code
RAMSES \citep{Teyssier02}, using cosmological parameters compatible with the constraints
from WMAP-7 \citep{Komatsu11}. The simulation includes the key processes relevant for galaxy formation:
cooling, heating and chemical enrichment of gas, the formation and evolution of stars and black holes
as well as feedback from stellar winds, supernovae and Active Galactic Nuclei (AGNs)
\citep{Dubois14}. Galaxies were identified in the distribution of stellar particles
as groups with more than $50$ members using the AdaptaHOP finder \citep{Aubert04}.
A key achievement of HAGN with respect to previous generations of cosmological hydrodynamic
simulations is the broad diversity of galaxy morphologies,
with realistic fractions of discy and elliptical objects \citep{Dubois16}.
A reasonable agreement with observed luminosity functions and color distributions
has been shown by \citet{Kaviraj17}, while moderate deviations from observed angular clustering have been found by \citet{Hatfield19}. 

For our analysis we select disc galaxies, which we define via the ratio $r_v = V_{\rm rot}/\sigma_v$
between the stellar rotation $V_{\rm rot}$, defined as the mean tangential velocity of stellar particles with
respect the galaxies' spin axis and the velocity dispersion $\sigma_v$ (see \citet{Chisari15} for details).
High values of $r_v$ indicate relatively high rotational velocities compared to those radial or
parallel to the spin axis.
For our analysis we select the $20\%$ of galaxies with the highest values of $r_v$, which corresponds
to a cut at $r_v>1.06$ at redshift $z=1.0$ (Fig. \ref{fig:morph_class_hydro}).
We attribute this low value of $r_v$ to the relatively low resolution of the simulation
and demonstrate in Section \ref{sec:method:P3Dmodel} that the selected galaxies are indeed
discy, although with ``puffed-up'' shapes, characterized by relatively high $C_{3D}/B_{3D}$ axis ratios.
We further limit the selection to galaxies with more than $500$ particles,
which corresponds to a stellar mass cut at $M_\star > 10^9 {\rm M}_\odot$,
to ensure reliable measurements of the morphological properties (see also Section \ref{sec:data:sims:axes_ratios}).
The final sample contains $14198$ disc galaxies.

\subsubsection{Illustris TNG100}
\label{sec:data:sims:TNG100}

The TNG100 \citep{Springel18, Naiman18, Nelson18, Marinacci18, Pillepich18b}
is a magneto-hydrodynamic simulation,
produced with the moving-mesh code
AREPO \citep{Springel10}, which was run with cosmological parameters from the \citet{Planck16}.
It includes the same  key processes for modeling galaxy formation as HAGN, although with significantly
different implementations which are described in \citet{Pillepich18a}. Galaxies are identified
in dark matter subhaloes with non-zero stellar components. These subhaloes are detected
in friends-of-friends groups by the SUBFIND algorithm \citep{Davis85, Springel01, Dolag09}.
While it has been calibrated to match the galaxy mass function at $z=0$, the simulation has been shown to predict main characteristics of observed galaxy populations reasonably well, such as morphological diversity \citep{Rodriguez19}, stellar mass functions at higher redshift \citep{Pillepich18b},
the color bimodality \citep{Nelson18} as well as color-dependent two-point clustering
statistics \citep{Springel18}. The basic properties of the simulation are compared to those
of the HAGN simulation in Table \ref{tab:sim_params}.

To identify discs in the TNG100 simulation we use each galaxy's fraction of stellar mass in the disc component
with respect to the total stellar mass. For this purpose stellar disc particles have been defined via their
circularity parameter $\epsilon \equiv J_z / J(E)$, where $J_z$ is the specific angular momentum around a
selected z-axis and $J(E)$ is the maximum specific angular momentum possible at the specific binding energy $E$
of a given stellar particle.

\begin{figure}
\centering\includegraphics[width=8 cm, angle=0]{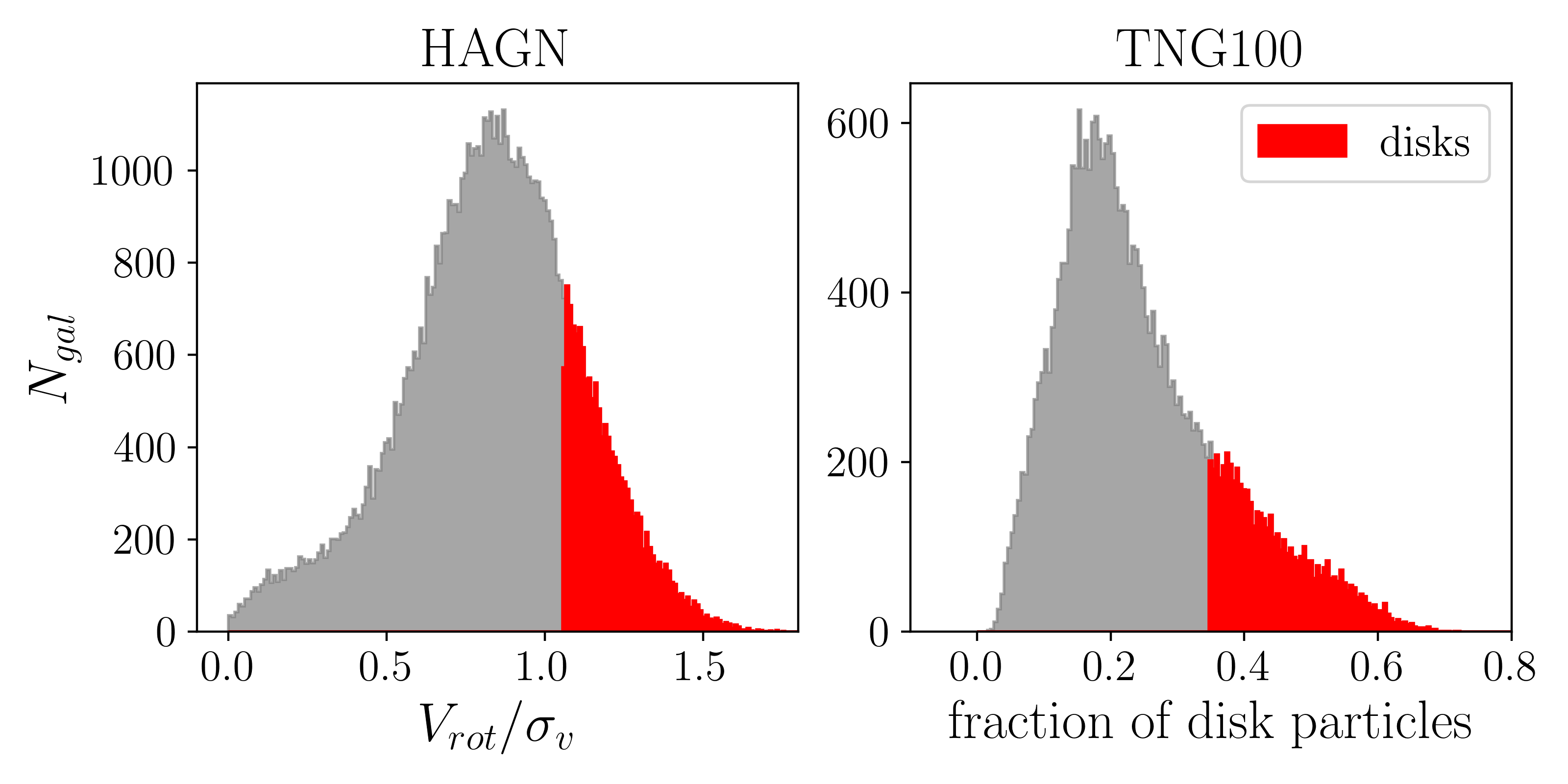}
\caption{{\it Left}: Distribution of the ratio between stellar rotation $V_{\rm rot}$ and the
velocity dispersion $\sigma_v$ for galaxies in the HAGN simulation. {\it Right}: Distribution
of the fraction of disc particles in each galaxy of the TNG100 simulations. The populations
of galaxies which we classify as discs are marked in red at the tails of the distributions
and make up $20$ \% of the entire sample. Results are shown for $z=1.0$.}
\label{fig:morph_class_hydro}
\end{figure}

The $z$-axis is thereby given by the principle angular momentum vector of the star forming gas,
or the stellar particles, if there is no star forming gas in the system \citep{Vogelsberger14b, Genel15}.
Particles with $\epsilon>0.7$ are considered to belong to the disc component \citep{Abadi03}.
The circularity parameter is provided on the Illustris database for all
subhalos with $M_\star > 3.4 \times 10^8 {\rm M}_\odot$ within the half mass
diameter $2 \ R_{1/2,\star}$ and at least $100$ stellar particles.
We classify TNG100 galaxies as discs if their fraction of disc particles is above $0.35$, which
corresponds to the upper $20$\% of the distribution at $z=1.0$, as shown in Fig.~\ref{fig:morph_class_hydro}.
As for HAGN, we attribute this low fraction of disc particles used for the cut to the limited resolution
of the simulation and demonstrate in Section \ref{sec:method:P3Dmodel} that it is a reasonable choice for
selecting discy objects. Note that we choose the disc particle fraction for the morphological classification
in TNG100 since this quantity is provided for the $z=1.0$ snapshot on the public TNG database in contrast to $r_v$.
Besides this practical motivation, using two different morphological classification schemes in both
simulations allows for drawing more robust conclusions regarding the validation of our analysis methods.
The final sample contains $5,674$ disc galaxies.

\begin{table}    \centering
 \begin{tabular}{ l c c c }
     & & TNG100 & HAGN    \\ \hline \hline
     $\Omega_{\Lambda}$ & - & $0.6911$ & $0.728$ \\
     $\Omega_m$ & - & $0.3089$ & $0.272$ \\
     $\Omega_b$ & - & $0.0486$ & $0.045$ \\
     $H_0$ & $[{\rm s}^{-1}{\rm km}]$ & $67.74$ & $70.4$ \\
     $\sigma_8$ & - & $0.8159$ & $0.81$ \\
     $n_s$ & - & $0.9667$ & $0.967$ \\  \hline
     $L_{\rm box}$ & $[h^{-1}\text{Mpc}]$ & $75$ & $100$ \\
     $m_{\star}$ & $[{\rm M}_\odot]$ & - &  $2 \times 10^6$\\
     $m_{\rm baryon}$ & $[{\rm M}_\odot]$ & $1.4 \times 10^6$ & - \\
     $m_{\rm dm}$ & $[{\rm M}_\odot]$ & $7.5 \times 10^6$ & $8 \times 10^7$
    \end{tabular}
    \caption{Main properties of the hydro-dynamic simulations Illustris TNG 100 and Horizon-AGN.}
    \label{tab:sim_params}
\end{table}

\subsubsection{Axis ratio measurements}
\label{sec:data:sims:axes_ratios}

The galaxies' axis ratios provided for the HAGN and TNG100 simulations are computed from the moment of inertia 
of their stellar mass distributions
\eq{
 I_{i,j} \equiv \frac{1}{M_{\star}} \sum_{n}^{N_\star} m_{\star}^n r^{n}_i r^{n}_j,
 \label{eq:def_Mi}
}
where $N_\star$ is the number of stellar particles in the galaxy,
$m_\star^n$ is the stellar mass of the $n^{\rm th}$ particle, $M_\star = \sum_n^{N_\star} m_\star^n$,
and $r_i^n$ are the components of the particle position vectors, defined with respect to the center
of mass \citep[][]{Chisari15, Genel15}.

The moment of inertia can be defined via the particle positions in 3D as well as in 2D.
In the latter case the positions are projected along one coordinate axis of the simulation, assuming
an observer at infinity. In the 3D case the square roots of the three absolute eigenvalues
$\lambda_1 \ge \lambda_2 \ge \lambda_3$ provide a measure of the 3D major, intermediate and minor axis lengths,
respectively, i.e. $(A_{3D}, B_{3D}, C_{3D}) = \sqrt{(\lambda_1,\lambda_2,\lambda_3)}$. Accordingly,
the 2D major and minor axis lengths are given by  $(A_{2D}, B_{2D}) = \sqrt{(\lambda_1,\lambda_2)}$ respectively.
The axis ratios are defined according to Equation (\ref{eq:3D_axes_ratios}) and (\ref{eq:2D_axes_ratio}).
We expect that the bias on such axis ratio measurements, which results from the discreteness
of the particle distributions \citep[e.g.][]{Joachimi13a, Hoffmann14, Chisari15}, is
negligible, due to the lower mass limits imposed on our sample
(see Section \ref{sec:data:sims:HAGN} and \ref{sec:data:sims:TNG100}).
The 2D axis ratios of the projected stellar mass distributions in HAGN are used to test
our method for reconstructing the 3D axis ratio distribution.
For TNG100 such measurements are currently not publicly available. However, we use 2D axis ratios measured
from second-order moments in synthetic images from TNG100 galaxies from \citet{Rodriguez19}. These images
have been produced using the SKIRT radiative transfer code\footnote{https://skirt.ugent.be} \citep{Baes11,Camps15}
by the Sloan Digital Sky Survey in the $g$ and $i$ broad band filters.
The axis ratios are obtained from the second-order moments of the pixels in these images that belong to the source.
Since these measurements are not PSF-corrected, we do not use them for testing the shape reconstruction
method. Instead, we use them to obtain a rough estimation of how strongly the observed galaxy shapes
depend on the wavelength range in which they are measured in order to interpret the COSMOS observations in Appendix \ref{app:shape_filter_effect}.


\section{Reconstructing 3D galaxy shape distributions from 2D observations}
\label{sec:method}

We now present the method used for constraining the distribution of 3D galaxy axis ratios from
the observed distribution of 2D axis ratios in COSMOS and test this method using galaxy shapes from the TNG100
and HAGN simulations.
The method is based on the assumption that each galaxy can be represented by an absorption-free,
self-similar, coaxial ellipsoidal stellar system, to which we refer to as 3D ellipsoid in the following.
The shape of its 3D luminosity density is fully described by the two axis ratios \qddd\ and \rddd,
given by Equation (\ref{eq:3D_axes_ratios}).
We expect such a one-component model to be a simplistic, but useful description for the objects in our observed sample:
disc-dominated galaxies, discarding bulge-dominated and irregular objects (see Section \ref{sec:data:cosmos:morph}).
However, substructures such as spiral arms could bias our results (see Fig. \ref{fig:acs_images}).
The isodensity contours of the projected luminosity density (i.e. the isophotes) of such model galaxies are self-similar,
coaxial ellipses \citep{Stark77}, whose 2D axis ratios can be obtained from the 3D axis ratios analytically, as detailed in
Appendix \ref{app:project_ellipsoid}. This allows for an efficient prediction of the 2D axis ratio distribution,
\pdd, for an ensemble of randomly oriented model galaxies with a given distribution of 3D axis ratios, \pddd.
This distribution can hence be constrained by comparing the corresponding \pdd\ prediction to observations.
This approach relies on a physically meaningful model for \pddd, which we motivate in Section \ref{sec:method:P3Dmodel}.
The free model parameters are obtained from  the COSMOS data using Bayesian interference as detailed in
Section \ref{sec:method:infer_params} and \ref{sec:method:test_reconstruction}. In the latter section we also study
the impact of inaccuracies of the employed \pddd\ model on the inferred 3D axis ratio distribution.

\subsection{Model for the 3D axis ratio distribution}
\label{sec:method:P3Dmodel}

The reconstruction of 3D axis ratio distributions relies on a physically meaningful
model for those distributions. Several of such models have been proposed in the literature.
\citet{Sandage70} found that the \qdd\ distribution of spiral galaxies
can be fitted reasonably well with a simple oblate disc model  according
to which $q_{3D}=1$ for all objects while \sddd\ is normal distributed
around $\langle s_{3D}\rangle \simeq 0.25$. Later studies based on larger samples
found that the $q_{2D}$ distribution of spirals is better fitted
using slightly triaxial disc models which describe the absence of
perfectly circular face-on spirals in observations \citep[e.g.][]{Binney81,Fasano93}.
Using normal distributions for \qddd\ and \sddd\
\citet{Lambas92} obtained good fits to the observations from the APM Bright Galaxy Survey.
The good performance of this model has been confirmed by recent results from
\citet{Satoh19} based on COSMOS data. \citet{Ryden04} found that a normal
distribution for \sddd\ and a log-normal distribution for the ellipticity
$\epsilon_{q_{3D}} \equiv 1 - q_{3D}$ delivers good fits to the \qdd\
distribution of spirals in SDSS observations, which was
confirmed by later studies
\citep[e.g.][]{Padilla08, Rodriguez13}. An alternative model based on
normal distributions of the ellipticity $\epsilon_{s_{3D}} \equiv 1 - s_{3D}$
and the triaxiality $T \equiv (1-q_{3D}^2)/(1-s_{3D}^2)$ proved to match the
\qdd\ distribution from SDSS, 3D-HST, COSMOS and CANDELS observations \citep{Chang13, Wel14}.
These different models were motivated by the fact that their prediction for the
2D axis ratios (or ellipticities) fit observations reasonably well, which indicates that
observations do not provide a strong constrain on the functional form of the 3D axes
ratio distribution.

In our analysis we use an approach that overcomes this limitation by employing
a new model for the 3D axis ratio distribution which we validate using the
HAGN and TNG100 simulations, described in Section \ref{sec:data:sims}.
Since cosmological hydrodynamical simulations reproduce observed galaxy morphologies
to varying degree of success and with a known dependence on simulation techniques,
sub-grid models and resolution
\citep{Snyder15,Dubois16,Correa17,Park19,Rodriguez19, Tacchella19}, we only use
them as a tool to validate our reconstruction method.

\begin{figure}
\centering\includegraphics[width=8.0 cm, angle=0]{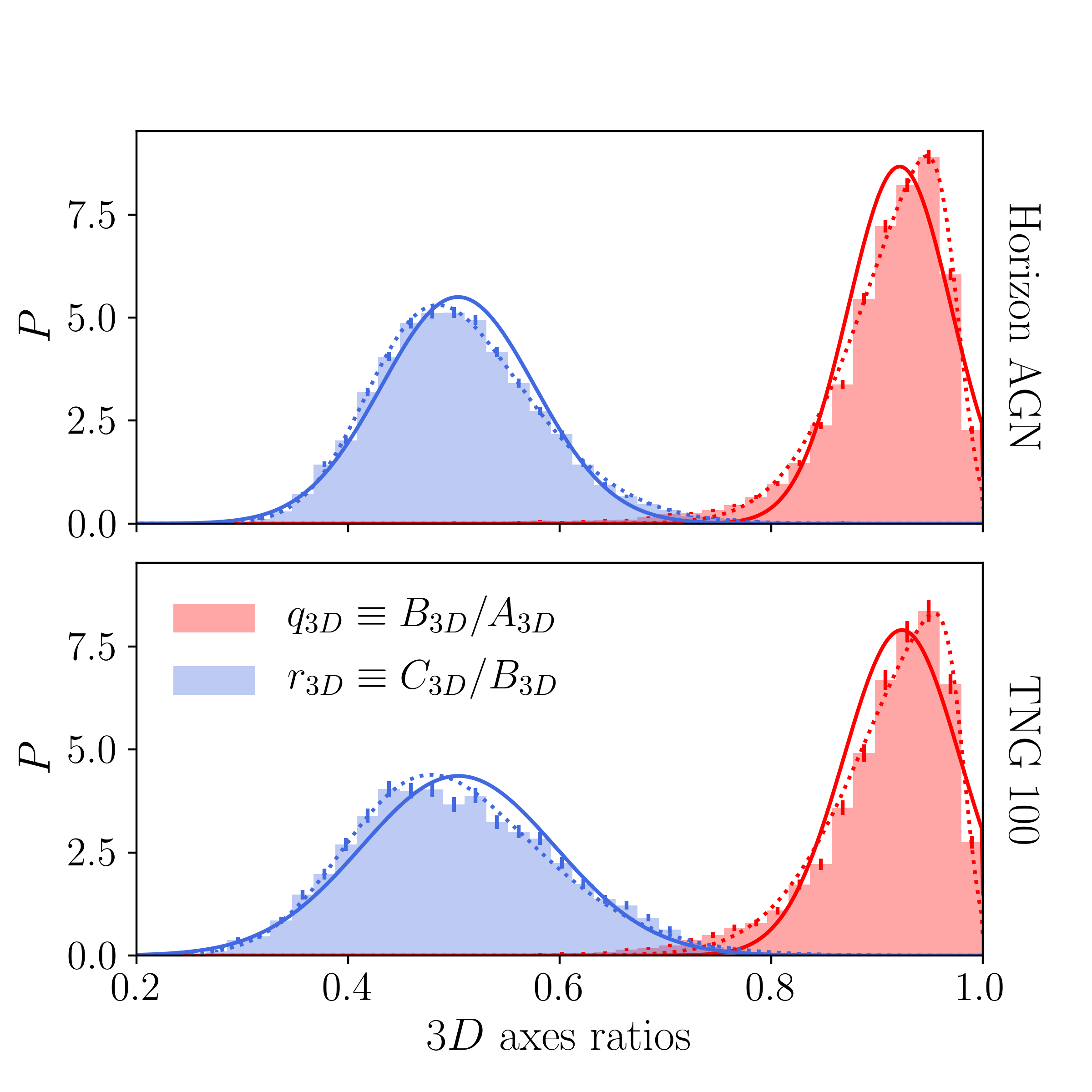}
\caption{Distributions of 3D axis ratios for disc galaxies in hydrodynamic simulations at $z=1.0$, obtained
from the simple moment of inertia. Solid and dotted lines show fits to a normal
and a skew normal distribution respectively.}
\label{fig:Pqr_hydrosims}
\end{figure}

In Fig. \ref{fig:Pqr_hydrosims}
we show the marginalized probability distributions of \qddd\ and \rddd\ for galaxies
classified as discs in the HAGN and TNG100 simulations (see Section \ref{sec:data:sims} for details).
We find very similar distributions for both simulations, which is interesting to note
given a) the significant differences in the physical models employed in the simulations
and b) the different parameters used for the morphological classification. The latter appears to
be reasonable, as the discs tend to be oblate with $r_{3D} < q_{3D}\lesssim 1$.
Note that the absence of thin discs with $r_{3D} << 1$ can be attributed to the
effective pressurized Equation of state employed in these simulations and their
relatively low resolution. Thin discs are present for instance in
the TNG50 run \citep[see][]{Pillepich19} or the New Horizon Simulation \citep{Park19},
which are higher resolution, lower volume versions of the runs analysed here.

We fit the probability distributions with normal and skew normal distributions,
given by $G(x) \propto \exp\{- \delta_x^2 \}$, and $G_s(x) = G(x) (1+{\rm erf}\{\gamma \delta_x\})$
respectively, with $\delta_x \equiv (x-x_0) / ( \sqrt{2} \sigma)$
\footnote{The fits are obtained by minimizing the $\chi^2$ deviation between model
and data, assuming shot noise errors on the latter.}. Both functions
are truncated outside of the interval $(0,1]$ to ensure that $ \infty > A_{3D} \ge B_{3D} \ge C_{3D}$.
We show in Fig. \ref{fig:Pqr_hydrosims} that the fit of the skew normal distribution
matches the axis ratios slightly better than the normal distribution, in particular \qddd. 
However, we decide to neglect the skewness in our modeling, since the improvement
in the 3D axis ratio fits is relatively small.
Furthermore we show in Appendix \ref{app:P3d_model_validation}
that a skewness on \qddd\ has a very minor impact on the corresponding distribution of \qdd\ and could
hence be constrained very poorly by observations. In the same appendix
we also show that \qddd\ and \rddd\ are uncorrelated in HAGN as well as in TNG100.
Based on these findings we approximate the joint 3D axis ratio distribution with the product
$G(q_{3D})G(r_{3D})$, i.e.
\eq{
\label{eq:P3d}
\tilde P(q_{3D},r_{3D}) =
\exp \Biggl\{
-\frac{1}{2}
\left[
\left( \frac{q_{3D} - q_0}{\sigma_q} \right)^2 +
\left( \frac{r_{3D} - r_0}{\sigma_r} \right)^2
\right]
\Biggr\},
}
where $q_0$, $\sigma_q$, $r_0$, $\sigma_r$ are the free model parameters.
The normalized truncated distribution is then given by
\eqa{
\label{eq:P3dnorm}
    P =
    \begin{cases}
     \tilde P_{3D} / \mathcal{N} & \text{if} \ q_{3D},r_{3D} \in (0,1] \\
    0 & \text{else}
    \end{cases}
\label{eq:qr_pdf}
}
with $\mathcal{N} = \int_{0}^1 \int_{0}^1 \tilde P_{3D}(q_{3D},r_{3D}) \ dr_{3D} \ dq_{3D}$.

\subsection{Error estimation and parameter inference}
\label{sec:method:infer_params}

We obtain constraints on our model parameter vector $\boldsymbol{\theta} = (q_0, \sigma_q, r_0, \sigma_r)$
from the observed data using Bayesian inference. We assume a multivariate normal distribution
of the likelihood for observing the data vector {\bf d} given the parameters $\boldsymbol{\theta}$, 
\eq{
\ln \mathcal{L}({\bf d}|\boldsymbol{\theta}) =
-\frac{1}{2}\chi^2({\bf d}|\boldsymbol{\theta}) + const.
}
with
\eq{
\chi^2({\bf d}|\boldsymbol{\theta}) = [{\bf d} - {\bf m(\boldsymbol{\theta})}]^{\mathsf T}C^{-1}[{\bf d} - {\bf m(\boldsymbol{\theta})}].
}
The data vector is the observed distribution \pdd,
measured in $25$ bins of equal width within the \qdd\ interval $(0,1]$ and $m(\boldsymbol{\theta})$
is the corresponding prediction from our \pddd\ model for the parameters $\boldsymbol{\theta}$. The posterior distribution of
the parameters $\boldsymbol{\theta}$ given the data ${\bf d}$ is given by Bayes' theorem as
\eq{
P(\boldsymbol{\theta}|{\bf d}) \propto \mathcal{L}({\bf d}|\boldsymbol{\theta})\Pi(\boldsymbol{\theta}),
}
where is $\Pi(\boldsymbol{\theta})$ is the prior, which we set to be flat in the intervals
$(0,1]$ and $[0.01,1]$ for the parameters $(q_0,r_0)$ and  $(\sigma_q$, $\sigma_r)$ respectively.
We thereby include parameter combinations in our sampling
that describe axis ratio distributions not only for discs
($r_0<<q_0$), but for any kind of ellipsoids, e.g. $r_{3D}\gtrsim q_{3D}$.

The covariance matrix $C$ is assumed to be given by $C_{ij} = \delta_{ij} \ \sigma_{i}^2$, with
Poisson shot noise variance $\sigma_i^2 \propto {N_i}$, where $N_i$ are the counts
of galaxies in a given bin $i$. The model prediction $m(\boldsymbol{\theta})$ is obtained by
drawing a sample of 3D axis ratio pairs $(q_{3D}, r_{3D})$ from the \pddd\ distribution
for a given set of parameters $\boldsymbol{\theta}$. The corresponding
\qdd\ axis ratios are then computed for a random orientation, as detailed
in Appendix \ref{app:project_ellipsoid}. The prediction for \pdd\ is measured from the resulting
\qdd\ sample in the same bins as the observed data.
For generating the predictions we choose a sample size of $10^5$, which is a compromise
between a fast computation, needed for efficiently estimating the posterior, and
having errors on the prediction which are negligible compared to those on the data vector.
The latter condition is satisfied, as the number of galaxies in our six COSMOS sub-samples
is more than two magnitudes smaller than the samples used for generating the predictions
(see Table \ref{tab:params_rec_cosmos}).
We estimate $P(\boldsymbol{\theta}|{\bf d})$ by sampling the parameter space with the
Markov-Chain-Monte-Carlo (MCMC) method using the code \verb|emcee|\footnote{emcee.readthedocs.io}
\citep{Foreman13}. For each posterior we run $32$ independent chains with at last $1,000$ steps each.
The best fit parameters are obtained from the position of the maxima of the marginalized
posterior distribution.

\subsection{Testing the reconstruction method}
\label{sec:method:test_reconstruction}

The method for reconstructing the distribution of 3D axis ratios \pddd\ from the
distribution of 2D axis ratios \pdd\ is now tested using the HAGN simulation. We
begin by validating two assumptions on which this method is based. Those are
1) that the galaxies' 3D stellar mass isodensities are coaxial, self-similar
3D ellipsoids, whose 3D axis ratios can be related analytically to the 2D axes
ratios of the projected 2D stellar mass isodensities as described in Appendix \ref{app:project_ellipsoid},
and 2) that our Gaussian model for the  3D axis ratio distributions from Equation (\ref{eq:qr_pdf})
is sufficiently accurate to provide good predictions for the 2D axis ratio distribution.
Subsequently, we test how well the distribution of 3D axis ratios in the simulation
can be recovered from the corresponding distribution of 2D axis ratios.

\begin{figure}
\centering\includegraphics[width=8.0 cm, angle=0]{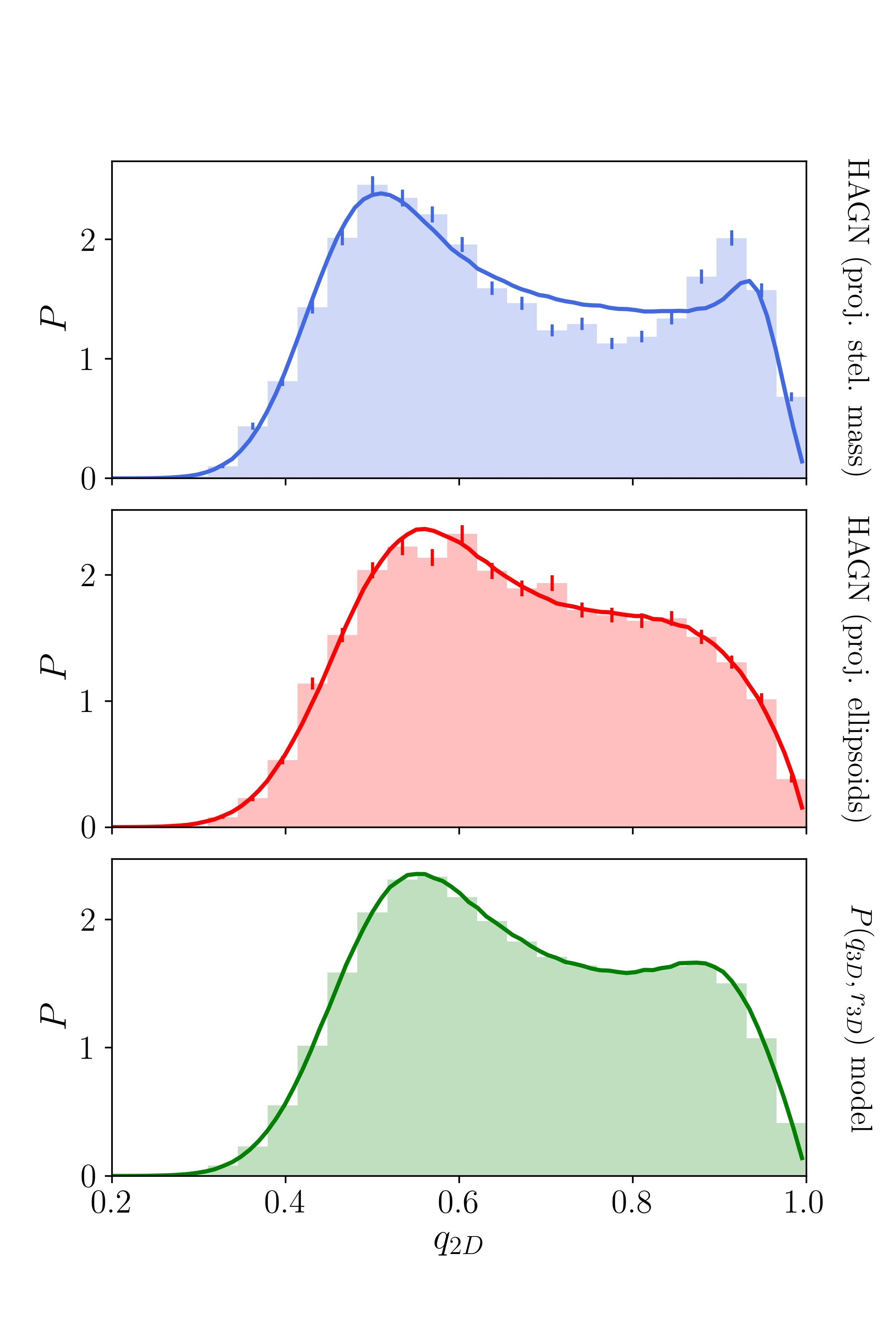}
\caption{Probability distribution of 2D axis ratios. 
Histograms in the top
and central panels show results from the
HAGN simulation, measured from the projected stellar mass distributions and analytic
projections derived from 3D axis ratios, assuming coaxial, self-similar ellipsoidal
stellar mass isodensities respectively. The histogram in the bottom panel shows
the 2D axis ratio distribution for a Gaussian model of the 3D axis
ratio distribution with parameters from the 3D fits, shown in Fig. \ref{fig:Pqr_hydrosims}.
In each panel we show fits of the model prediction as solid lines. The corresponding
parameter constraints and best fit values
are compared in Fig. \ref{fig:Pqr_params_corner} and Table \ref{tab:params_rec_hagn}.
Poisson shot-noise estimates of the standard deviation are shown as error bars on
the histograms. Note that errors are smaller than the line width in the bottom panel.}
\label{fig:q2D_fits}
\end{figure}

To test assumption 1) we compare in Fig. \ref{fig:q2D_fits} the \qdd\ distribution
in the HAGN simulation, measured directly from the projected distributions of each
galaxy's stellar particles as described in Section \ref{sec:data:sims:axes_ratios} (top panel),
to the distribution of 2D axis ratios, obtained analytically from each galaxy's 3D axes
ratios assuming ellipsoidal 3D stellar mass isodensities (central panel).
We find that the distributions differ significantly from each other,
which indicates that the assumption of ellipsoidal stellar mass isodensities
can only serve as a very rough approximation.
This conclusion can already be expected from the images of face-on late-type
galaxies in our COSMOS sample, shown in Fig. \ref{fig:acs_images}. However, certain characteristics
are present in both distributions, namely the cut-offs at
$q_{2D}\simeq0.4$ and $q_{2D}\simeq0.9$, as well as a skewness in the distributions
towards low axis ratios.
To test assumption 2) we show in the bottom panel of Fig. \ref{fig:q2D_fits} the \qdd\
distribution predicted for an ensemble of $10^6$ 3D axis ratios,
drawn from our \pddd\ model fit to the HAGN simulation (see Fig. \ref{fig:Pqr_hydrosims}
and \ref{fig:qr_contours}). We find that this predicted distribution is similar to the
results obtained from the analytical projection of 3D axis ratios (shown in the central
panel of Fig. \ref{fig:q2D_fits}), which indicates that our Gaussian model for \pddd\ from
Equation (\ref{eq:qr_pdf}) is an appropriate approximation for our analysis.

\begin{figure}
\centering\includegraphics[width=8 cm, angle=0]{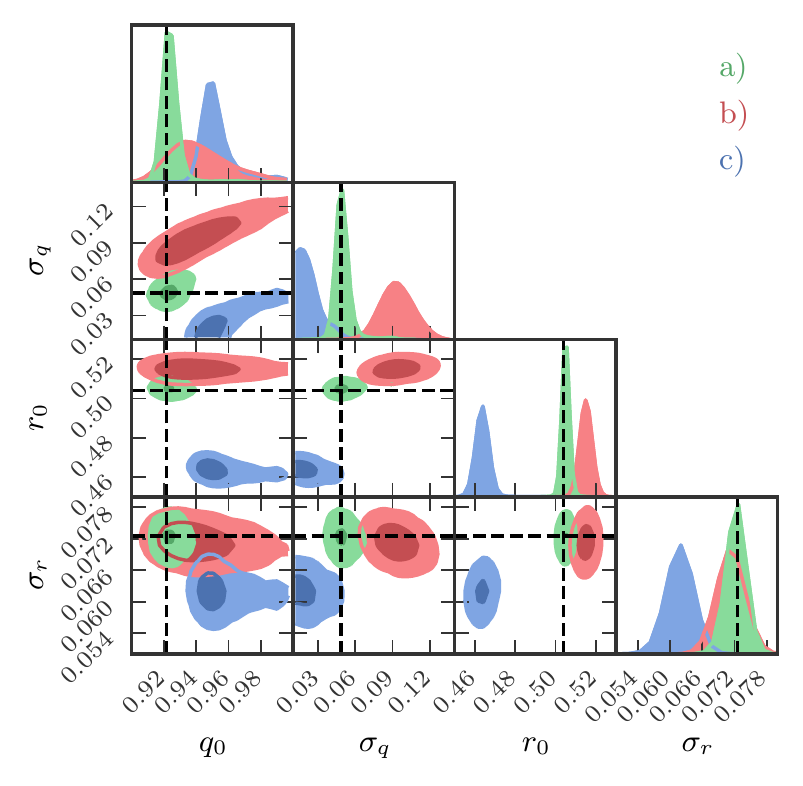}
\caption{
{Posteriors of the parameters of the Gaussian model for the 3D axis ratio distribution,
    given by Equation (\ref{eq:P3d}). Results are derived from the 2D axis ratio distributions of 
    a) an ensemble of projected ellipsoids with 3D axis ratios that follow the same Gaussian model
    (model input values are marked as black dashed lines),}
    b) projected 3D ellipsoids with 3D axis ratio
    distributions of disc galaxies in the HAGN simulation,
    c) projected stellar mass distributions of disc galaxies in HAGN.
    Light and dark areas mark $95$ and $68\%$ confidence levels respectively.
}
\label{fig:Pqr_params_corner}
\end{figure}

We study the performance of the reconstruction method, starting with a self consistency
test (a) in which we fit the \pddd\ model to match its own \pdd\ prediction, using parameters
from the HAGN fits shown in Fig. \ref{fig:q2D_fits}.
We find that the fit, shown as green line in the bottom panel of Fig. \ref{fig:q2D_fits},
is in good agreement with the reference measurements. The MCMC estimate of the posterior
distribution is displayed in Fig. \ref{fig:Pqr_params_corner} as light and dark green
contours, indicating the $68$\% and $95$\% confidence intervals, respectively. We find that the
parameter constraints are in good agreement with the input parameters of the \pddd\ model,
shown as black dashed lines in Fig. \ref{fig:Pqr_params_corner}. This result demonstrates that the 3D
axis ratio distribution can be reconstructed from the corresponding distribution of 2D
axis ratios for a sample of idealised disc galaxies, which satisfy the model assumptions
outlined above. Note that this is not necessarily the case for elliptical galaxies, which
are not subject of this work. In Appendix \ref{app:params_randsamp} we show that the
input parameters parameters are also recovered for samples that are as small as the observational
samples used on this work, but with larger confidence intervals.

We perform the same test under more realistic conditions by fitting the \qdd\ distribution
obtained by the projection of idealized 3D ellipsoids with axis ratios measured from HAGN galaxies (test b).
The fit, shown as red line in the central panel of Fig. \ref{fig:q2D_fits}, is in an good
agreement with the corresponding measurements. We attribute the remaining deviations to inaccuracies
of the \pddd\ model and not to the assumption of ellipsoidal stellar mass isodensities used
for obtaining the 2D axis ratios, since this assumption is employed for generating both, the
prediction as well as the reference measurements. The parameter constraints, shown as red contours
in Fig. \ref{fig:Pqr_params_corner}, are significantly biased with respect to the reference parameters
from the \pddd\ model fits to the 3D axis ratio distribution in HAGN. However, the relative deviations
of the best fits (defined as maxima of the marginalized probabilities) are at the percent level for
$q_0$, $r_0$ and $\sigma_r$ (see Table \ref{tab:params_rec_hagn}). The $\sigma_q$ parameter shows
the strongest deviation from the reference values of roughly $100$\%.

Our most realistic performance test of the reconstruction method consists in fitting the
model to the distribution of 2D axis ratios measured from each galaxy's projected stellar mass
distribution (test c). We find that the fit, shown as blue line in the top panel of Fig. \ref{fig:q2D_fits},
strongly deviates from the corresponding measurements. This finding can be expected from the
previously discussed deviation between the \qdd\ distribution obtained from the galaxies'
projected stellar masses and the one obtained analytically from the 3D axis ratio measurements
assuming idealized 3D ellipsoidal isodensities since the latter assumption is employed in the modeling.
However, some main characteristics, such as the cut-off of the distribution, the position
of the local maxima and the skewness towards low axis ratios, are captured by the fit.
We find that the parameter constraints, shown as blue contours in Fig. \ref{fig:Pqr_params_corner}
 are biased more strongly with respect to the reference
values from the direct \pddd\ model fits compared to the previous test case b)
(see also Table \ref{tab:params_rec_hagn}).

A potential shortcoming of these tests is that we defined discs in HAGN as the upper $20$\% of
galaxies with the highest values of $r_v = V_{rot}/\sigma_v$, which corresponds to a selection
cut at $r_v>1.06$ (see Section \ref{sec:data:sims:HAGN}). We thereby might include discs with
large bulges for which a one-component ellipsoidal model is inaccurate. In order to investigate
how robust our accuracy tests are towards variations of the disc selection criteria, we repeat
our analysis using
galaxies with $r_v>1.17$ and $r_v>1.27$.
This corresponds to selecting the upper $10$\% and $5$\% of galaxies with the highest
$r_v$ values respectively and should reduce the fraction of discs with large bulges
at the price of higher noise on the axis ratio distributions due to the smaller sample size.
Our results from these different samples show no significant improvement
in the accuracy of the reconstructed parameters when increasing the
minimum values of $r_v$. This finding speaks against the possibility that
discs with large bulges are the main reason for the inaccuracies
of the reconstruction method.
For test c), that is based on the projected 2D stellar mass distributions, we obtain
overall relative deviations between the reconstructed model parameters and those derived
from fits to the 3D stellar mass distributions of around
$\sim3$\%, $\sim60$\%, $\sim8$\% and $\sim20$\% for $q_0$, $\sigma_q$, $r_0$ and $\sigma_r$,
respectively (Table \ref{tab:reldiff_params_hagn}).

\begin{table*}
    \centering
    \begin{tabular}{l l l l l l l l l l}
    & & model & $q_0$ & $\sigma_q$ & $\gamma_q$ & $r_0$ & $\sigma_r$ & $\gamma_r$ & \\ \hline \hline
    fits to 3D axis ratio distribution: & & skew Gauss & $0.977$ & $0.081$ & $-5.009$ & $0.426$ & $0.112$ & $2.016$  \\
    & & Gauss & $0.921$ & $0.049$ & - & $0.504$ & $0.073$ & - \\ \hline
    reconstruction from 2D axis ratio distribution: & projected 3D ellipsoids & Gauss & $0.934 $ & $0.093 $ & - & $0.515$ & $0.071$ & -  \\
                                                   & projected stellar mass & Gauss & $0.949 $ & $0.017 $ & - & $0.464$ & $0.062$ & - \\
        \end{tabular}
    \caption{Best fit parameters of the model for the 3D axis ratio distribution, derived from disc galaxies in the HAGN simulation at $z=1.0$.
    The two top rows show results for the Gaussian and the skew Gaussian model from direct fits to the marginalised distributions of \qddd\ and \rddd\
    (see Section \ref{sec:method:P3Dmodel}), shown in Fig. \ref{fig:Pqr_hydrosims}.
    The two bottom rows show the parameters of the Gaussian model inferred from the distribution of 2D axis ratios using the reconstruction method.
    The 2D axis ratios used for the reconstruction are obtained either analytically from the galaxies' 3D shapes, assuming that the stellar mass distributions
    are perfect 3D ellipsoids, or directly from shape measurements of the projected stellar mass distribution (labelled as projected ellipsoids and projected stellar mass respectively, see Fig. \ref{fig:q2D_fits}).
    }
    \label{tab:params_rec_hagn}
\end{table*}

\begin{table} 
\centering
 \begin{tabular}{ c | c c c c}
    min($V_{rot}/\sigma_v$) & $\Delta q_0 [\%]$ & $\Delta \sigma_q [\%]$ & $\Delta r_0 [\%]$ & $\Delta \sigma_r [\%]$ \\ \hline
    $1.06$  & $3.0$  & $-64.1$  & $-8.0$  & $-14.6$ \\
    $1.17$  & $3.1$  & $-74.2$  & $-8.1$  & $-16.6$ \\
    $1.27$  & $3.6$  & $-51.2$  & $-8.3$  & $-24.3$ \\
    \end{tabular}
    \caption{
    Relative deviations between the parameters of the Gaussian
    \pddd\ model, reconstructed from the
    2D axis ratio distribution of the projected stellar masses in HAGN (test case c)
    in Section \ref{sec:method:test_reconstruction}) with respect
    to the reference parameters obtained from direct fits to the 3D axis ratio distribution.
    Results are shown for three samples, consisting of the $20\%$, $10$\% and $5$\% of galaxies
    at $z=1.0$ with the highest values of $V_{rot}/\sigma_v$ respectively. Each samples minimum
    value of $V_{rot}/\sigma_v$ is given in the left column.
    }
    \label{tab:reldiff_params_hagn}
\end{table}

\begin{figure}
\centering\includegraphics[width=8.0 cm, angle=0]{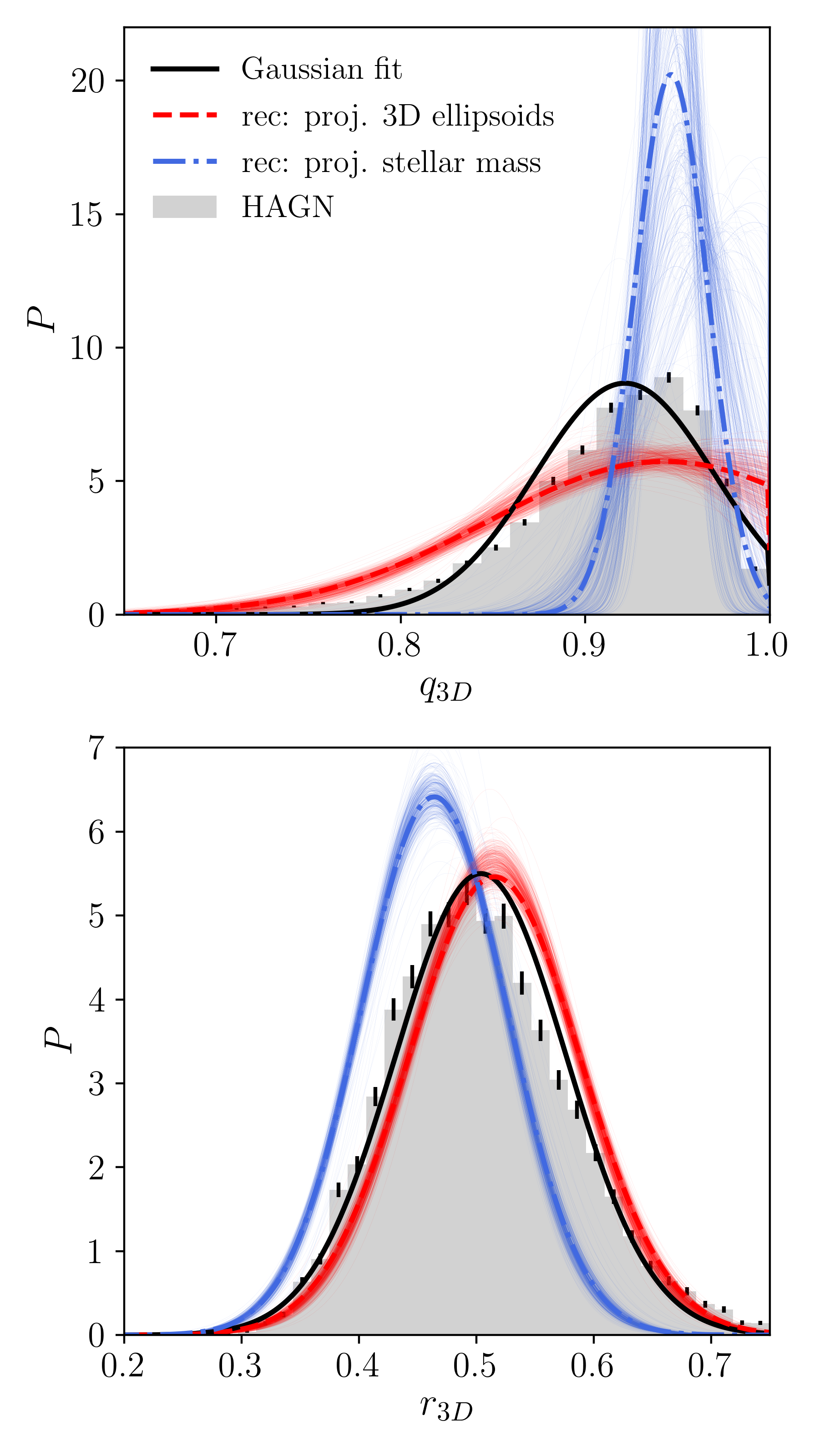}
\caption{Testing the reconstruction of the 3D axis ratio distribution of disc galaxies in the HAGN simulation at $z=1.0$.
Grey histograms show measurements in the simulation with $1\sigma$ shot-noise errors displayed as black bars for each bin.
Black solid lines show the fits to a truncated normal distribution, which are also shown as solid lines
in the top panel of Fig. \ref{fig:Pqr_hydrosims}. Red dashed lines show the distribution reconstructed
from fits to the 2D axis ratio distribution which was derived from the 3D axis ratios using the
ellipsoidal galaxy model (see central panel of Fig. \ref{fig:q2D_fits}). Dashed-dotted blue lines
show the reconstruction based on fits to the 2D axis ratios measured from the projected stellar mass
(see top panel of Fig. \ref{fig:q2D_fits}). Thin lines show predictions for $500$ random
sampling points of the posterior distributions and reflect the uncertainties expected on the prediction.}
\label{fig:Pqr_reconstruction_hagn}
\end{figure}

Refocusing on our sample of discs in HAGN with $r_v>1.06$, we compare in
Fig. \ref{fig:Pqr_reconstruction_hagn} the reconstructed distributions
for $q_{3D}$ and $r_{3D}$ directly with the HAGN measurements. This figure illustrates that
the location of the maxima of the reconstructed $q_{3D}$ and $r_{3D}$ distributions, given by
the model parameters $q_0$ and $r_0$, are close to the positions of the maxima in the measurements.
The width of the reconstructed \qdd\ distribution, quantified by the $\sigma_q$ parameter,
differs strongly from the HAGN measurements and is therefore not considered in the final
discussion of our results.

We emphasize that the conclusion drawn in this section could change when conducting these
tests under more realistic conditions. An ideal test would be based on synthetic images
including effects of dust extinction, lensing, PSF convolution
and pixelization, as well as image noise. These images would then need to be analysed
using the same software as used for the observations. Such a realistic test is beyond the scope
of this work, but would be an interesting objective for future investigations.

\section{3D axis ratio distributions in COSMOS}
\label{sec:results}

\begin{figure*}
\centering\includegraphics[width=16.0 cm, angle=0]{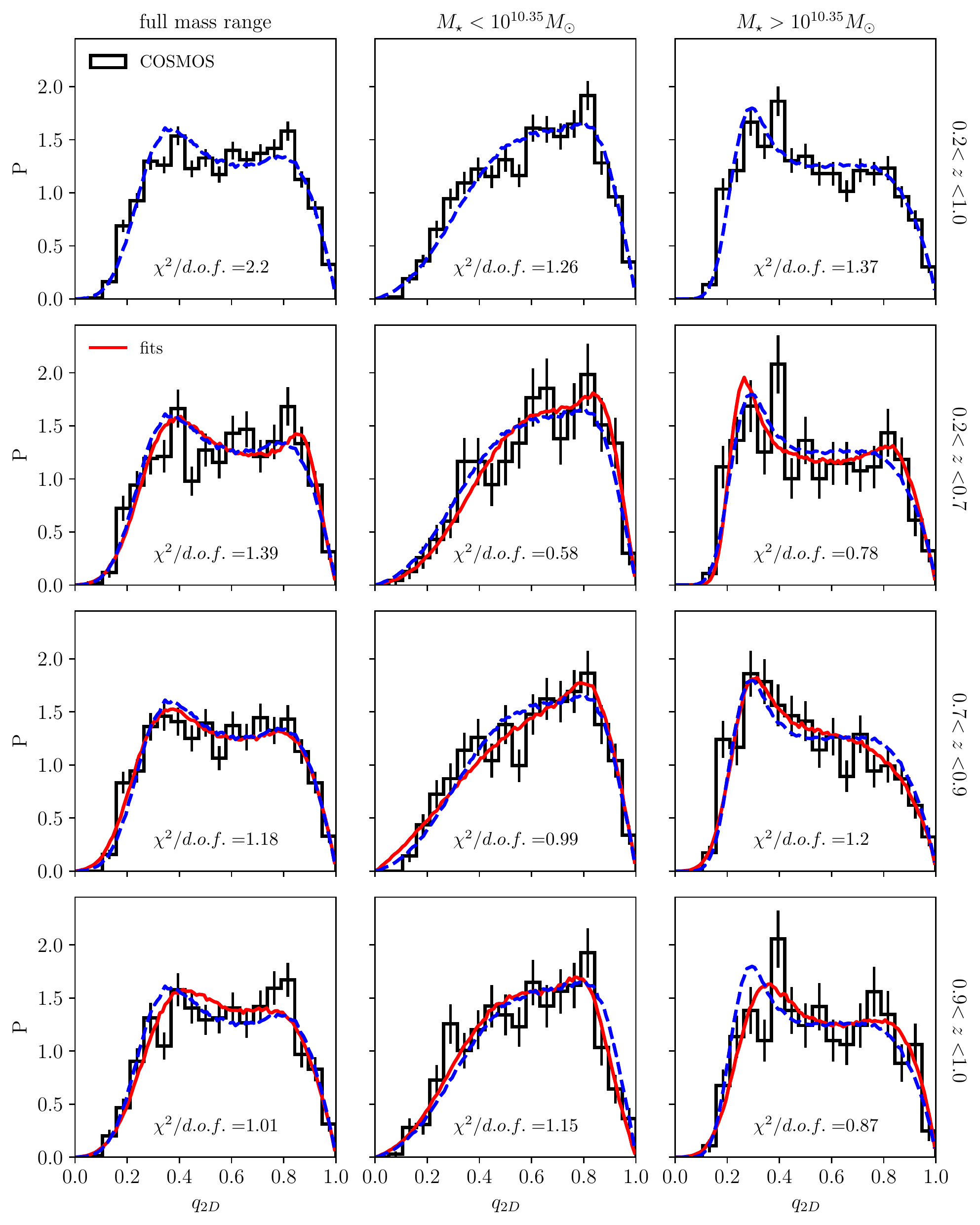}
\caption{Probability distribution of apparent axis ratios \qdd\ of disc-dominated galaxies
from our COSMOS samples in different redshift and stellar mass ranges, shown as black histograms.
Error bars show shot noise estimates of the standard deviation in each bin.
Red lines are model fits for each redshift sub-sample in a given stellar mass range.
Blue lines are fits for the stellar mass samples defined over the full redshift range (shown in the top panels).
The model parameters derived from these fits are summarized in Table \ref{tab:params_rec_cosmos}.
The corresponding posterior distributions are shown in Fig. \ref{fig:corner_cosmos}.
}
\label{fig:q2D_fits_cosmos}
\end{figure*}

\begin{table*}
    \centering
    \begin{tabular}{c c || c c c c c c c}
stellar mass range & redshift range & $N_{gal}$ & $q_0$ & $\sigma_q$ & $r_0$ & $\sigma_r$ & $s_0$ & $\chi^2 / d.o.f.$ \\ \hline  \hline 
full mass range  & $0.2 < z < 1.0$ & $3749$ & $0.88_{0.87}^{0.92}$ & $0.08_{0.07}^{0.12}$ & $0.29_{0.27}^{0.30}$ & $0.09_{0.09}^{0.12}$ & $0.25_{0.24}^{0.27}$ & $2.20$ \\
& $0.2 < z < 0.7$ & $971$ & $0.89_{0.87}^{0.91}$ & $0.04_{0.03}^{0.10}$ & $0.29_{0.28}^{0.31}$ & $0.10_{0.09}^{0.13}$ & $0.26_{0.25}^{0.28}$ & $1.39$ \\
& $0.7 < z < 0.9$ & $1551$ & $0.89_{0.87}^{0.92}$ & $0.09_{0.07}^{0.12}$ & $0.27_{0.26}^{0.28}$ & $0.10_{0.09}^{0.12}$ & $0.24_{0.23}^{0.25}$ & $1.18$ \\
& $0.9 < z < 1.0$ & $1217$ & $0.87_{0.86}^{0.93}$ & $0.10_{0.07}^{0.14}$ & $0.32_{0.30}^{0.33}$ & $0.12_{0.11}^{0.14}$ & $0.28_{0.26}^{0.29}$ & $1.01$ \\ \hline
$M_\star < 10^{10.35} M_\odot$  & $0.2 < z < 1.0$ & $1913$ & $0.87_{0.86}^{0.90}$ & $0.07_{0.06}^{0.10}$ & $0.37_{0.36}^{0.39}$ & $0.21_{0.20}^{0.23}$ & $0.33_{0.31}^{0.34}$ & $1.28$ \\
& $0.2 < z < 0.7$ & $443$ & $0.89_{0.82}^{0.91}$ & $0.05_{0.04}^{0.22}$ & $0.43_{0.41}^{0.63}$ & $0.21_{0.18}^{0.29}$ & $0.38_{0.36}^{0.53}$ & $0.58$ \\
& $0.7 < z < 0.9$ & $789$ & $0.86_{0.84}^{0.88}$ & $0.07_{0.06}^{0.11}$ & $0.28_{0.11}^{0.31}$ & $0.35_{0.30}^{0.52}$ & $0.23_{0.10}^{0.27}$ & $0.98$ \\
& $0.9 < z < 1.0$ & $681$ & $0.83_{0.80}^{0.86}$ & $0.06_{0.05}^{0.21}$ & $0.35_{0.33}^{0.46}$ & $0.20_{0.18}^{0.29}$ & $0.30_{0.28}^{0.40}$ & $1.13$ \\ \hline
$M_\star > 10^{10.35} M_\odot$  & $0.2 < z < 1.0$ & $1836$ & $0.90_{0.89}^{0.97}$ & $0.15_{0.12}^{0.18}$ & $0.26_{0.25}^{0.27}$ & $0.06_{0.05}^{0.06}$ & $0.24_{0.23}^{0.25}$ & $1.39$ \\
& $0.2 < z < 0.7$ & $532$ & $0.88_{0.86}^{0.94}$ & $0.07_{0.06}^{0.15}$ & $0.24_{0.24}^{0.26}$ & $0.04_{0.03}^{0.06}$ & $0.21_{0.21}^{0.23}$ & $0.79$ \\
& $0.7 < z < 0.9$ & $768$ & $0.98_{0.88}^{0.98}$ & $0.25_{0.20}^{0.29}$ & $0.27_{0.26}^{0.29}$ & $0.06_{0.04}^{0.07}$ & $0.26_{0.24}^{0.28}$ & $1.18$ \\
& $0.9 < z < 1.0$ & $536$ & $0.92_{0.88}^{0.97}$ & $0.12_{0.08}^{0.15}$ & $0.28_{0.27}^{0.30}$ & $0.09_{0.07}^{0.10}$ & $0.26_{0.25}^{0.28}$ & $0.87$
    \end{tabular}
    \caption{Parameters of the 3D axis ratio distribution model (equations (\ref{eq:P3d}) and (\ref{eq:P3dnorm})), inferred from fits
    to the 2D axis ratio distribution from the COSMOS survey, shown in Fig. \ref{fig:q2D_fits_cosmos}. Results are given for
    different stellar mass - redshift samples, which are selected as indicated in the two left columns. For each model parameter we provide the lower and upper
    limit of the $68$\% confidence level, obtained from the marginalized posteriors. The right column shows the $\chi^2$ goodness of fit per degree
    of freedom.}
    \label{tab:params_rec_cosmos}
\end{table*}
We apply the method for reconstructing 3D axis ratio distributions to the different samples
from the COSMOS data, that were selected by stellar mass and redshift as described in Section \ref{sec:data:cosmos:sub_samples}.
The fits to the observed distribution of 2D axis ratios from which we infer the parameters of
our model for the 3D axis ratio distribution are shown in Fig. \ref{fig:q2D_fits_cosmos}.
For the samples that are selected by both, stellar mass and redshift, we find that the fits
match the measurements well with $\chi^2$ values per degree of freedom around unity
(see Table \ref{tab:params_rec_cosmos}), capturing the skewness towards low and high values of \qdd\
for the high and low mass samples, respectively (as shown in Fig. \ref{fig:qrssfr_pdf}).

The fits to the samples that are selected only by stellar mass or only by redshift, shown in the top
row and left column of Fig \ref{fig:q2D_fits_cosmos} respectively, are less accurate as indicated by the
higher $\chi^2$ values per degree of freedom. This finding can be expected on one hand from the larger
sample size that decreases the errors on the measurements and hence increases the significance
of deviations between measurements and model with respect to the errors. On the other hand the underlying model assumption
that the 3D axis ratios, \qddd\ and \rddd\ , follow a Gaussian distribution may be less accurate
for samples that are more broadly defined. This could in particular be the case if sub-samples have very
distinct \qdd\ distributions, like the low and high mass sub-samples.

However, we note that the model fits the observed \qdd\ distributions overall better than corresponding
distributions from the projected stellar mass in the HAGN simulation, shown in the top panel of Fig. \ref{fig:q2D_fits}.

This finding might result from the fact that we selected all disc galaxies in HAGN, including those with
large bulges for which a one-component {ellipsoidal shape} model is clearly an inadequate approximation.
Additional reasons may be differences between the 2D shapes of the projected stellar mass and the projected
luminosity densities or shortcomings of the simulation, caused for instance by resolution effects
as discussed in Section \ref{sec:data:sims}.

In Fig. \ref{fig:corner_cosmos} we show the MCMC estimates for the posterior probability
distribution of the \pddd\ model parameters, $q_0$, $r_0$, $\sigma_q$ and $\sigma_r$,
quantifying the average disc circularity, the average relative disc thickness and the corresponding
dispersions, respectively. The best fit parameters, defined as the positions of the maxima
of the posteriors are summarized in Table \ref{tab:params_rec_cosmos}. 
In this table we also provide results for the parameter $s_0 \equiv q_0 r_0$,
whose posterior distribution we obtain by multiplying the coordinates
of the sampling points of the posterior distributions of $q_0$ and $r_0$. It specifies
the position of the maximum in the $P(q_{3D}, s_{3D})$ distribution and quantifies the disc
thickness relative to the major axis, facilitating a comparison to the disc circularity parameter $q_0$
and the estimation of the absolute disc thickness (see Section \ref{sec:disc_thickness}).
The posteriors shown in Fig. \ref{fig:corner_cosmos} and the corresponding model parameters in
Table \ref{tab:params_rec_cosmos} were derived from \qdd\ distributions measured in $25$ bins.
When repeating the same analysis using $20$ bins we find only marginal changes of the
posteriors and no significant change in the corresponding model parameters.
However, larger fluctuations may occur for larger variations of the binning.
In Appendix \ref{app:params_randsamp} we show that the posteriors can be strongly affected
by noise in the data. Varying the binning would change the noise and therefore the posteriors.
However, a rigorous investigation of how the binning impacts the posteriors is beyond the scope of
our analysis.

\subsection{Redshift dependence}

We start examining the redshift dependence of the galaxy shape distribution by
comparing the fits to the different redshift sub-samples with those for the entire redshift range
(shown as red solid and blue dashed lines in Fig. \ref{fig:q2D_fits_cosmos} respectively).
We find that within a given mass range, these different fits are close to each other, which is a
first indication that the redshift evolution of the axis ratio distribution in our samples is weak.

This visual impression lines up with the relatively small variation of the
best fit parameters for the average disc circularity and the relative disc thickness ($q_0$ and $r_0$)
with redshift. For the full mass redshift sub-samples we find the values of $q_0$ and $r_0$ to vary
by $\sim 1 \%$ and $\sim 10 \%$ around the results for the full redshift range respectively (see Table \ref{tab:params_rec_cosmos}).
For the higher and low mass sub-samples these variation increase up to $\sim 10 \%$ and $\sim 25 \%$ 
for $q_0$ and $r_0$ respectively. The redshift variations
of the corresponding dispersions, $\sigma_q$ and $\sigma_r$ are much higher
with up to $\sim 50\%$.
However, the amplitudes of these different redshift variations are overall consistent with the
$68\%$ confidence intervals from the marginalized posteriors.

In order to obtain a more detailed insight into how the parameter variations with redshift
compare to the uncertainties, we show the posterior probability distributions of
the \pddd\ model parameters from the different sub-samples in Fig. \ref{fig:corner_cosmos}.
The $68$\% confidence levels of the posteriors for the full mass samples in the different redshift bins,
shown in the top left of Fig. \ref{fig:corner_cosmos}, are mostly overlapping. A weak indication for a
redshift dependence in the data might be given by the fact that the $68$\% confidence levels of the
intermediate and the high redshift samples are slightly disconnected in the $r_0$ direction. Nonetheless,
the $95$\% confidence levels still overlap clearly in that case.
More significant differences between the posteriors from different redshift bins are present for the
low and high mass sub-samples, as shown in the bottom left and right of Fig. \ref{fig:corner_cosmos} respectively.
However, for the low mass sample the $68$\% confidence levels still overlap plainly.
For the high mass sample the $68$\% confidence level of the central redshift bin is slightly
disjoint from those of the other bins, although the $95$\% confidence levels from all redshifts still overlap.
Overall our Bayesian analysis does not provide convincing evidence for
a redshift dependence of the \pddd\ model parameters
that is significant with respect to the uncertainties.
It is thereby important to note that the differences in the shapes of the posteriors are not an indication for such
a dependence since they can be expected from the sampling noise in the \qdd\
distributions, as we demonstrate in Appendix \ref{app:params_randsamp}.

It is further important to note that the insignificant redshift dependence
of the parameters with respect to the uncertainties does not imply that the shapes of
disc-dominated galaxies do not evolve since $z=1.0$. A
potential evolution could just be too weak to be detected reliably with the Bayesian analysis
presented here. In particular the large uncertainties
on the dispersions of the circularity and relative thickness may obscure
even a relatively strong redshift dependence of these parameters.
We therefore employ the two-sample Kolmogorov–Smirnov test
\citep[e.g.][hereafter referred to as K-S test]{Hodges58}
as an independent tool for testing the null hypothesis that the axis ratio distributions from two
different redshift samples are drawn randomly from the same redshift independent distribution.
We apply this test directly on the observed \qdd\ distributions and are therefore
independent of any model assumptions. 
In Table \ref{tab:ks_q2d_cosmos} we show the $p-$values of the K-S test,
which is the probability of obtaining a K-S test result that is as large or larger than
the one obtained for the two samples under the null hypothesis.
A typical, but arbitrary rejection criteria is $p < 0.05$.
For the redshifts samples defined over the full mass range the K-S test does not reject the
null hypothesis. This result lines up with the strong overlap of the model parameter contours
for these samples, which we discussed above.
The situation is less conclusive for the $p$-values from the low and high mass
sub-samples.
For the low mass sub-sample the K-S test suggests, that the \qdd\ distributions from the lowest and
highest redshift bins are drawn from different underlying distributions, since the $p$-value of $0.0273$
is below $0.05$. This finding is interesting, given that the parameter contours from these two redshift
bins are strongly overlapping (see bottom left panel of Fig. \ref{fig:corner_cosmos}). Given this apparent disagreement
and that fact that i) our arbitrarily chosen rejection criteria just slightly surpassed for this redshift bin combination
and ii) the null hypothesis has not been rejected for the other redshift bins combinations, 
we conclude that the low mass sub-sample provides no strong indication for a substantial
redshift evolution of the galaxy shapes.

For the high mass sub-sample the K-S test suggests, that the \qdd\ values from the intermediate and
the highest redshift bin are drawn from different underlying distributions. In this case the
rejection criteria is clearly fulfilled by the $p$-value of $0.0053$. This finding lines up with
with the weaker overlap of the corresponding parameter contours, shown in the bottom right of Fig. \ref{fig:corner_cosmos},
and could be an indication for a redshift dependence of the shapes of high mass disc-dominated galaxies.
However, it could also result from systematic effects, such as potential misclassifications of galaxy
morphologies in the ZEST catalogue.
Another explanation for a redshift dependence might be cosmic variance, as galaxy clusters in the COSMOS field
(see Fig. \ref{fig:vol_lim_samp}) could change the mean values and dispersions of the disc thickness and circularity
in a given redshift bin.
It is thereby interesting to note that the \qdd\ distributions from the lowest and the highest redshift
sample are consistent with each other, according to both, the K-S test as well the overlap of the parameter
contours for these two samples shown in Fig. \ref{fig:corner_cosmos}, which speaks against a physical evolution
of the shapes.
However, the same systematic effects may also change the \qdd\ distributions in
such a way that a detection of a redshift evolution is prevented. Samples with more
objects that are distributed over larger areas and morphological classifications based on
different techniques would be needed to better understand the impact of misclassification or
cosmic variance on our results.

\begin{table}    \centering
 \begin{tabular}{ c | c | c | c }
                        & full mass range &$M_\star < 10^{10.35}$ $1$    & $M_\star > 10^{10.35}$ \\ \hline
     $z_0$ - $z_1$      & 0.3796 & 0.2177 & 0.2496 \\
     $z_1$ - $z_2$      & 0.1684 & 0.1648 & 0.0053 \\
     $z_2$ - $z_0$      & 0.8569 & 0.0273 & 0.1221 \\
    \end{tabular}
    \caption{$p-$values of the two-sample Kolmogorov–Smirnov test. The test
    is applied on pairs of 2D axis ratio distributions measured in three redshift bins
    (Fig. \ref{fig:q2D_fits_cosmos}). The redshift bins $z_0$, $z_1$, $z_2$
    correspond to the ranges $[0.2,0.7], [0.7,0.9], [0.9,1.0]$ respectively.
    }
    \label{tab:ks_q2d_cosmos}
\end{table}

\subsection{Mass dependence}
{We now proceed with studying the dependence of the galaxy shapes on stellar mass.
Given that the evidence for a redshift evolution of the shapes is overall weak,
we will thereby focus on the mass samples that are defined over the entire redshift
range ($0.2 < z < 1.0$) to maximize the statistical power of the samples.

\subsubsection{Intrinsic disc circularity}
\label{sec:disc_circularity}

\begin{figure*}
\centering\includegraphics[width=7.5 cm, angle=0]{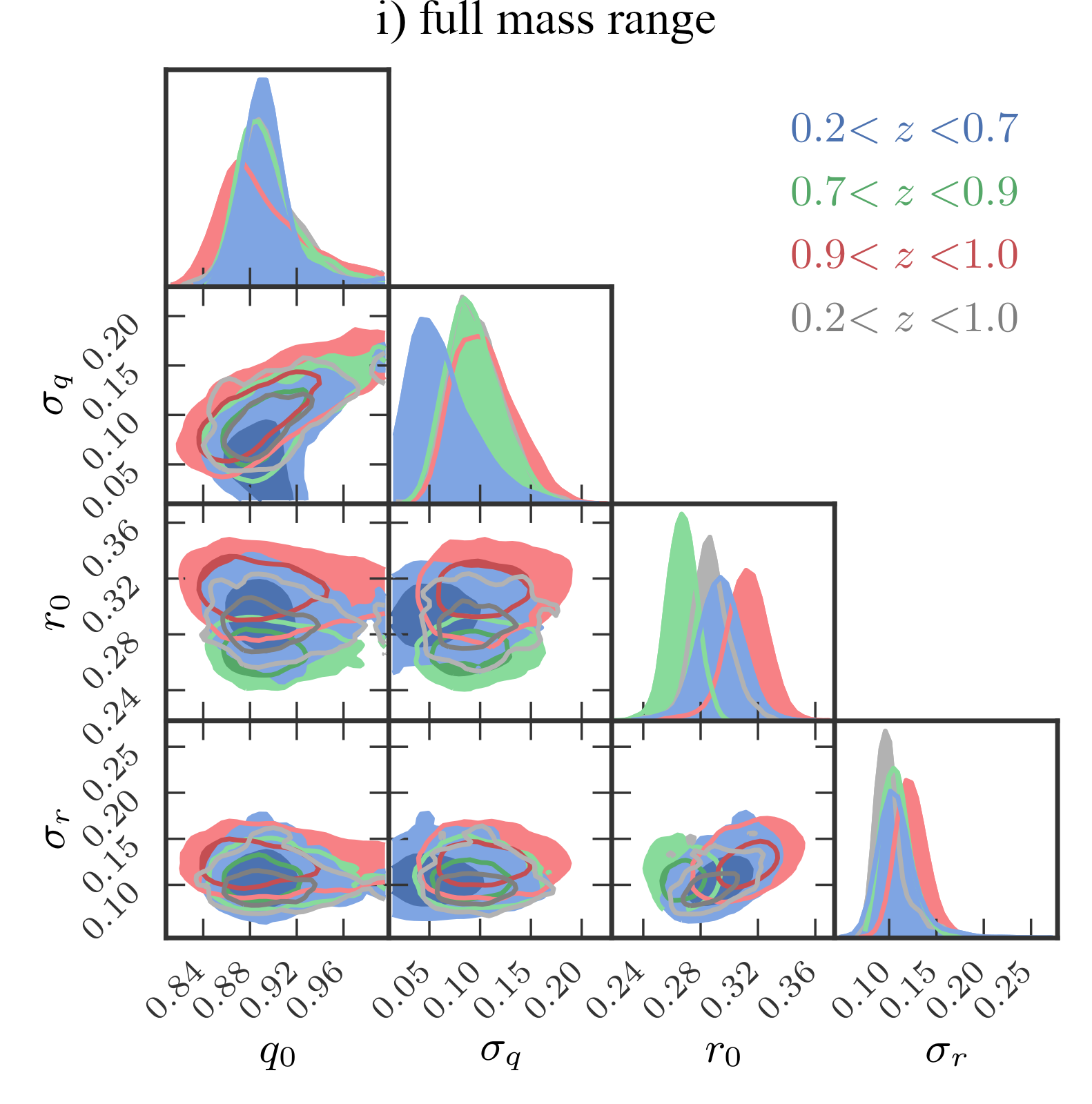}
\centering\includegraphics[width=7.5 cm, angle=0]{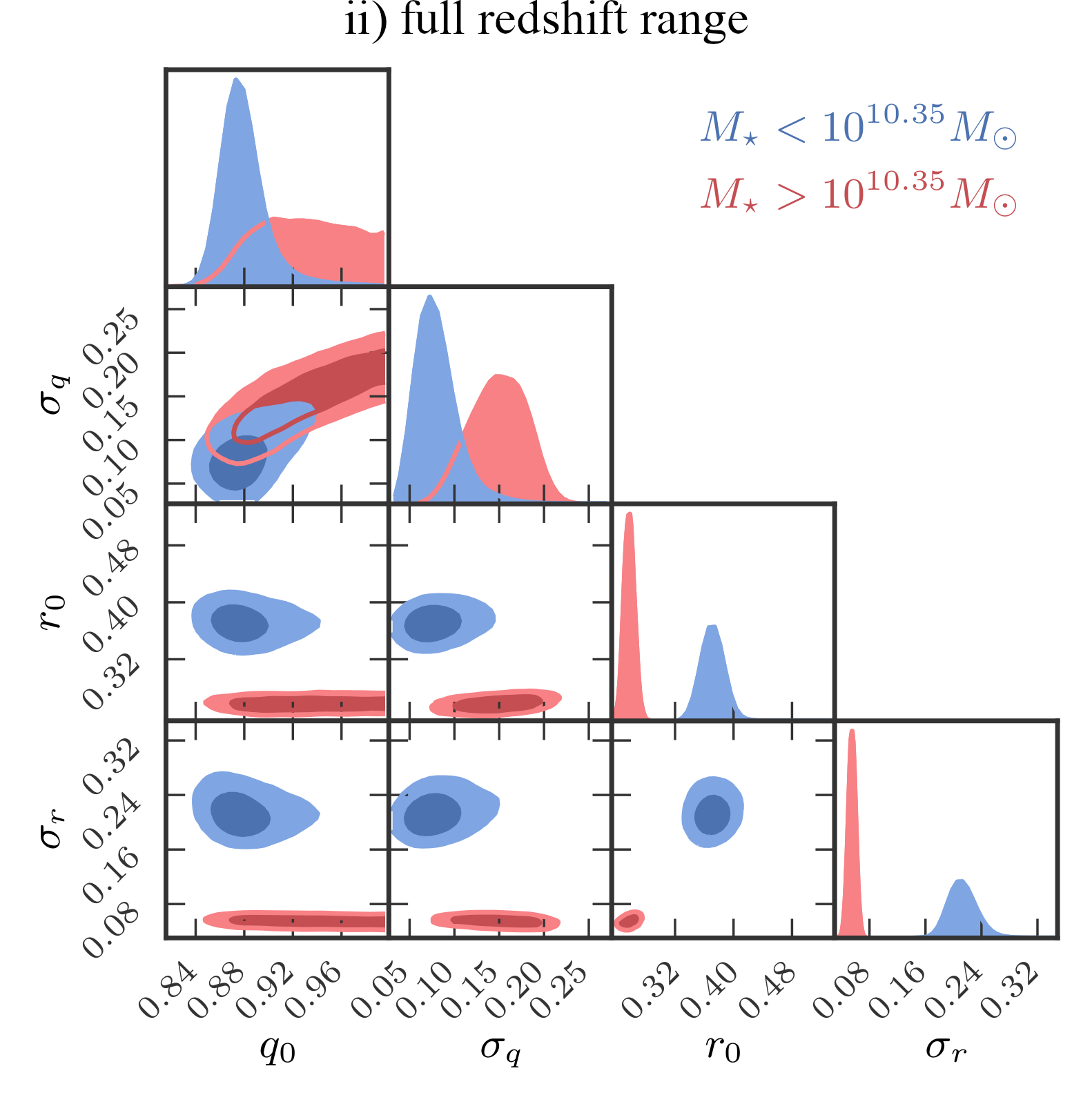}
\centering\includegraphics[width=7.5 cm, angle=0]{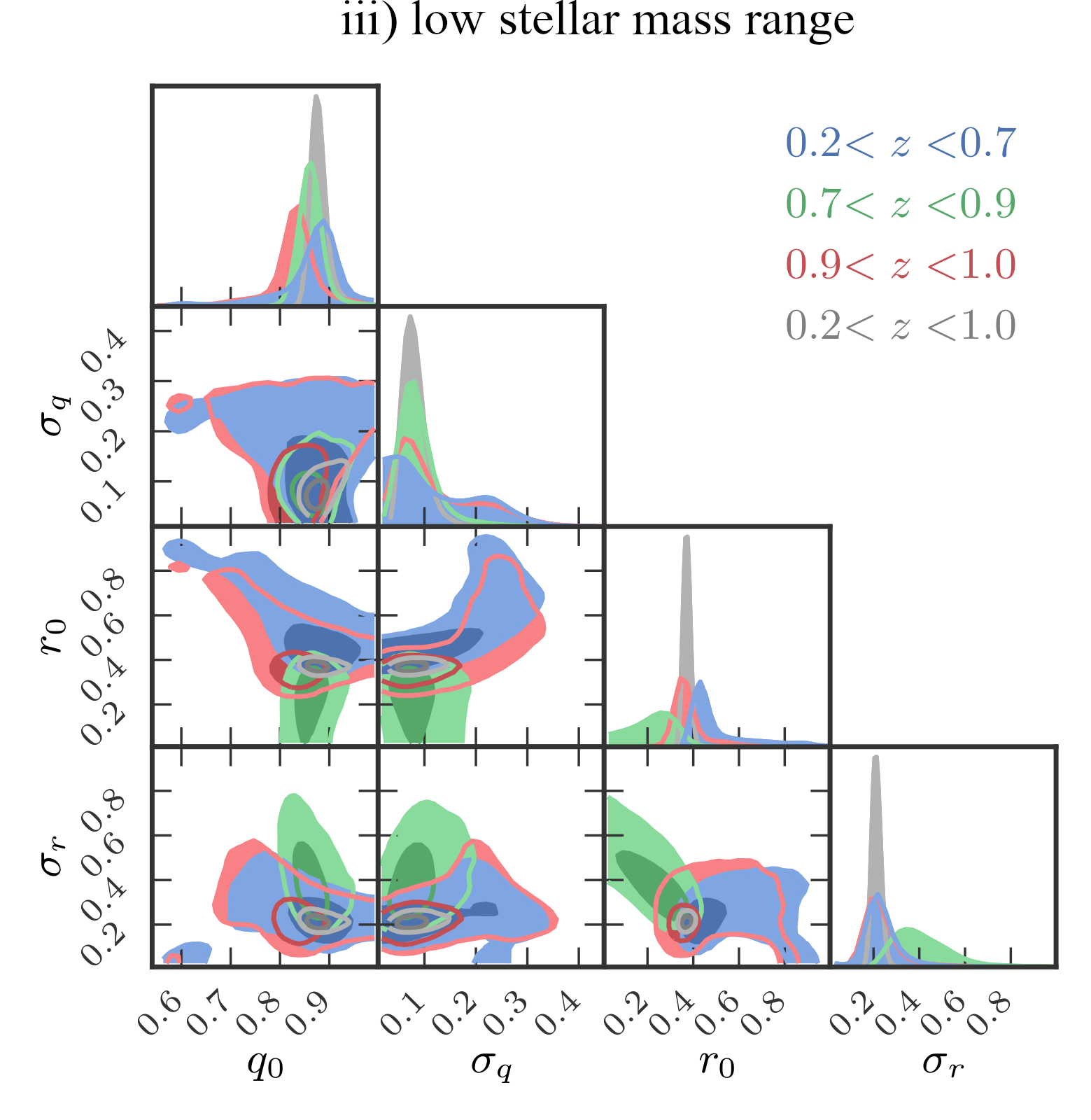}
\centering\includegraphics[width=7.5 cm, angle=0]{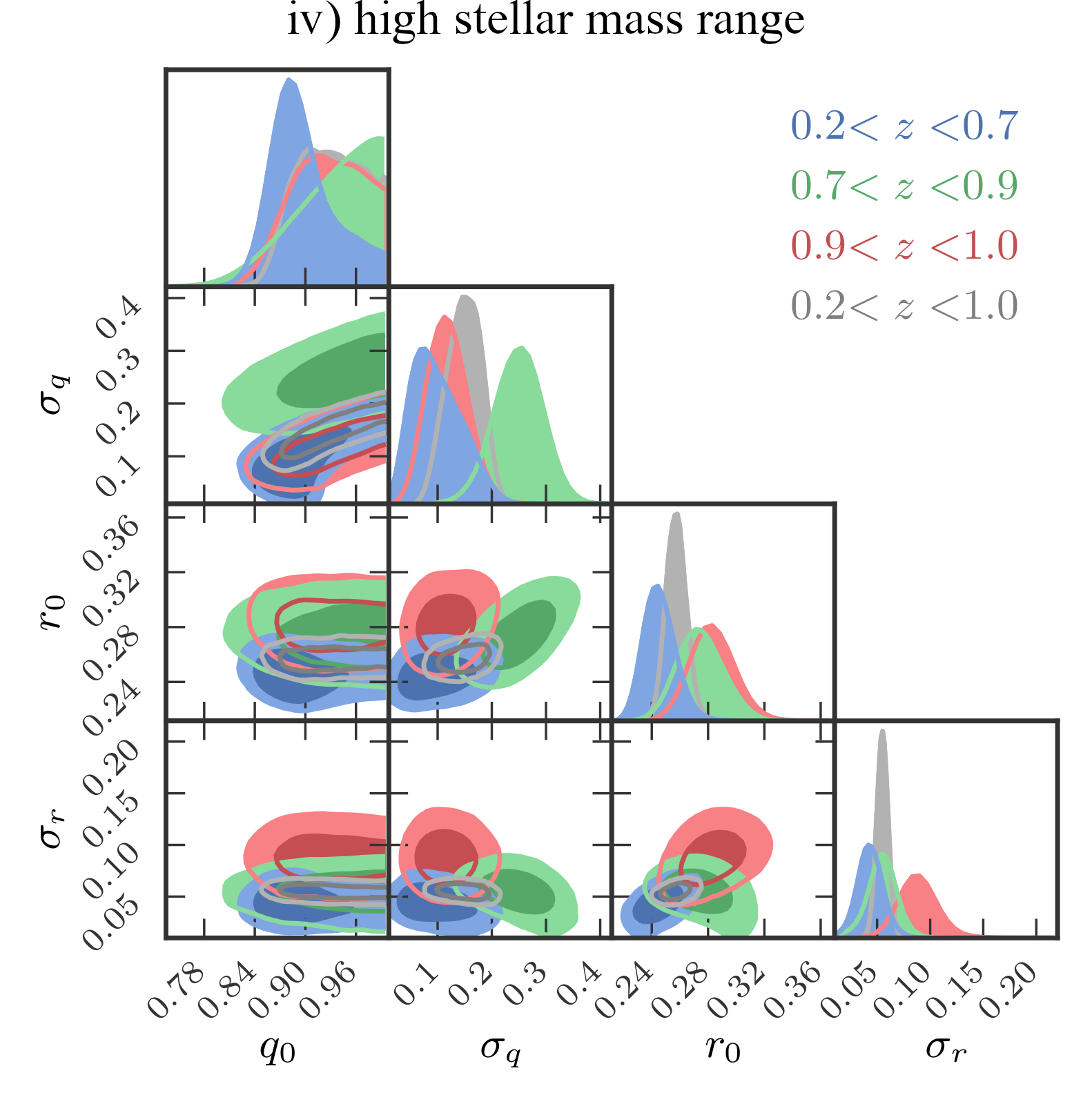}
\caption{
Posteriors of the 3D axis ratio model parameters from Equation (\ref{eq:P3d}),
derived from the 2D axis ratio distributions in COSMOS (Fig. \ref{fig:q2D_fits_cosmos}).
Light and dark areas mark $95$ and $68$ \% confidence levels respectively.
Results for
i) redshift samples defined over the full stellar mass range (top left),
ii) stellar mass samples, defined over the full redshift range (top right),
iii,iv) redshift samples with stellar masses below and above $10^{10.35} M_\odot$ (bottom left and right respectively).
The marginalized posterior distributions are shown on the diagonal panels of each sub-figure.
}
\label{fig:corner_cosmos}
\end{figure*}

The intrinsic circularity of the discs in our samples is quantified
by the axis ratio \qddd. The maximum and the width of the \qddd\
distribution are described by the parameters $q_0$ and $\sigma_q$
of our \pddd\ model. In the top right of Fig. \ref{fig:corner_cosmos} 
we compare the constraints on these model parameters from the
low and high mass samples in the entire redshift range.
We find that the high mass sample prefers slightly higher values of
$q_0$ and $\sigma_q$ compared to the low mass samples. However,
the corresponding confidence intervals from both mass samples
are overlapping significantly, which indicates that our data provides
no evidence for a mass dependence of the disc circularity. 
}

We find $q_0$ values of around $0.9$ for both samples (see Table \ref{tab:params_rec_cosmos}), which is consistent with
estimates reported for discs at lower redshifts in previous studies
\citep{Sandage70, Fasano93, Ryden04, Ryden06, Rodriguez13} and describes the deficit of
circular face-on discs with $q_{2D}=1.0$ in the observed data (see Fig. \ref{fig:q2D_fits_cosmos}).
Different reasons for this deficit have been discussed in the literature
\citep[e.g.][]{Bertola91, Huizinga92, Rix95, Bernstein02, Joachimi13a}.
On the one hand, it is argued that deviations from perfect circularity could result from
observational effects, such as noisy isophotes, artifacts like cosmic rays in the galaxy images,
or simply the fact that the images are pixelized. However, we expect the impact of such
effects on our measurements to be minor, since i) we selected objects with ``good'' fits
to the \sersic\ profile (see Section \ref{sec:data:cosmos:shapes}), which should remove
objects whose images are heavily distorted by artefacts from the sample and ii) the ACS pixel scale
of $0.03"$ is well below our lowest limit for the effective angular radius of $0.17"$ at $z=1.0$
(see Fig. \ref{fig:vol_lim_samp}). However, since the deviations from perfect circularity
are predicted to be relatively small, even minor systematics might be relevant.
On the other hand, one could expect that the disc galaxies are intrinsically not perfectly circular due to
patchy star formation activity and substructures, such as spiral arms (see Fig. \ref{fig:acs_images}) or galactic warps
\citep[see e.g.][for the latter]{Binney92, Gomez17}. This expectation lines up with the non-circularity
of discs in the HAGN and TNG100 simulations, which we see in Fig. \ref{fig:Pqr_hydrosims} and \ref{fig:q2D_fits}
as the lack of galaxies with \qddd\ and \qdd\ close to unity. 

Note that we omit entering a detailed discussion of our $\sigma_q$ constraints, since the test of our
reconstruction method in Section \ref{sec:method:test_reconstruction} and Appendix \ref{app:params_randsamp}
suggested that these constraints may be strongly biased.

\subsubsection{Intrinsic disc thickness}
\label{sec:disc_thickness}

The relative disc thickness is quantified by the minor to intermediate axis ratio \rddd.
The position of the maximum and width of the \rddd\ distribution are described in our \pddd\ model
by the parameters $r_0$ and $\sigma_r$. In the top right of Fig. \ref{fig:corner_cosmos}
we show that $r_0$ exhibits a significant dependence on stellar
mass. It is predicted to be around $0.3$ with lower and higher values
for the high and low mass sample, respectively (see Table \ref{tab:params_rec_cosmos}).
The corresponding constraints on the relative thickness $s_0\equiv q_0 r_0$
(quantifying the position of the maximum of the $s_{3D}\equiv C_{3D}/A_{3D}$ distribution) are $0.33^{0.34}_{0.31}$ and $0.24^{0.25}_{0.23}$ for the
low mass and high mass sample respectively, where the upper and lower values here
are the limits of the $68\%$ confidence intervals. These values are consistent with
results reported for discs at low redshifts which were obtained using similar reconstruction
techniques as used in this work \citep{Fasano93, Ryden04, Ryden06, Rodriguez13} or from direct measurements based on
edge-on oriented discs \citep{Mosenkov15, Reshetnikov16, Mosenkov20}. 
The dispersion of the relative disc thickness, $\sigma_r$, shows a significant mass dependence
as well, as it takes lower and higher values for the high and low mass sample respectively.
This mass dependence of the thickness dispersion explains why the cuff-off on the left side
of the \qdd\ distribution in Fig. \ref{fig:q2D_fits_cosmos} is sharper for high than for low mass discs,
while we expect also this latter parameter to be significantly biased by $\sim20$\% (Section \ref{sec:method:test_reconstruction}).

\begin{figure}
\centering\includegraphics[width=7.5 cm, angle=0]{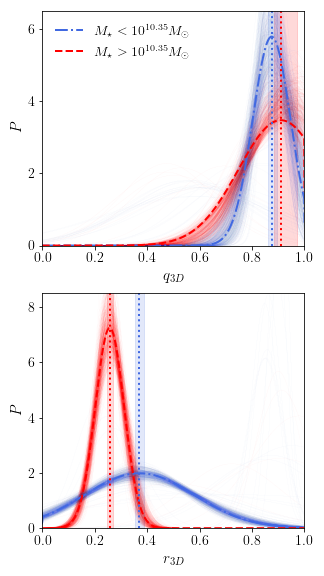}
\caption{Distributions of galaxy 3D axis ratios, reconstructed from the distribution of 2D galaxy
axis ratio in the COSMOS survey. Results are shown for the low and high mass sample, defined within $0.2<z<1.0$.
Vertical lines mark the positions of the maxima, described by the best fit model parameters $q_0$ and $r_0$ from Table \ref{tab:params_rec_cosmos}.
Vertical shaded areas indicate the corresponding $68\%$ confidence intervals.
Thin lines show predictions for $500$ random sampling points of the posterior distribution.}
\label{fig:Pqrs_reconstruction_cosmos}
\end{figure}
%
In Fig. \ref{fig:Pqrs_reconstruction_cosmos} we show the reconstructed marginalized distributions of the
3D axis ratios \qddd\ and \rddd\ for the low and high mass sample over the full redshift range,
as predicted from the best fit parameters given in Table \ref{tab:params_rec_cosmos}. The uncertainties on the prediction is
illustrated as the density of thin lines which are obtained from randomly selected sampling points of the posterior
in the \pddd\ model parameters space.
The corresponding values $q_0$
and $r_0$ are shown as vertical lines. They are enclosed by shaded areas which
indicate the $68\%$ confidence intervals for these parameters, given in Table \ref{tab:params_rec_cosmos}.
Note that these uncertainties are similar to the systematic bias of $\sim3$\% and $\sim8$\%
for $q_0$ and $r_0$ respectively, which we found in our systematic tests (Section \ref{sec:method:test_reconstruction}).

When comparing the distributions of both mass samples, we see that the values
of \rddd\ are well below those of \qddd\, as expected for disc galaxies. This indicates that our
results carry physically meaningful information, despite the expected biases.
Fig. \ref{fig:Pqrs_reconstruction_cosmos} further illustrates that
i) high mass discs tend to be thinner with respect to their diameter than low mass discs,
ii) high mass discs tend to be slightly more circular than low mass discs,
iii) the dispersion of the circularity is larger for high mass than for low mass discs and
iv) the dispersion of the relative thickness is larger for low mass than for high mass discs.
However, the mass dependence of the peak of the circularity distribution, quantified by $q_0$,
is statistically not significant, since the $68\%$ confidence intervals for $q_0$ overlap
(see also discussion in Section \ref{sec:disc_circularity}).
The lower relative thickness for high mass discs could indicate that these galaxies tend to be more
relaxed than low mass discs, which are more prone to perturbations by feedback, merging and tidal interactions.
This interpretation is supported by the higher dispersion of the relative thickness of low mass discs
and lines up with the overall higher star formation rates for low mass discs, which we see in the right panels
of Fig. \ref{fig:qrssfr_pdf}.

\begin{figure}
    \centering\includegraphics[width=7.5 cm, angle=0]{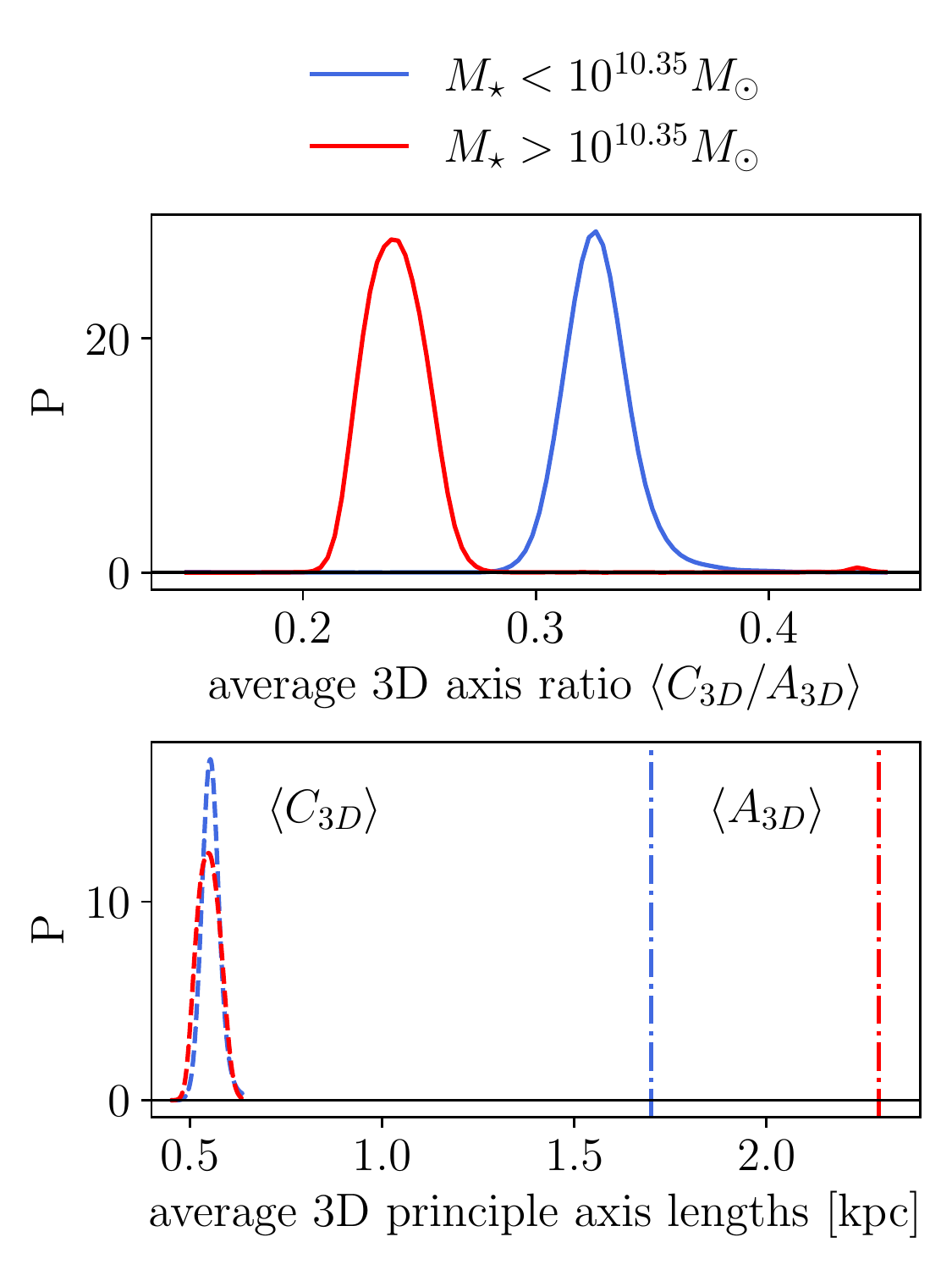}
\caption{
{\it Top}: Marginalized posterior of the parameter $s_0 \equiv q_0 r_0$
that serves as an approximation for the average 3D axis ratio $\langle s_{3D} \rangle = \langle C_{3D}/A_{3D} \rangle$.
{\it Bottom}: Dash-dotted vertical lines mark the average major axes length
$\langle A_{3D} \rangle$ that is approximated as the average comoving effective radius
$R_\perp$ of discs with $q_{2D}>0.5$.
Dashed lines show the marginalized posterior of the average minor axes length $\langle C_{3D} \rangle$ which is approximated as $\langle{C}_{3D} \rangle \simeq s_0 \times \langle A_{3D} \rangle$.
The different approximations are discussed in Section \ref{sec:disc_thickness}.
Results for the low and high mass sample, defined within $0.2<z<1.0$, are shown as blue and red
lines respectively.
}
\label{fig:Psac_cosmos}
\end{figure}
However, another reason to expect a lower relative thickness of massive discs with respect to their radius
is that the radius is on average larger for discs in the high mass sample, as indicated by the mass dependence
of the effective radii shown in the central panels of Fig. \ref{fig:qrssfr_pdf}. Such an increase of the radius
with mass could lead to a decrease of the relative thickness, even if the size of the absolute thickness is mass independent.

In order to test if this scenario is consistent with our data we derive an estimate of the expectation value for the absolute
thickness, $\langle C_{3D} \rangle$, for the low and high mass sample via our constraints on the peaks of
the $s_{3D}\equiv C_{3D}/A_{3D}$ axis ratio distributions, quantified by $s_0$.
Note that we use $s_{3D}$ instead of $r_{3D} \equiv C_{3D}/B_{3D}$, since we can approximate $A_{3D}$
directly from the observed 2D radii. We estimate the expectation value for $C_{3D}$ using the approximation
$\langle C_{3D} \rangle \simeq
\langle A_{3D} \rangle \langle C_{3D}/A_{3D} \rangle =
\langle A_{3D} \rangle \langle s_{3D} \rangle \simeq
\langle A_{3D} \rangle s_0$.
This approximation is based on two assumptions:
1) We assume $\langle s_{3D} \rangle \equiv \langle C_{3D} / A_{3D} \rangle \simeq \langle C_{3D}\rangle / \langle A_{3D} \rangle$.
This approximation corresponds to the zero-order term in a perturbative expansion of $f(A,C) = C/A$ around the
$\langle A \rangle$ and $\langle C \rangle$. We validated this relation using our sample of disc galaxies
in the HAGN simulation and found it to be accurate at the sub-percent level.
2) We assume that the expectation of the $s_{3D}$ distribution is located at its maximum,
i.e. $\langle s_{3D} \rangle \simeq s_0$. This implies that the $s_{3D}$ distribution
is symmetric, which is commonly assumed to be the case for observed discs \citep[e.g.][]{Ryden04}.
Again, we also verified that this is a good approximation using the disc galaxies in our HAGN sample.

We obtain the posterior distribution of $s_0$ from the posteriors of $q_0$ and $r_0$, using
$s_0 = q_0 r_0$ as detailed in the beginning of this section.
It is shown in the top panel of Fig. \ref{fig:Psac_cosmos} for the low mass
and high mass disc samples, defined over the entire redshift range. The distributions differ
significantly from each other, as it can be expected from the constraints on $r_0$ shown in Fig. \ref{fig:corner_cosmos}.
We proceed by estimating the posterior for the expectation value of the minor axis size from the posterior of $s_0$,
using the aforementioned relation $\langle C_{3D} \rangle \simeq \langle A_{3D} \rangle s_0$, which
we apply on each MCMC sampling point.
For this last step we need an estimate for the average comoving size of the major axis $A_{3D}$.
This size is approximated with the mean effective comoving radius $R_\perp$ of the disc galaxies
in our sample, using objects with $q_{2D} > 0.5$. This value is shown in the central panels of
Fig. \ref{fig:qrssfr_pdf} as vertical red and blue solid lines for the high and low mass sample
respectively. The cut on $q_{2D}$ was thereby chosen to mitigate the impact the projection effects
on the effective radius as discussed in Section \ref{sec:data:cosmos:main_sample}.

The estimated posterior of $\langle C_{3D} \rangle$ is shown for the samples of high mass and low mass
discs in the bottom panel of Fig. \ref{fig:Psac_cosmos}. The two posteriors overlap each other almost
completely, which means that our approximation of the average 3D disc thickness shows no significant mass dependence.
These posterior distributions are compared
to the estimates of the mean major axis sizes $\langle A_{3D} \rangle$ which are shown as vertical lines in the same panel.
The strong mass dependence, which we find for $\langle A_{3D} \rangle$ suggests that the mass dependence in the relative
3D thickness \sddd\ is driven by the mass dependence of the major axis sizes.

\section{Summary and Conclusions}
\label{sec:summary}

We studied the 3D shapes of disc-dominated late-type galaxies from the COSMOS survey
in different mass and redshift ranges, tackling the question of how these galaxies could grow without
forming a large central bulge. 
%
We approximated the 3D light distribution of these galaxies as 3D ellipsoids described by the two ratios $q_{3D} \equiv B_{3D}/A_{3D}$ and $r_{3D} \equiv C_{3D}/B_{3D}$,
which quantify the circularity and relative thickness of the discs, respectively.
We inferred the distribution of these 3D axis ratios from the observed distribution of 2D axis ratios,
using a reconstruction method based on the assumption that the distribution
of \qddd\ and \rddd\ is well approximated by a two-dimensional Gaussian.
This Gaussian is characterized by the average disc circularity $q_0$,
the average disc thickness $r_0$ and the corresponding dispersions $\sigma_q$
and $\sigma_r$.

Variations of this method have been widely used in the literature, but their accuracy
as well as the assumptions they employ remained to be tested. We began our analysis by
performing such tests for the first time using two state-of-the-art
hydrodynamic simulations of galaxy formation in cosmological volumes, Horizon-AGN and Illustris TNG100.
We demonstrated that the 3D axis ratio distributions of disc galaxies in these simulations
are adequately described by a Gaussian model. Reconstructing the 3D axis ratio distribution from the distribution
of 2D axis ratios, we found that the inferred parameters of the Gaussian model are biased with respect to
those derived directly from fits to the 3D distributions. For our most realistic test,
based on 2D galaxy shapes from projected stellar mass distributions, we find this bias to be
moderate for the parameters $q_0$ and $r_0$ ($\sim 3$\% and $\sim 8$\% respectively)
but strong for the corresponding dispersions $\sigma_q$ and $\sigma_r$ ($\sim 60$\% and $\sim 20$\% respectively,
see Table \ref{tab:reldiff_params_hagn}).
We concluded that the bias is mainly driven by an inaccuracy of the ellipsoidal model
for the stellar mass distribution, which we use for relating 3D to 2D galaxy shapes
analytically during the reconstruction (see Section \ref{sec:method:test_reconstruction}).
The strong simplification implied when approximating late-type galaxies as 3D ellipsoids becomes obvious
in the COSMOS images of such objects, shown in Fig. \ref{fig:acs_images}.
Nevertheless, our bias estimates derived from the HAGN simulation may be overly pessimistic, since
our test is based on all disc galaxies found in this simulation, including those which have a bulge
for which a one-component ellipsoidal galaxy model is expected to be inadequate. More realistic tests
should be based on synthetic galaxy images, take into account the effect of dust extinction and be analysed
in the same way as the reference observational data.

After having tested the shape reconstruction method, we applied it on a volume limited sample
of disc-dominated galaxies in COSMOS which is limited in redshift ($0.2 < z < 1.0$), absolute
magnitude ($M_i < -21.5$) and transverse comoving size ($R_\perp > 0.64$ kpc).
We demonstrated that a conservative cut on $M_i$ is required in order to minimize the
impact of dust extinction on the observed 2D axis ratio distribution. Otherwise,
this effect can lead to an apparent redshift evolution of the distribution
(Fig. \ref{fig:qcm_cosmos}) and consequently to incorrect physical interpretations of the observations.
%
We found that the ellipsoidal galaxy shape model in conjunction with the Gaussian model
for the 3D axis ratio distribution provides good fits to the distribution of the observed
2D axis ratios (Fig. \ref{fig:q2D_fits_cosmos}).

The constraints on the parameters $q_0$ and $r_0$ inferred from these fits show
a variation with redshift around $\sim 1\%$ and $\sim 10\%$ respectively when considering
the full mass range (Table \ref{tab:params_rec_cosmos}). This finding indicates that the redshift evolution of the shapes is relatively weak.
Splitting the sample into a high and a low mass sub-sample leads to larger variations
of $\sim 10\%$ and $\sim 25\%$ for $q_0$ and $r_0$ respectively.
The parameters $\sigma_q$ and $\sigma_q$ vary up to $~50\%$ across the different
redshift samples. However, overall the variations with redshift lie within or close
to the $68\%$ confidence intervals on these parameters.
Studying the joint posteriors on the different shape model parameters, we find no clear
indication for a redshift evolution that is significant with respect to the uncertainties
as the $95$\% confidence intervals of these parameters from different
redshift samples overlap (Fig. \ref{fig:corner_cosmos}). We crosschecked this result using the Kolmogorov–Smirnov (K-S) statistic
to test the null hypothesis that the 2D axis ratio distributions from two redshift bins are drawn from the same underlying
redshift independent distribution. Studying all redshift bin combinations for different mass samples,
we find that the K-S test is mostly consistent with the results from the Bayesian analysis. 
As an exception the K-S test clearly rejects the null hypothesis that the 2D axis ratios of the high mass samples
in our intermediate and high redshift bins are drawn from the same distributions. This finding appears to be inconsistent
with the agreements between the 2D axis ratio distributions from the high mass samples in the low and high redshift bins
and could be an indication for cosmic variance or misclassified objects contaminating the high mass sample at
intermediate redshifts. In any case, the fact that the Bayesian analysis revealed no significant redshift dependence
indicates that any dependence detected by the K-S test is relatively weak compared to the errors on the data.

This weak dependence on redshift is contrasted by a relatively strong dependence on mass (Fig. \ref{fig:corner_cosmos}). 
In particular, the parameter $r_0$ is significantly higher for discs in our sample
with stellar masses below $10^{10.35} M_\odot$ compared to discs with higher masses, even within the expected
uncertainties of the reconstruction method (e.g. Fig. \ref{fig:Pqrs_reconstruction_cosmos}). This finding
indicates that the relative disc thickness decreases with mass. However, the absolute disc thickness,
estimated from the relative thickness and a proxy for the disc diameter, shows no mass dependence
(Fig. \ref{fig:Psac_cosmos}), which suggests that the relative thickness of low mass discs is higher
mainly because of their smaller diameters (e.g. Fig. \ref{fig:qrssfr_pdf}).

In summary, we found indications that the distribution of the 3D axis ratios of disc-dominated
galaxies in our sample is not (or if, then just weakly) redshift dependent in the range $0.2<z<1.0$ and the absolute
disc thickness does not depend significantly on mass. These findings speak against mass accretion by major mergers
and a subsequent suppression of bulge formation by strong feedback after $z\lesssim1.0$, since one could expect that
such disruptive events would decrease the discs circularity and increase their thickness
due to vertical heating \citep[e.g.][]{Quinn93, Grand16, Park21}.
The decrease of the specific star formation rates of massive discs with redshift, shown in
Fig. \ref{fig:qrssfr_pdf}, suggests that feedback occurs in these galaxies, suppressing star formation
by removing cold gas, but may not be sufficiently strong to significantly affect the observed shapes.
The absence of major mergers in disc-dominated galaxies since $z\lesssim 1$ is further supported by the
fact that the mean comoving sizes, displayed in Fig. \ref{fig:qrssfr_pdf}, show no redshift evolution
for low mass galaxies and a very weak increase with redshift for high mass galaxies.
We conclude that disc-dominated galaxies accreted most of their mass before $z=1.0$ and lived
preferentially in isolation ever since.
This picture lines up with the results from \citet{Grossi18}, who find that the star formation rates
of disc-dominated galaxies show no significant dependence on the density of their environment, which
indicates the absence of major interactions \citep[see also][]{Sachdeva16}. A tranquil evolution is further supported by the
weak redshift dependence of the transverse comoving size distribution (Fig. \ref{fig:Pqrs_reconstruction_cosmos}),
which has also been reported for disc galaxies in the GOODS fields by \citet{Ravindranath04}.

However, we emphasize that the validity of our results is limited for several reasons.
The small sizes of our samples lead to large statistical errors, which may obscure
a weak evolution of the galaxy shapes with redshift and could hence affect our conclusions.
This limitation can be overcome in future analysis by using measurements of galaxy shapes in
larger volumes from upcoming weak lensing surveys like {\it Euclid} or the Legacy Survey of Space
and Time on the Vera Rubin Observatory. 
Such high precision measurements will require an improved accuracy of the shape reconstruction
method, based for instance on corrections which can be developed using realistic mock images.

Our results may further be affected by inaccuracies of the assumptions on which the reconstruction method
is based. One inaccurate assumption is the aforementioned ellipsoidal model for galaxy morphologies,
which differs strongly from the complex shapes of real disc galaxies. Another potential inaccuracy
may result from the Gaussian model, that we assume for the 3D axis ratio distribution.
Our choice for this model was based on 3D axis ratio distributions measured in two hydro-dynamic simulations.
However, we also demonstrated that the 2D axis ratio distributions in these simulations differ
significantly from the COSMOS observations due to the lack of thin discs with $q_{2D}<<1$,
which we attribute to resolution limits and approximations in the simulation method
(see Sec. \ref{sec:data:sims}). More realistic simulations could reveal that our Gaussian model
is inadequate, which could bias the outcome of the reconstruction method.

Potential errors in our analysis may further result from a miss-classification
of galaxies in the ZEST catalogue. Although this catalogue has been carefully calibrated
and passed rigorous validations, it may contain a fraction of objects that
are miss-classified as disc-dominated late-type galaxies and contaminate our samples.
By inspecting the color-color distribution of the objects in our catalogue we find no
indication of a significant fraction of miss-classified objects. However, visual
inspection of randomly selected galaxies revealed the presence of some objects,
in particular at high redshifts and low masses, whose morphological class is ambiguous
to us (Appendix~\ref{app:discs_zest}).
In order to assess the uncertainty from potential miss-classifications on our results
one could compare the morphological classification in ZEST against alternative
classification methods that are for instance based on machine learning techniques.

More detailed interpretations of the redshift evolution of observed galaxy shapes may also
require a characterization of two observational systematic effects which we assumed to
be insignificant with respect to the errors on our measurements.
A first potential systematic could result from the fact that
galaxy shapes are observed in the same filter at different redshifts.
High redshift galaxies are therefore seen at bluer rest frame wavelengths
than galaxies at lower redshifts. This may affect the observed shapes for two reasons.
The first is that the age of stellar populations is not uniformly distributed, as
star formation takes place in distinct regions such as spiral arms
\citep[e.g.][]{Martin01}. The second reason is that extinction and reddening by dust
in the source galaxy has a stronger impact on observations in blue than in red rest-frame
wavelengths and may hence distort the observed shapes more strongly at high than at lower
redshifts. This effect is further complicated by the fact that the distribution
and overall density of dust evolves with time. In fact, we find strong indications
of dust extinction in our data (Fig. \ref{fig:qcm_cosmos}). However, studying shapes measured
from synthetic galaxy images at $z=0.0$, we find no strong dependence of the 2D axis ratio distribution
on the filter band. This finding suggests that observing shapes in the same filter at
different redshift has only a mild effect on the observed shape distribution (see Appendix \ref{app:mcut_qapp}).
A second systematic effect can be expected from gravitational lensing, while this effect
is typically weak with a contribution of less than $1$\% to the observed ellipticity
\citep[e.g.][]{Kirk15}.

An important outcome of our analysis is that the model for the 3D galaxy shapes and their
distribution provides good fits to the observed distribution of 2D shapes for disc-dominated
galaxies, despite the expected inaccuracies. This model can hence
be employed to generate mock catalogues of galaxy shapes in large cosmological dark matter-only simulations
in which stellar mass distributions of individual galaxies are not provided. In order to construct such mocks,
\citet{Joachimi13a} approximated disc galaxies by flat opaque cylinders
and showed that the resulting 2D ellipticity distribution differs significantly
from COSMOS observations. An improvement on this aspect is crucial for building
precision mock galaxy catalogues with intrinsic alignments for the preparation
of future weak lensing surveys. Disc galaxies are expected to dominate the lensing
source samples at high redshifts in which intrinsic alignments contaminate
the gravitational shear induced by the large scale structure at lower redshifts
(e.g. Fig. \ref{fig:ztypefrac}). Building and testing such improved mocks is subject of our ongoing work.

\section*{Acknowledgements}

KH acknowledges support by the Swiss National Science Foundation (Grant No. 173716, 198674)
and the Forschungskredit grant of the University of Zurich (Projekt K-76106-01-01).
CL acknowledges support from the Programme National Cosmology and Galaxies (PNCG) of CNRS/INSU with INP and IN2P3.
This work is part of the Delta ITP consortium, a program of the Netherlands Organisation for Scientific Research (NWO) that is funded by the Dutch Ministry of Education, Culture and Science (OCW).   
The research of JD is supported by the Beecroft Trust and STFC.
Our analysis is based on data products from observations made with ESO Telescopes at the LaSilla Paranal Observatory under ESO programme ID 179.A-2005 and on data products produced by TERAPIX and the Cambridge Astronomy Survey Unit on behalf of the UltraVISTA consortium.
Our analysis was performed using {\sc Jupyter Notebook} \citep{Kluyver2016jupyter},
with the Python packages
{Colossus} \citep{Diemer18},
{emcee} \citep{Foreman13},
{Matplotlib} \citep{Hunter:2007},
{NumPy} \citep{harris2020array},
{pandas} \citep{mckinney-proc-scipy-2010},
{pyGTC} \cite{Bocquet2016},
{SciPy} \citep{2020SciPy-NMeth}.

\section*{Availability of data}
The galaxy catalogues from COSMOS (COSMOS2015, ACS-GC, ZEST) and Illustris TNG
used in this work are publicly available at the links provided in Section \ref{sec:data}. 
The matched COSMOS catalogue described in that section is available
upon request to K. Hoffmann (kai.d.hoffmann@gmail.com). The Horizon AGN catalogue
is available upon reasonable request to N. E. Chisari (n.e.chisari@uu.nl).

\bibliographystyle{mnras}
\bibliography{references}


\appendix
\section{Testing the classification of disc-dominated late-type galaxies in the ZEST}
\label{app:discs_zest}

We extend our visual inspection from Fig. \ref{fig:acs_images},
now including objects of any orientation and any redshift from our
volume limited samples to further validate if these objects have been classified
correctly as disc-dominated late-type galaxies.

\begin{figure*}
\centering\includegraphics[width=0.9 \textwidth, angle=0]{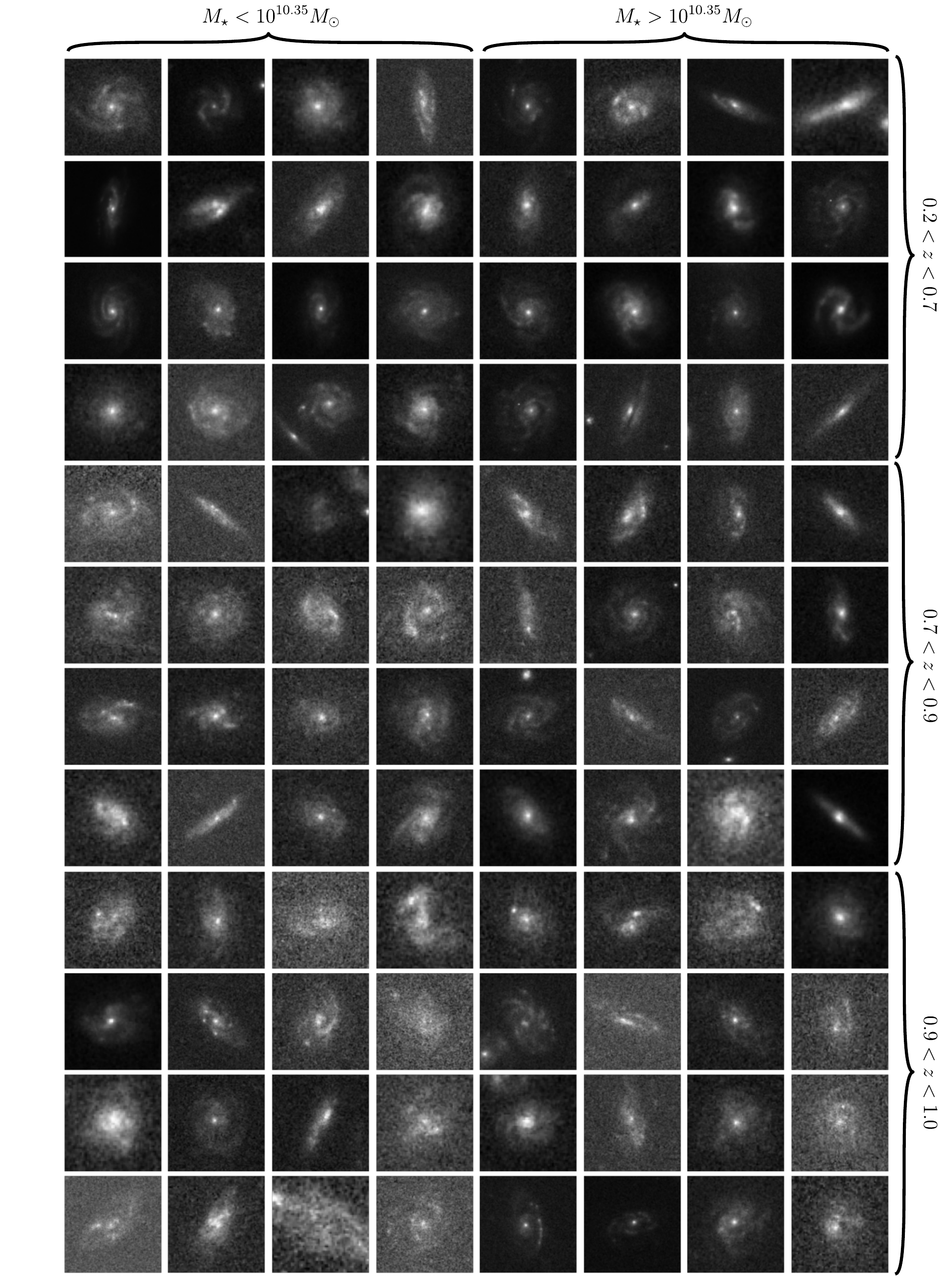}
\caption{
Object classified as disc-dominated late-type galaxies
in the ZEST catalogue of the COSMOS field. The figure shows $16$ randomly
selected galaxies for each of our $6$ redshift-stellar mass samples.
}
\label{fig:acs_images2}
\end{figure*}

In Fig. \ref{fig:acs_images2} we show $16$ randomly selected examples of
galaxies for each of our $6$ stellar mass-redshift sub-samples. We find visually that
the majority of objects appears to be classified correctly. However, we find
several objects with irregular shapes, in particular in the high redshift bins,
which could potentially be misclassified.

\begin{figure}
\centering\includegraphics[width=0.45 \textwidth, angle=0]{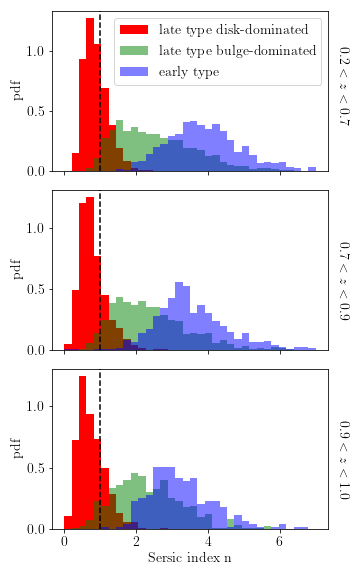}
\caption{
\sersic\ indices $n$ of galaxies in our volume limited main sample with different morphological
classifications provided in the ZEST catalogue. The different panels show results from the different
redshift bins studied in our analysis. The vertical line is drawn at $n=1$. }
\label{fig:sersic_index}
\end{figure}

As an additional validation of the morphological classification we 
show in Fig. \ref{fig:sersic_index} the distributions of the \sersic\ index
$n$ from Equation (\ref{eq:sersic_profile}), as provided in the ZEST catalogue.
Results are shown for objects which are classified as late-type disc-dominated,
late-type bulge-dominated and early-type galaxies (see Section \ref{sec:data:cosmos:morph}
for details on the classification). We find that the \sersic\ indices of
late-type disc-dominated galaxies are distributed in the range $0\lesssim n \lesssim 2$,
while the peak of the distribution is located at $n\simeq0.5$, which is consistent
with values that are typically found for disc-dominated galaxies. The \sersic\ indices
of early-type galaxies are distributed between $2\lesssim n \lesssim 7$
and are hence clearly separated from those of late-type disc-dominated galaxies.
The \sersic\ indices of late-type, bulge-dominated galaxies lie in between the two other
distributions with significant overlaps in both directions. Note that such
overlaps can be expected as the \sersic\ index alone is not sufficient
for a detailed morphological classification.

In summary we conclude that the low \sersic\ indices for late-type disc-dominated galaxies
and the larger \sersic\ indices of late-type bulge-dominated and early-type galaxies
are consistent with the morphological ZEST classification. However,
the image inspection revealed some poorly resolved objects at high redshifts,
which are potentially misclassified as late-type disc-dominated galaxies and could hence
contaminate the samples on which our investigation is based. Addressing this
problem will require higher resolution imaging surveys of high redshift galaxies,
which will become achievable with future space based telescopes, such as the James Webb Space Telescope.
The reliability of the ZEST morphological classification could further
be tested with a comparison against alternative classification methods already
with current data \citep[e.g.][]{Martin20,Cheng21}.

\section{from 3D to 2D axis ratios}
\label{app:project_ellipsoid}
We obtain the 2D axis ratios of a projected 3D ellipsoidal stellar system following
\citet{Joachimi13a}, under the assumption that the 3D system is absorption-free,
self-similar and coaxial. A 3D ellipsoid is thereby expressed in a coordinate system defined by
the two orthogonal unit vectors $\{\vek{\hat e}_u, \vek{\hat e}_v \}$ which span the projection plane and the unit vector $\vek{\hat e}_\parallel$, which is orthogonal to $\{\vek{\hat e}_u, \vek{\hat e}_v\}$, pointing along the observer's line of sight.
In this new coordinate system the principal axes are given by
${\bf \tilde S}_\mu = \{
\langle \vek{\hat e}_u \vek{S}_\mu \rangle,
\langle \vek{\hat e}_v \vek{S}_\mu \rangle,
\langle \vek{\hat e}_\parallel \vek{S}_\mu \rangle
\}^\tau \equiv
\{\tilde S_{u,\mu},\tilde S_{v,\mu},\tilde S_{\parallel,\mu}\}^\tau$
with ${\bf S}_\mu \in \{\vek{A}_{3D}, \vek{B}_{3D}, \vek{C}_{3D}\}$.
The projected 2D ellipse is given by all points $\vek{r}$
in the projection plane which fulfil $\vek{r}^\tau {\mathbf W}^{-1} \vek{r} = 1$, where
\eq{
\label{eq:ellipseprojection1}
{\mathbf W}^{-1} \equiv \sum_{\mu=1}^3 \frac{\vek{\tilde S}_{\perp,\mu} \vek{\tilde S}_{\perp,\mu}^\tau}{\tilde S_\mu^2} - \frac{\vek{k} \vek{k}^\tau}{\alpha^2}\;,
}
with
\eq{
\label{eq:ellipseprojection2}
\vek{k} \equiv \sum_{\mu=1}^3 \frac{\tilde S_{\parallel,\mu} \vek{\tilde S}_{\perp,\mu}}{\tilde S_\mu^2} ~~\mbox{ and }~~
\alpha^2 \equiv \sum_{\mu=1}^3 \br{ \frac{\tilde S_{\parallel,\mu}}{\tilde S_\mu} }^2\;
}
and $\vek{\tilde S}_{\perp,\mu} \equiv \bc{\tilde S_{u,\mu}, \tilde S_{v,\mu}}^\tau$
is the principal axes component in the projection plane.
The 2D ellipticity vector of the projected ellipsoid is then given by
\eq{
\begin{split}
\label{eq:polar_w}
\begin{pmatrix} \epsilon_1\\ \epsilon_2\end{pmatrix} = 
\frac{1}{\mathcal{N}}
\begin{pmatrix} W_{11}-W_{22}\\ 2\, W_{12} \end{pmatrix}
\end{split}
}
with $\mathcal{N} \equiv W_{11}+W_{22} + \sqrt{\text{det} \, {\mathbf W}}$.
The absolute value of the ellipticity, $\epsilon = \sqrt{\epsilon_1^2 + \epsilon_2^2}$,
is related to the 2D axis ratio $q_{2D} \equiv B_{2D} / A_{2D}$ of
the projected ellipsoid as
\begin{equation}
q_{2D}  = \frac{1-\epsilon}{1+\epsilon},
\label{eq:q_2d}
\end{equation}
where $A_{2D}$ and $B_{2D}$ are the principle axes of the 2D ellipse.
An alternative approach for obtaining $q_{2D}$ has been derived
by \citet{Binney85} \citep[see also][]{Stark77, Benacchio80}.
However, we find this latter calculation to
be computationally less efficient.

\section{Validating the model for the 3D axis ratio distribution}
\label{app:P3d_model_validation}

We use the shapes of disc galaxies, measured in the HAGN and TNG100 simulations
to further validate the approximations that are implied by using the Gaussian model
for the 3D axis ratio distribution, \pddd, introduced in Section \ref{sec:method:P3Dmodel}. 

\begin{figure}
\centering\includegraphics[width=7.0 cm, angle=0]{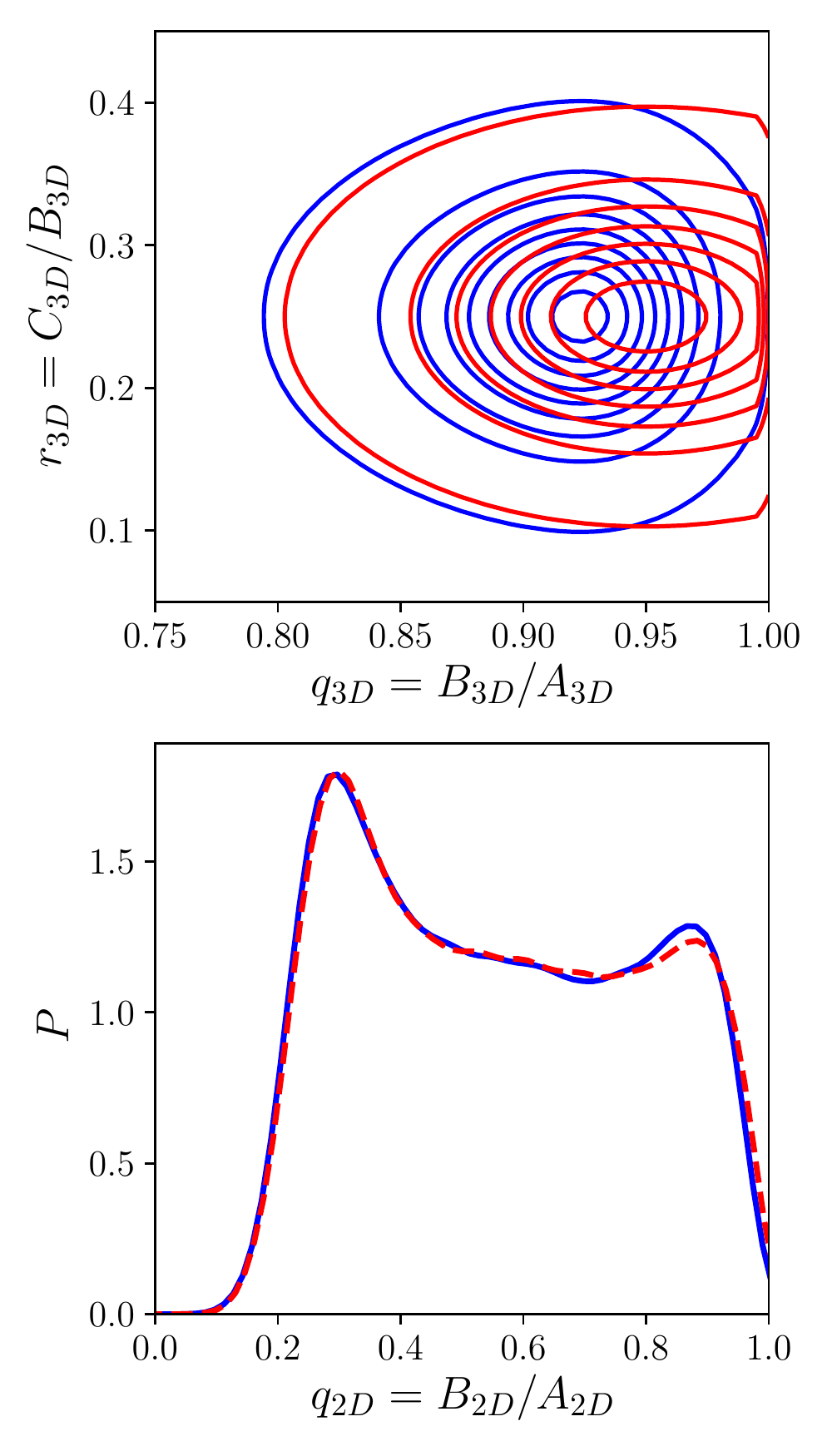}
\caption{
{\it Top}:model for the 3D axis ratio distribution with and without skewness
in the \qddd\ dimension (blue and red contours respectively).
{\it Bottom}: Corresponding predictions for the 2D axis ratios
distribution. The skewness as a very small affect on the 2D axis ratios.}
\label{fig:P2D3D_skew}
\end{figure}

A possible improvement of that model could be the inclusion of a skewness in the \qddd\ dimension,
as demonstrated in Fig. \ref{fig:Pqr_hydrosims}. In the top panel of Fig. \ref{fig:P2D3D_skew}, we
show the joint distribution of \qddd\ and \rddd\ for the truncated Gaussian model with
and without skewness, using parameters which are typical for the disc galaxies
in our analysis. The corresponding distribution of the apparent 2D axis ratios
are shown in the bottom panel of the same figure. We find that the effect of the
skewness is small compared to the errors on our measurements (see Fig. \ref{fig:q2D_fits_cosmos}).
As a consequence the skewness in the \qddd\ distribution cannot be constrained from our
observed data and is therefore not included in our \pddd\ model.
%

\begin{figure}
\centering\includegraphics[width=7.0 cm, angle=0]{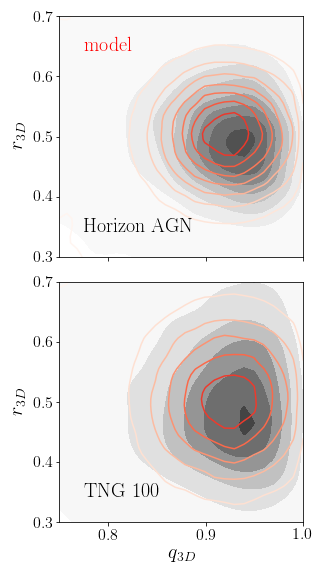}
\caption{Distributions of the axis ratios $q_{3D}\equiv C_{3D} / A_{3D}$ and
$r_{3D}\equiv C_{3D} / A_{3D}$ of disc galaxies in the HAGN and TNG100 simulations at $z=1.0$.
These distributions are compared to the \pddd\ model from Equation (\ref{eq:P3d}), which
was fitted to the marginalized distributions of \qddd\ and \rddd\, as shown in
Fig. \ref{fig:Pqr_hydrosims}.}
\label{fig:qr_contours}
\end{figure}

A second possible improvement of our model could be the inclusion of a correlation between 
$q_{3D}$ and $r_{3D}$ which could be described by a covariance matrix in Equation (\ref{eq:P3d}).
In order to asses the necessity for such a model extension we inspect the kernel density estimates
of the \pddd\ distribution from HAGN and TNG100 in Fig. \ref{fig:qr_contours}. We find no evidence for
a significant correlation between \qddd\ and \rddd\ in both simulations. For a visual impression of the
accuracy of our Gaussian model we in compare it to the measurements in Fig. \ref{fig:qr_contours},
using the model parameters from the fits to the marginalized 3D axis ratio distributions, shown in Fig. \ref{fig:Pqr_hydrosims}.
The model describes the simulation data reasonably well. while the relatively weak deviations from the
measurements appear to result from the neglected skew rather from a neglected covariance.

\section{Bias in the 2D axis ratio distributions from apparent magnitude cuts}
\label{app:mcut_qapp}

\begin{figure*}
\centering\includegraphics[width=15.0 cm, angle=0]{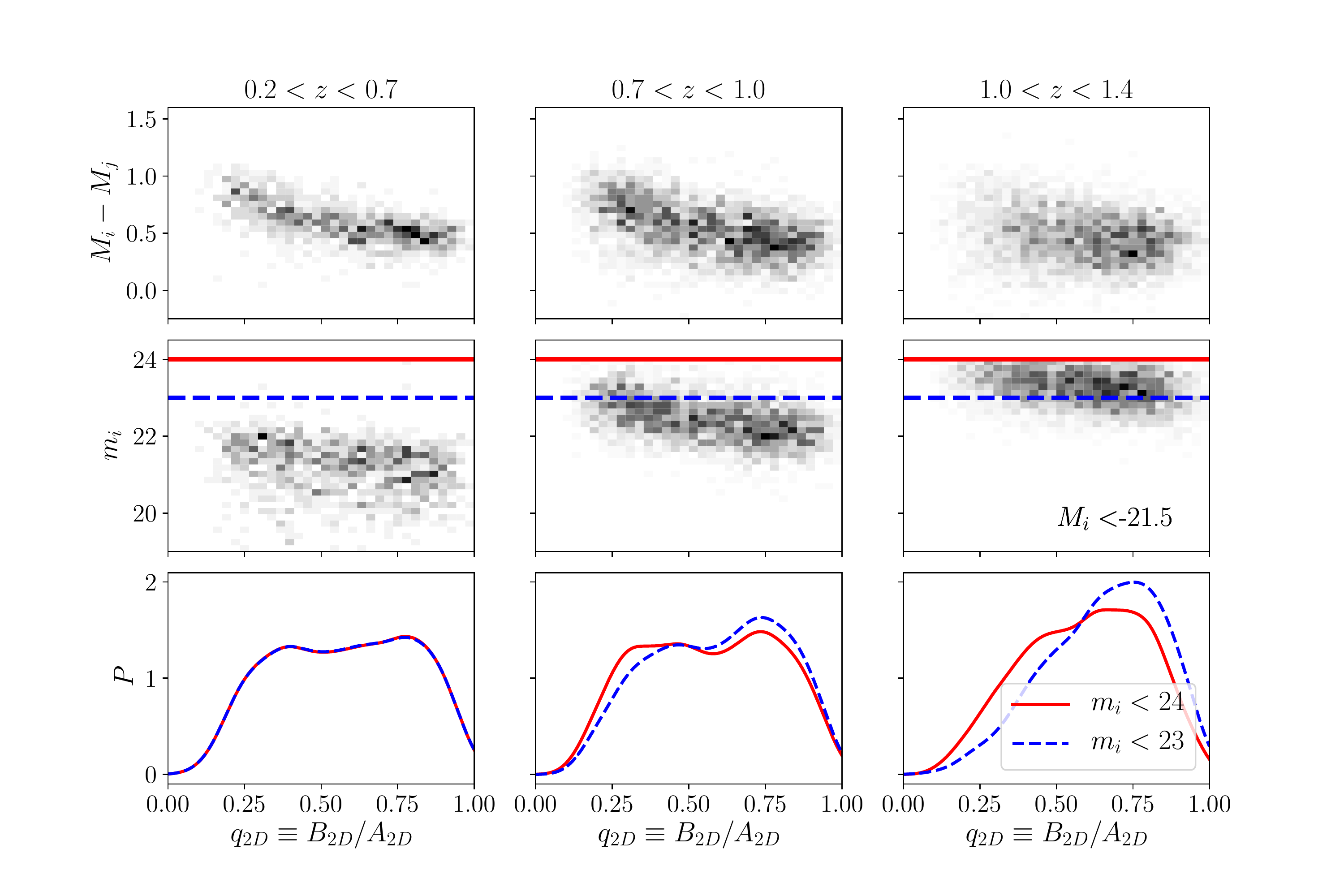}
\caption{Apparent 2D axis ratios \qdd\ of disc-dominated galaxies in COSMOS
with absolute $i$-band magnitudes brighter than $M_i=-21.5$ versus the rest-frame color index
and the apparent $i$-band magnitude (top and central panels respectively). The solid
red lines in the central panels indicate the apparent magnitude cut of our volume
limited sample at $m_i=24$. The dashed blue lines mark a cut at $m_i=23$.
The corresponding \qdd\ distributions, shown in the bottom panels,
display how the apparent magnitude cut can bias the \qdd\ distribution,
in particular at high redshifts.}
\label{fig:qcm_cosmos}
\end{figure*}

We study the effect of dust extinction on the color and apparent magnitude of
the disc-dominated galaxies from our matched catalogue with $M_i < -21.5$ in
Fig. \ref{fig:qcm_cosmos}. In this figure we display the apparent $i$-band magnitude
and the $i-j$ color index against the apparent 2D axis ratio \qdd\ in three redshift bins.

We see at all redshifts that galaxies with small apparent axis ratios are significantly redder (i.e.
higher color index) and dimmer than those with apparent axis ratios close to unity. These effects can
be expected from the extinction by dust in the interstellar medium of the source galaxy, as the pathway
of light through the dust of the source towards the observer is longer for discs which are seen edge-on
than for face-on objects (i.e. $q_{2D} << 1$ and $q_{2D}\simeq 1$ respectively).
As a consequence dust extinction can shift discs with inclined orientations below the apparent magnitude limit
(marked as solid red horizontal line in Fig. \ref{fig:qcm_cosmos}), in particular at high redshifts \citep[e.g.][]{Binney81, Huizinga92}.
This effect can introduce a redshift-dependent bias in the observed distribution of axis ratios
towards apparently round (face-on) discs, which could be mistaken for an evolution in the
intrinsic shape distribution, if not taken properly into account. We demonstrate this
effect by applying a cut at $m_i=23$, shown as dashed blue horizontal line in Fig. \ref{fig:qcm_cosmos}.
The apparent evolution of the axis ratio distribution introduced by this cut can be seen in the bottom
panels of Fig. \ref{fig:qcm_cosmos}.

An additional problem caused by this selection effect is that orientations of galaxies in samples
affected by the apparent magnitude cut cannot be expected to be randomly distributed with respect
to the observer, which violates a basic requirement for the 3D shape reconstruction method employed in this work.
In order to mitigate these biases we select objects with absolute $i$-band magnitudes
below $-21.5$, which appear not to be affected by the apparent magnitude cut of $m_i<24$
for $z<1.0$, used for selecting our volume limited sample.

\section{Dependence of galaxy shapes on color}
\label{app:shape_filter_effect}

\begin{figure}
\centering\includegraphics[width=8 cm, angle=0]{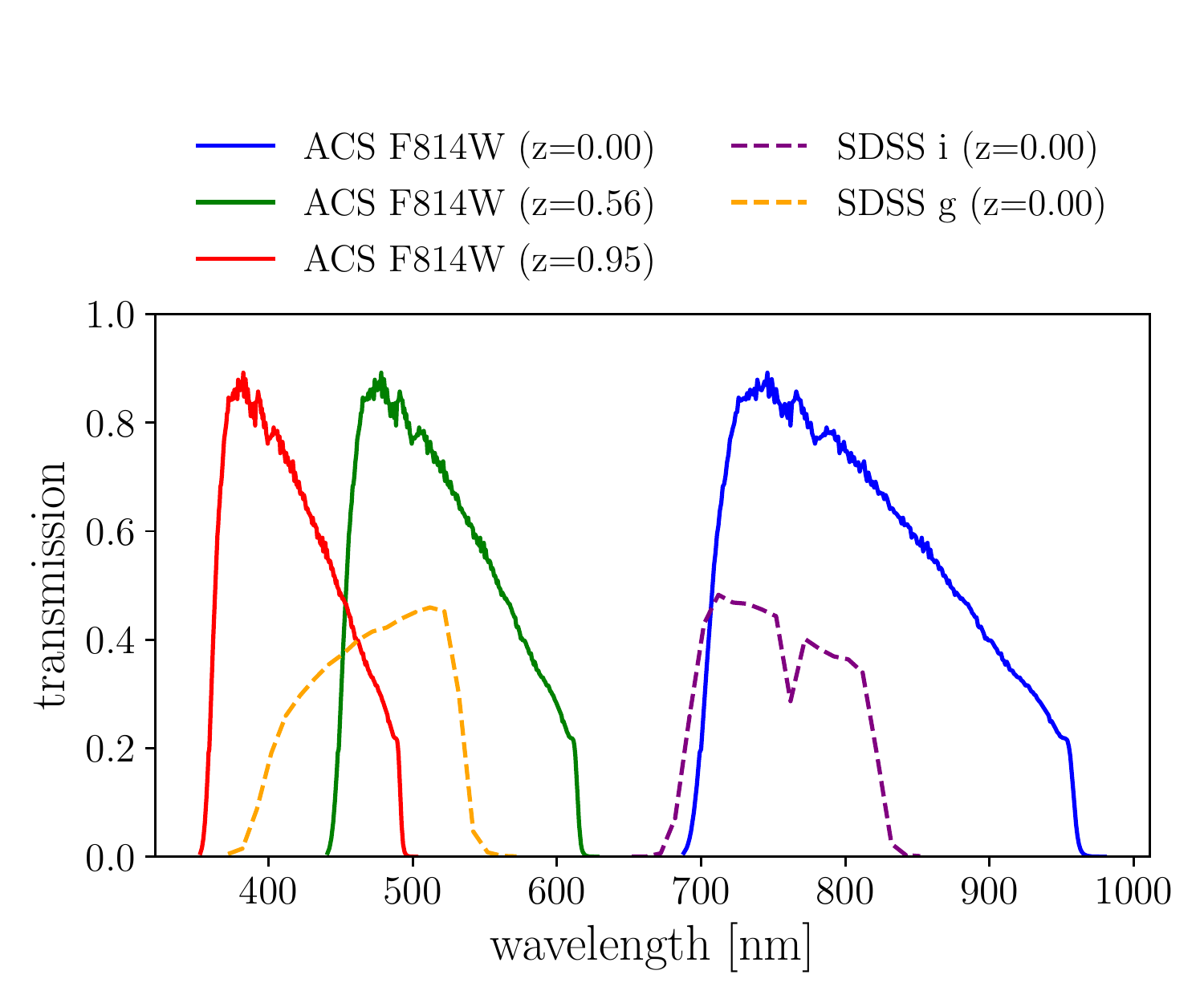}
\caption{Filters of the SDSS $g$- and $i$-bands and the ACS F814W filter
used for the imaging in the COSMOS survey (dashed and solid lines respectively).
The F814W filter covers the near infrared wavelengths at $z=0.0$, which corresponds to
red and green wavelengths at the lowest and highest redshift bin used in our analysis
(with mean redshifts of $z=0.56$ and $z=95$ respectively).}
\label{fig:filters}
\end{figure}

\begin{figure}
\centering\includegraphics[width=7.0 cm, angle=0]{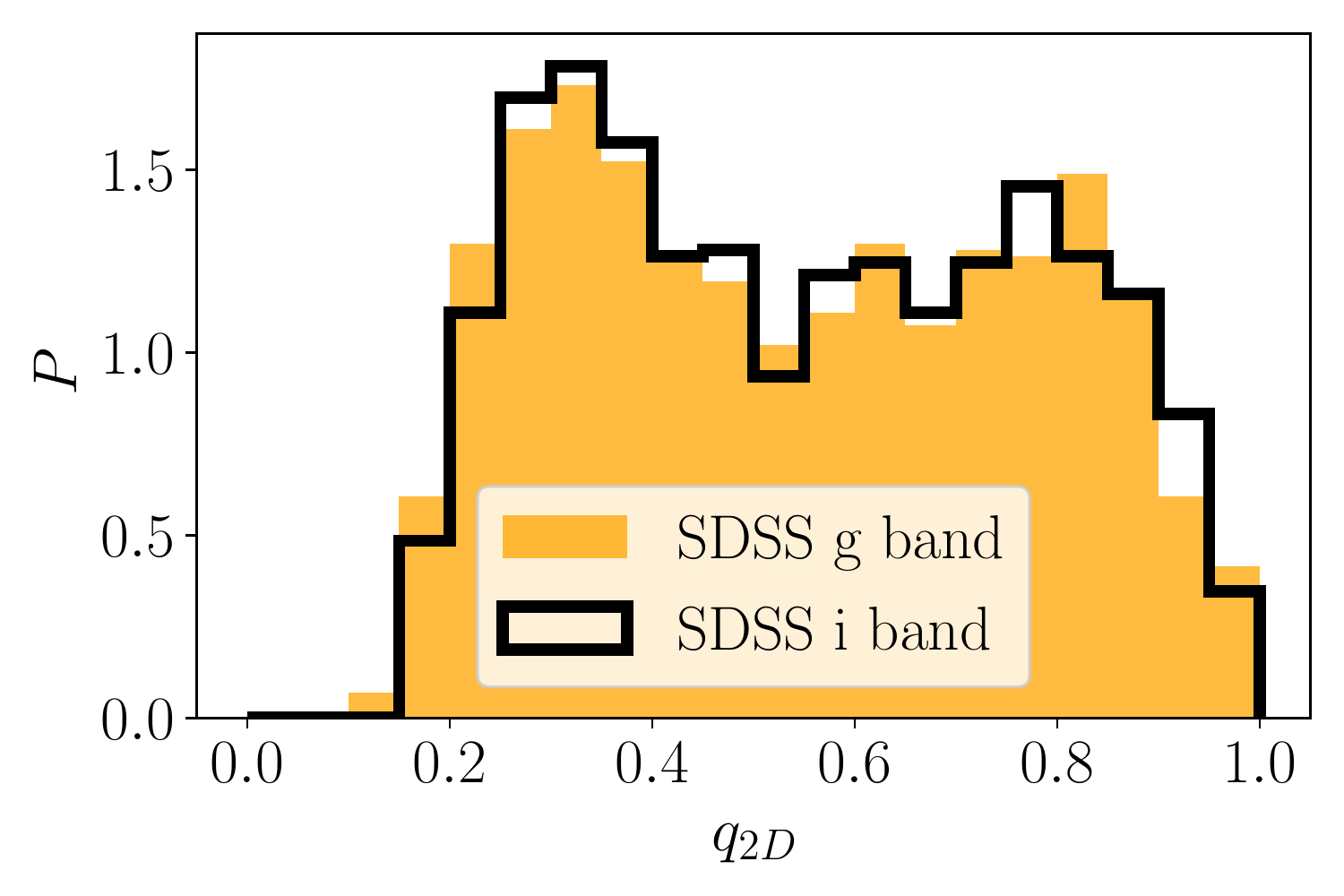}
\caption{Distribution of apparent 2D axis ratios, measured from synthetic images of disc galaxies
in the Illustris TNG100 simulation at $z=0.0$ in the SDSS $g$- and $i$-bands.}
\label{fig:q2d_filter}
\end{figure}

The galaxy shape measurements used in this work are based on ACS imaging in the F814W filter.
This filter corresponds to a different rest-frame wavelength range at each source redshift,
which can affect the observed shape of a given galaxy if its color is not uniformly
distributed (for instance, due to extinction and reddening by dust or patchy star formation).
We study the impact of this systematic effect on the distribution of apparent axis ratios
measured from second-order moments in synthetic images of disc galaxies from the TNG100
simulations at $z=0.0$, produced by \citet[][see Section \ref{sec:data:sims:axes_ratios}]{Rodriguez19}.
These measurements were performed in the SDSS $i$- and $g$-bands,
which correspond roughly to the ACS F814 band at $z=0.0$ and $0.5 < z < 1.0$
respectively, as illustrated in Fig. \ref{fig:filters}.
In Fig. \ref{fig:q2d_filter} we show that the change of the apparent axis ratio distribution is weak,
compared to the shot-noise errors which we expect for our COSMOS samples (see Fig. \ref{fig:q2D_fits_cosmos}).
We find the same result when using PSF corrected axis ratios obtained by \citet{Rodriguez19}
from \sersic\ profile fits to the same synthetic images.
These findings line up with those from \citet{Georgiou19}, who report a minor difference
between galaxy ellipticities measured in different filters of the KiDS survey. They
are further consistent with from \citet{Ryden06}, who shows that axis ratios measured in the $K_s$ and $B$ band
are strongly scattered, but not biased. We therefore do not expect
our results to be significantly affected by the fact that the ACS F814W filter probes
different rest-frame wavelengths at different redshifts.

\section{Dependence of parameter contours on noise in the 2D axis ratio distribution}
\label{app:params_randsamp}

We aim to test in this section if the strong variations of the
posteriors from different redshifts samples, shown in the central and bottom
panels of Fig. \ref{fig:Pqr_params_corner}, can be expected if the axis ratios
of galaxies in these samples are drawn randomly from the same redshift independent
axis ratio distribution.

\begin{figure}
    \centering\includegraphics[width=8 cm, angle=0]{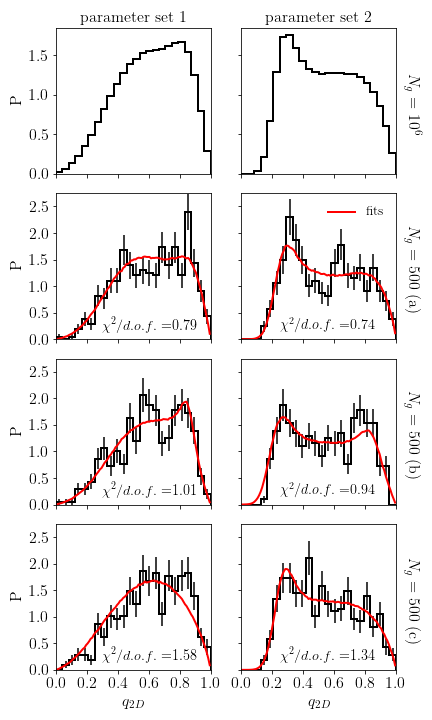}
    \caption{2D axis ratio distributions, generated from our model for the 3D axis ratio distribution
    for two different sets of parameters.
    The top panels show results for the parent samples, consisting of $10^6$ objects.
    The bottom panels show results from three random sub-samples (a,b,c) of the parent sample, consisting of
    $500$ objects each. Red lines show fits from 3D axis ratio distribution model to each sub-sample.
    }
    \label{fig:q2D_fits_randsamp}
\end{figure}

\begin{figure}
\centering\includegraphics[width=8 cm, angle=0]{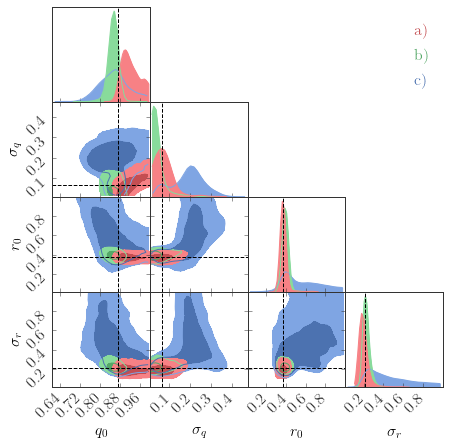}
\centering\includegraphics[width=8 cm, angle=0]{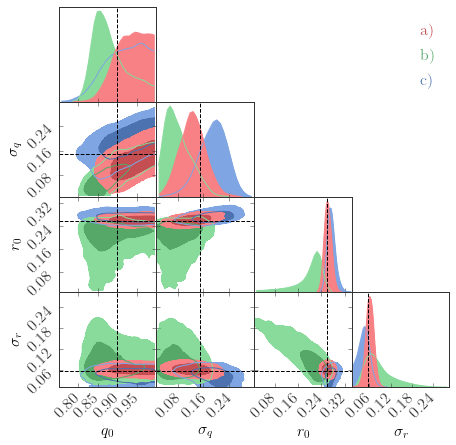}
\caption{
Posteriors of the parameters of the Gaussian model for the 3D axis ratio distribution,
derived from the 2D axis ratios of projected ellipsoids, whose 3D axis ratios
follow the same Gaussian model (Fig. \ref{fig:q2D_fits_randsamp}). Results are
shown for the three random sub-samples (a,b,c). Light and dark areas mark $95\%$ 
and $68\%$ confidence levels respectively. The parameter values used
for generating the parent samples are marked as
horizontal and vertical lines. Top and bottom panels show results for
the model parameter sets $1$ and $2$ respectively.
}
\label{fig:contours_randsamp}
\end{figure}

We tackle this question with a numerical experiment which is based on two sets of $10^6$
artificial ellipsoids whose 3D axis ratio distributions we generate with our Gaussian model
from eq. (\ref{eq:P3d}) and (\ref{eq:P3dnorm}). The parameters used to generate
these two samples are the best fit values for the high and low mass samples in the
entire redshift range of the COSMOS main sample, given in Table \ref{tab:params_rec_cosmos}.
These low and high mass parameters are referred to as parameter set $1$ and $2$ respectively.
The distributions of 2D axis ratios from the corresponding samples are
derived assuming random orientations and are shown in the top panels of
Fig. \ref{fig:q2D_fits_randsamp}.
From each of the two sets we draw three sub-sets with $500$ randomly selected objects,
which corresponds roughly to the size of our observational samples.
The 2D axis ratio distributions of these sub-sets are shown in the three lower panels
of Fig. \ref{fig:q2D_fits_randsamp}.
We proceed by performing the same MCMC parameter inference as for the observational samples,
described in Section \ref{sec:method:infer_params}.
The best fit model prediction for each sub-set is shown as red line in Fig. \ref{fig:q2D_fits_randsamp}.
The corresponding posterior distributions are shown in Fig. \ref{fig:contours_randsamp}.
As for the observational data we find strong variations across results for the different sub-sets.
This finding can be expected, since the posterior is computed directly from the data.
Variations in the data hence translate into variations in the posterior.
The true values of the input model parameters, shown in Fig. \ref{fig:contours_randsamp}
by horizontal and vertical lines, lie within the $95 \%$ confidence intervals, but often
outside of the $68 \%$ intervals.
This test shows that strong variations of the posteriors can be expected for
samples that are drawn from the same underlying distribution if the sample size is
similar to the sizes of our observational redshift-stellar mass samples.

\bsp	
\label{lastpage}
\end{document}